%% file: main.tex
\documentclass[format=acmsmall, review=false, screen=true]{acmart}

\usepackage{booktabs} 
\usepackage{enumerate}
\usepackage{bbding}
\usepackage[ruled]{algorithm2e} 

\SetAlFnt{\small}
\SetAlCapFnt{\small}
\SetAlCapNameFnt{\small}
\SetAlCapHSkip{0pt}
\IncMargin{-\parindent}
\definecolor{lightgray}{gray}{0.9}
\usepackage{xcolor}

\usepackage{colortbl}
\usepackage{pifont}
\usepackage{subcaption}
\usepackage{graphicx}
\usepackage{rotating}
\usepackage{url}
\usepackage{longtable} 
\usepackage{adjustbox}
\usepackage{multirow} 
\usepackage{booktabs} 
\usepackage{forest}
\usepackage{tikz}
\usetikzlibrary{arrows.meta}

\begin{document}
\title[Algorithmic Fairness for Influence Maximization]{Algorithmic Fairness for Influence Maximization: A Survey of Models, Metrics, and Optimization Techniques}

\author{Shashank Sheshar Singh}
\orcid{0000-0003-0909-2258}
\affiliation{%
  \institution{Department of Computer Science and Engineering, Faculty of Engineering and Technology, South Asian University}
  \state{New Delhi}
  \postcode{110068}
  \country{India}}
\email{shashank.sheshar@gmail.com}

\author{Sunil Kumar Meena}
\orcid{0000-0002-7623-7712}
\affiliation{%
  \institution{School of Computer Science Engineering and Technology, Bennett University, Greater Noida}
  \country{India}}
\email{csesunil120meena@gmail.com}

\author{Kuldeep Singh}
\orcid{0000-0001-5289-362X}
\affiliation{%
  \institution{Department of Computer Science, University of Delhi, New Delhi}
  \country{India}}
\email{Corresponding author's email: ksingh@cs.du.ac.in}

\author{Akrati Saxena}
\orcid{0000-0002-7151-6309}
\affiliation{%
  \institution{Leiden Institute of Advanced Computer Science, Leiden University, Leiden}
  \country{Netherlands}}
\email{a.saxena@liacs.leidenuniv.nl}

\begin{abstract}
Influence maximization (IM) has become a basis of modeling information spread across social networks, with applications in viral marketing, public health, and political campaigns. However, classical IM algorithms often prioritize the spread of overall influence and ignore its fairness, resulting in disadvantages for marginalized or underrepresented communities. This survey provides a detailed look at fairness-aware influence maximization, focusing on important challenges, models, and methods designed to ensure fair influence maximization. We categorize existing methods into five groups based on their fairness objectives, including parity, diversity, and utility, and discuss how they handle biased network structure, protected attributes, and seeding constraints. We also examine how fairness-aware influence maximization methods perform on real-world data to highlight the importance of combining fairness with classical influence maximization methods. By bridging the gap between influence spread and social equity, this survey lays the foundation for designing fair algorithms that ensure fair representation in large-scale diffusion processes. The survey concludes by outlining emerging trends, open challenges, and prospective future directions in the domain, highlighting the need for expertise spanning multiple disciplines.
\end{abstract}



\maketitle
\renewcommand{\shortauthors}{S. S. Singh et al.}
\input{samplebody-journals}

\end{document}

%% file: samplebody-journals.tex
\section{Introduction}
\label{Introduction}
Social networks are crucial in today's world due to their capability to shape opinions, share information, and enable people to act together \cite{S1.behaviour-aware}. Social network analysis (SNA) provides tools to identify patterns, communities, and the ways in which people can influence each other in these complex webs of interactions \cite{S3.ye2022influence}. One of the most studied problems, influence maximization (IM), focuses on identifying a small group of key individuals whose acceptance of an idea, innovation, or behavior might lead to the spread quickly throughout the network \cite{10.1145/956750.956769}. This can be used in various real-life applications, including political campaigns \cite{brass2012power}, viral marketing \cite{tang2017profit}, health initiatives \cite{del2020social}, and public awareness \cite{budak2011limiting}. However, as IM methods become increasingly automated and data-driven, issues about fairness, representation, and unintended bias have arisen \cite{G1.10.1145/3308560,G2.10.1145/336642,G3.inproceedings,G4.10.1007/11523468_91}. For instance, classical IM methods generally ignore the minorities, less-connected nodes, or less active communities, which leads to skewed influence spread \cite{G5.10.1145/34596}. 

Some of the challenges are listed below to highlight the importance and need of fair IM methods that focus beyond maximizing influence spread alone.
\begin{enumerate}
    \item Classical IM methods usually favor influential nodes that have a high connectivity or visibility. This leads to ignorance of minority or less-connected communities, even if they are crucial for spreading influence \cite{S3.ye2022influence,S4.jaouadi2024survey}.
    \item Most of the time, classical methods do not consider attributes like a user's ethnicity, gender, age, or financial background. This can lead to campaign methods that mistakenly benefit or exclude some demographic groups, increasing existing disparities \cite{G6.abs-1903-00967,G7.becker2023cost,G8.10.1613/jair.1.13367}. Therefore, classical IM methods try to optimize for maximal influence spread without considering fair distribution among user groups or communities, with limited seed constraints \cite{G20.doi:10.1137/1.9781611976236.69,G21.10.14778/3551793.3551824,G23.rahmattalabi2021fair}.
    \item Most classical methods assume influence probabilities are uniform or randomly distributed across nodes \cite{10.1145/956750.956769}. This simplification ignores the complex processes that people engage in diverse social or cultural settings, especially involving marginalized users \cite{G9.10.5555/3491440}.
    \item Fairness-agnostic IM methods do not consider fairness evaluation metrics to compute equality in influence spread. Therefore, it is hard to find fair outcomes in applications that are based on influence metrics without measuring representational balance or influence disparity \cite{G12.MEENA2025108045,G11.GONG2023119138,G14.10.1145/3664618}.
\end{enumerate}

To tackle these challenges, ensuring that influence is fairly allocated across users and communities has become increasingly important. Fairness-aware IM methods focus beyond the usual goal of influencing the most people, by additionally considering fairness, inclusiveness, and representational balance in influence spread \cite{G24.10.1145/3580305.3599847,G25.10161600,G26.zhang2025fairness}. Therefore, it is necessary to ensure that marginalized or less-connected groups are not systematically overlooked during seed selection or influence campaign \cite{G27.10.1145/3701737}. Recent methods use demographic information, community structures, and individual exposure histories to make decisions that are more fair and inclusive \cite{G28.xiao2025fairness}. 
In addition, the notion of contextual influence maximization emphasizes that fairness is not universal across applications; rather, what constitutes a fair solution depends on the social, cultural, and ethical context of the problem \cite{S2.singh2021influence,S7.singh2022influence}. For instance, fairness requirements in a public health setting may differ substantially from those in social media platforms such as Instagram. In public health campaigns, fairness may entail ensuring that information reaches an equal proportion of individuals across all communities. In contrast, social media information dissemination may prioritize selecting an equal number of seed users from each group, as initial influencers are often compensated or receive early-bird benefits. 



\begin{figure*}[!ht]
        \centering \includegraphics[height=3.3in,width=1.0\textwidth]{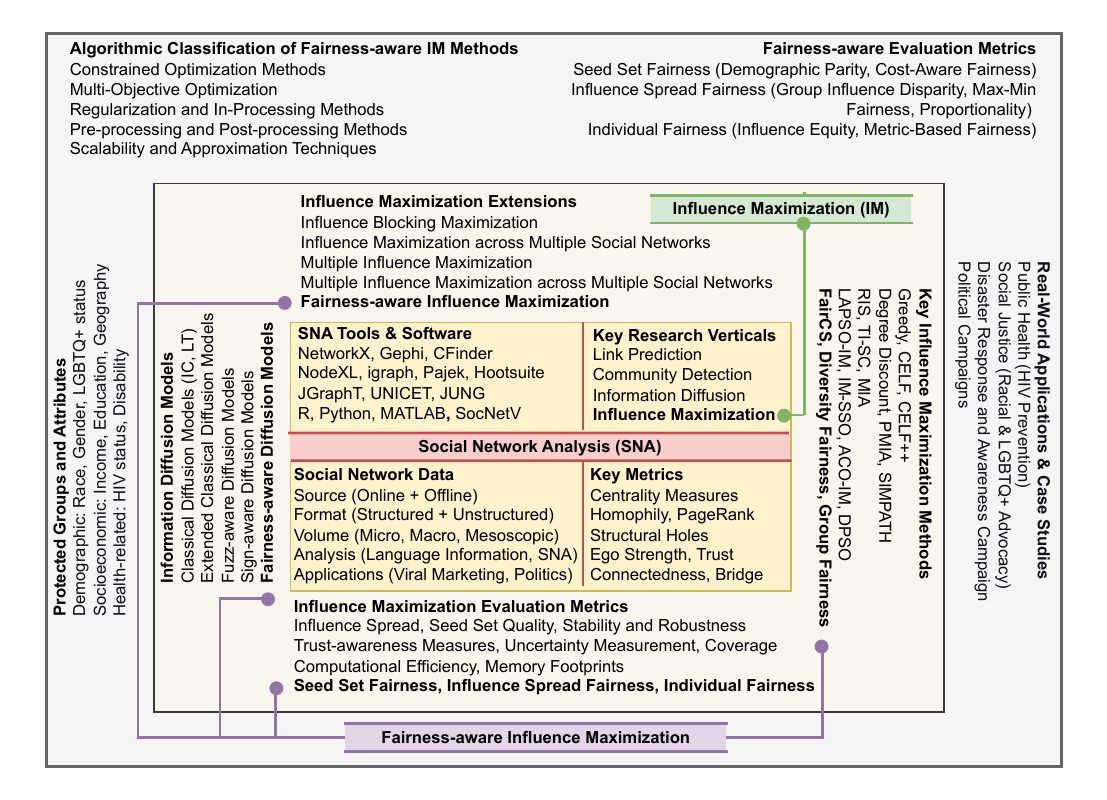}
        \caption{Illustration of the Overview of Fairness-aware Influence Maximization.}
        \label{fig:Overview}
\end{figure*}

In many real-world scenarios, fair IM methods can help in decision making that are more inclusive to individuals and communities. In public health \cite{del2020social}, they can assist in ensuring that critical information like disease prevention and immunization awareness is shared fairly. In education \cite{10.1145/3707463}, they can help to ensure that students from different backgrounds get fair recommendations for scholarships or mentors. In politics and civic life \cite{brass2012power}, they may prevent echo chambers and encourage a variety of opinions in ways that derive people to vote. In marketing \cite{tang2017profit}, fairness-aware models may ensure that product exposure is not limited only to groups of users who are already in charge. Other applications include analyzing criminal behavior and locality \cite{11037728}, job recruitment platforms \cite{hosain2020recruitment}, disaster response systems \cite{jaeger2007community}, credit or loan distribution in financial services \cite{patacchini2017social}, and even countering the spread of harmful misinformation across social media platforms \cite{budak2011limiting}. 

\subsection{Key Contributions}
The overview of the paper is illustrated in Fig. \ref{fig:Overview}, and key contributions of the manuscript are listed as follows.
\begin{enumerate}
    \item This survey is the first to provide a comprehensive overview of how algorithmic fairness has been utilized to achieve inclusive and fair outcomes of classical IM methods. It highlights how fairness-aware attributes can help to go beyond a focus on maximizing reach but also account for equity, inclusion, and representational balance.
    \item The paper presents a taxonomy that categorizes fairness-aware IM methods based on the benefits of algorithmic fairness, including constrained optimization methods, multi-objective optimization methods, regularization and in-processing methods, pre-processing and post-processing methods, and scalability and approximation methods. It also presents a detailed comparative discussion on how fairness-aware IM methods adapt to ensure seed set fairness, influence spread fairness, and individual fairness.
    \item The survey discusses the datasets, evaluation methods, and reproducibility of fair outcomes. It also outlines emerging trends, open challenges, and future directions for fair solutions.
\end{enumerate}

\subsection{Difference from Existing Surveys}
Several surveys have examined IM from algorithmic \cite{S2.singh2021influence,S7.singh2022influence}, computational \cite{SNA3.singh2024quantum,S8.peng2016social,SNA2.singh2025big}, or application-specific viewpoints \cite{S1.behaviour-aware,SNA1.singh2024social,SNA4.10.1145/3539732}, but they generally ignore social diffusion fairness. Prior research has focused on optimizing influence spread or decreasing computing overhead, but it ignores dispersing influence among varied user groups. In \cite{10.1145/3772061}, authors present a discussion on influence maximization on signed and polarized networks. In contrast, this survey examines fairness-aware IM by analyzing recent advances in fairness metrics, equitable seed selection methods, and bias-mitigation methods. Both the surveys \cite{FAIRSNA.10.1145/3653711,Meena.10.1145/3779058} narrowly discussed the fairness of IM. The survey \cite{FAIRSNA.10.1145/3653711} discussed algorithmic fairness with the different aspects of SNA, like fairness in link prediction, IM, community detection, etc. Another survey \cite{Meena.10.1145/3779058}, aimed to discuss the algorithms of IM and also provide a few fairness-based IM algorithms while ignoring the key aspect of fair IM, like metric, type, etc. In the proposed work, we discuss seed set fairness, influence spread fairness, and individual fairness to analyze how classical methods can unintentionally amplify social inequality. To the best of our knowledge, this paper is the first to provide a comprehensive review of fairness IM methods, balance trade-offs between influence spread and fair distribution, recent trends, and open issues. 


The rest of the paper is organized as follows. Section \ref{sec:Background} discusses the fundamental concepts like IM, diffusion models, and algorithmic fairness in SNA. Section \ref{sec:Fair-IM} presents the taxonomy of fairness-aware IM, including objective formulations and evaluation metrics. Fairness-aware IM methods are discussed, reviewed, and compared in Section \ref{sec:Fair-IM_Method}. Section \ref{sec:Datasets} presents a discussion on datasets, evaluation, and reproducibility. Section \ref{sec:Datasets1} presents empirical analysis of the state-of-the-art methods. Section \ref{sec:Challenges} discusses the emerging trends, open challenges, and future directions. Finally, Section \ref{sec:Concluding Remarks} presents concluding remarks on the theoretical and empirical insights of existing works and future directions. 

\section{Preliminaries}
\label{sec:Background}
\subsection{Influence Maximization (IM)}
Influence maximization is a well-known SNA problem that aims to identify a subset of users (often called seed nodes) who can maximize the influence spread in the network \cite{10.1145/956750.956769}. Given a social network represented as a directed graph $G = (V, E)$, where $V$ is the set of nodes (users) and $E$ is the set of edges  (relationships), the objective of IM is to select a seed set $S \subseteq V$ of size $k$ (where $k$ is a seed budget), that maximizes the expected influence spread $\sigma(S)$. Mathematically, the problem is formulated as:
$
S^* = \arg\max_{S \subseteq V, |S| = k} \sigma(S)
$,
where $\sigma(S)$ denotes the expected number of nodes that will be influenced from the initial seed set $S$, based on a given diffusion model like independent cascade (IC) and linear threshold (LT). Since the problem is NP-hard, most solutions rely on approximation algorithms, heuristics, or learning-based techniques to find near-optimal seed sets. These models work well with the perspective of influence spread, but generally ignore the complexities of real-world social systems, like diversity, community imbalance, and differences in socio-demographics.
\vspace{-12pt}
\begin{figure*}[!ht]
        \centering \includegraphics[height=2.5in,width=1.0\textwidth]{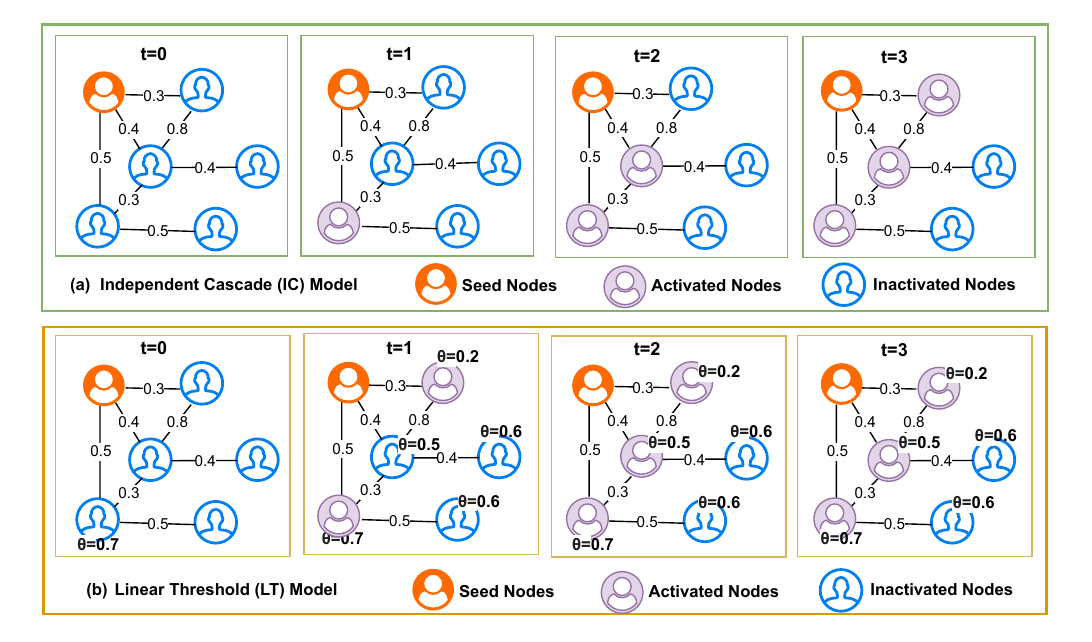}
        \caption{Illustration of the Classical Diffusion Models.}
        \label{fig:IC}
\end{figure*}
\vspace{-18pt}
\subsection{Classical Diffusion Models}
\subsubsection{Independent Cascade Model}
The independent cascade (IC) model is a probabilistic diffusion model often used for analyzing influence diffusion in the social network \cite{10.1145/956750.956769}. This model has an initial set of \textit{seed nodes} activated at timestamp $t=0$. Every activated node at timestamp ($t-1$), has one chance to activate its neighboring nodes with a specified probability at timestamp $t$. Upon successful activation of a neighbor, it may next attempt to activate its own neighbors in the subsequent timestamp ($t+1$). This process repeats until no additional activations are possible. The IC model is extensively employed to simulate the spread of ideas, rumors, or viral marketing initiatives. Fig. \ref{fig:IC}(a) illustrates an example of the diffusion process of the IC model.

\subsubsection{Linear Threshold Model}
The linear threshold (LT) model is a diffusion model utilized to simulate the spread of influence within a network \cite{10.1145/956750.956769}. In contrast to the IC model, the LT model relies on a threshold mechanism. Each node has a defined activation threshold (ranging from 0 to 1), indicating its susceptibility to activation. An initial set of \textit{seed nodes} is activated at timestamp $t=0$. A node activates when a sufficient proportion of its neighbors are active, surpassing its threshold. The process continues until no further activations are possible. The LT model is effective for modelling situations such as decision-making in social networks, where individuals require a specific degree of peer affirmation prior to accepting a new idea or behavior. Fig. \ref{fig:IC}(b) illustrates an example of the diffusion process of the LT model.

\subsection{Classical IM Algorithms}
Classical methods for maximizing influence have used algorithmic methodologies that find a balance between influence spread and computational cost. The greedy algorithm \cite{10.1145/956750.956769} with Monte Carlo simulations is an example of a simulation-based strategy that can guarantee a wide spread of influence under classical diffusion models like IC and LT. However, these methods can be very time-consuming. Kempe et al. \cite{G4.10.1007/11523468_91} present a greedy algorithm for maximizing the spread of influence in a social network using the decreasing cascade model. The decreasing cascade model is a framework for studying influence propagation in networks, where the probability of a node activating another node decreases as more nodes attempt to activate it. This model is characterized by the property that the functions $p_v(u, S)$ are non-increasing in $S$, meaning that if a set of nodes $S$ has already attempted to activate a node $v$, the probability of activation decreases with the addition of more nodes attempting to activate $v$. The greedy algorithm is shown to be a ($1 - \frac{1}{e}$) approximation for selecting a target set of size $k$. Some extensions of the greedy algorithm, like CELF \cite{CELF_Leskovec_2007} and CELF++ \cite{CELF_PP_2011_Goyal}, utilize marginal gain and submodular property to reduce simulation cost. Heuristic approaches like SIMPATH \cite{SIMPATH_2011_Goyal}, DD \cite{DEGREE_DISCOUNT_Chen_2009}, PMIA \cite{MIA_PMIA_Wang_2012_Scalable}, SP1M \cite{SP1M_SPM_2006}, and PageRank \cite{PAGERANK_1999} were added to make solutions more scalable. They use local network features to make quicker influence estimates. Sampling methods like reverse influence sampling (RIS)  \cite{RIS2014Borgs} and TI-SC \cite{2018_TI_SC_CommunityBased_Based} utilize a balance approach to lowering the number of simulations needed and maintaining the influence spread comparable. Additionally, metaheuristic methods like ACO-IM \cite{2020_IM_NatureInsp_AntColony2}, LAPSO-IM \cite{2019_Singh_IM_NatureInsp_LAPSO}, and IM-SSO \cite{2020_Singh_IM_NatureInsp_IMSSO_Spider} have been used to look into large search spaces in complex networks, especially when influence spread functions are not submodular or are limited by other real-world factors. Despite their applicability, these methods assume neutral conditions, overlooking critical factors like fairness, trust, and diversity in influence spread.

Beyond classical methods, IM research has expanded to address the complex and varied characteristics of real-world social networks. Advanced approaches consider the structural complexities and behavioral dynamics. Signed-aware IM methods like PIMSN \cite{2020_Weijia_PIMSN_SignedIM_FirstSignedUnderIC_ProbBasedOnPath_Based} and IMMA \cite{2021_Guo_A_IMMA_SignedIM_UnwillingofInfNodes}  examine both positive and negative relationships. The budgeted IM like BIM \cite{2013_FirstBudgeted_BIM_Nguyen_SOTA_BIM_WeightedDegree}, IMAGE \cite{2020_Song_BudgetedIM_IMAGE_FrameworkForBiggernetwork}, and BCT \cite{2017_Nguyen_IM_BudgetedBIM_BCT_Cost_aware_Targeted_Viral_Marketing_CTVM_GeneralIMBIM_CostAndBudgetBIM} work with limited resources, while fuzzy-based models like CSO \cite{2024_MineCSODynamicNetwork} and CNL \cite{2020_Fuzzy1_CNL_UsingStructuralPropoerty_Logistic_Sparsification_Disadv_AI_Regularisation} handle uncertainty and vagueness in user behavior. Topical methods like C2IM \cite{2019_C2IM_Shashank}, TCIC \cite{DynamicTopic_TCIC_CFDI_FeatureBased_2022_Qin}, and NEW \cite{2020_TopicIM_NEW_UserIntrestAndCharacterstics}, along with location-aware IM methods like DSSA \cite{2019_LocationIM_DSSA_TheoreticalGurantee}, Lazy-Sower \cite{2020_LocationIM_LazySower_IntroGeodemographic}, and LIBM-H/C\cite{2018_LocationIM_LIBM_HC_BlockingInf} consider user's interest and their demographic regions, which makes them more effective to target marketing or promoting events. 
Competitive IM models like RNVGA \cite{2020_RNVGA_CompetitiveIM_Follower_based}, GCW \cite{2021_CompetitiveIM_GCW_GameTheoryBased_AwarenessModel}, and DRHGA \cite{2021_CompetitiveIM_DRHGA_DeepLearingCompetitiveProduct} examine different sources of influence, such as competing products and conflicting ideologies. Multi-product models like MIMGreedy \cite{2016_Sun_MIM_Greedy}, GCW \cite{2021_CompetitiveIM_GCW_GameTheoryBased_AwarenessModel}, and MIM2 \cite{2019_MIM2_IM_Shashank} move a step further by regulating the spread of many products at the same time. 
Multiplex methods like DeepIM \cite{2020_MultiplexIM_DeepIM_PathBased}, T*one \cite{DynamicAndMultiNWIM_TONE_DMNIMP}, and MPRPF \cite{2024_MultiplexIM_MPRPF_UsingMetaPaths} consider different kinds of relationships across layers (like retweet, mention, and follower) to compute influence spread. They give us a better way to model influence. These advanced models push beyond influence spread to address diversity, realism, and effectiveness in influence diffusion across heterogeneous and dynamic settings \cite{DYNAG_dynamic_top_k,DynamicIM_TTG_TTH_TopicBased_Min_2020}. 


\subsection{Defining Fairness-aware Influence Maximization}
Fairness-aware influence maximization extends the classical objective of maximizing information spread by incorporating fairness-focused constraints and values into the diffusion process \cite{G1.10.1145/3308560,G4.10.1007/11523468_91,G12.MEENA2025108045}. Fairness-aware models not only select the nodes with the most influence to maximize reach, but also ensure that influence is spread fairly among different social groups, including communities based on demographics, location, or behavioral attributes \cite{G4.10.1007/11523468_91,G6.abs-1903-00967}. This perspective addresses the concern that classical IM methods may unfairly favor groups who are well-connected or majority groups, unknowingly amplifying existing biases. 

\section{Taxonomy of Fairness-aware Influence Maximization}
\label{sec:Fair-IM}
\subsection{Fairness Notions}
Algorithmic fairness aims to make sure that automated decision-making systems, such as IM methods, treat people or groups fairly because of sensitive characteristics. Some of the key fairness notions are discussed below with their applicability to IM using real-life examples.

\subsubsection{Protected Attributes and Groups}
This fairness can be determined by the question, "What makes a group vulnerable to bias in influence spread?" Protected attributes \cite{SensitiveAttributeOrthogonal,ML_SurveyOnBiasAndFairness} include attributes like race, gender, age, and socioeconomic status that should not adversely affect outcomes. In IM, these attributes may be associated with a person's position in the network. Therefore, marginalized groups have fewer connections, which may lead to disparities in influence spread. For example, if an IM method selects seeds from cities more often than from rural regions because they are better connected, rural populations (a protected group) may not get as much information about medications.

\subsubsection{Group Fairness}
This fairness can be determined by the question, "Are different groups receiving equitable opportunities to be influenced?" Group fairness \cite{FairGraphMining_ASurvey} ensures that outcomes are the same for all protected groups. The possibility of being influenced should be the same for all groups, like if influence is a benefit, like job ads, then all groups should have the same chance to receive it. For example, a viral marketing campaign for a loan program should not favor high-income neighborhoods over low-income ones to ensure demographic parity. Similarly, a scholarship announcement should reach both male and female students equally to ensure equal opportunities.

\subsubsection{Individual Fairness}
This fairness can be determined by the question, "Are similar people treated similarly, regardless of group membership?" Unlike group fairness, individual fairness \cite{IndvidualFairness,MeasureIndvidualFairness} says that persons with identical features, such as, their network position or interests should have the same chance of being influenced. This is usually defined using node similarity metrics. For example, two users with the same number of followers and engagement levels should have same chance of being selected as seeds, even if they are from different demographic groups. 

\subsubsection{Causality-based Fairness}
This fairness can be determined by the question "Is influence spread free from discriminatory mechanisms?" Causal fairness looks at whether protected attributes lead to disparities in influence \cite{G6.abs-1903-00967}, eliminating false correlations. For example, if a social platform's algorithm shows political content to users in some regions more than others (because of past biases), then causal fairness examines whether this user would have seen this content less if they lived in some other places.

\subsubsection{Distributional Fairness}
This fairness can be determined by the question "Is influence allocation balanced across the entire population?" Distributional fairness \cite{10.1145/3652854,FairGraphMining_ASurvey,2025_TNSE_MineMultiFairness_PDTF_IM,DistributionalFairness} does not consider groups or individuals; rather, it looks at the total distribution of influence, usually measured by economic inequality. There are measures like max-min fairness, proportionality, and the Gini index that are used to evaluate distributional fairness. For example, a disaster alert system should prioritize the regions with the least access to information. Similarly, a political campaign should reach demographic groups in the same proportion as their voting population.

\subsection{Fairness-aware IM Objectives}
Formulating fair IM requires translating fairness metrics into actionable optimization goals. Unlike classic IM, which maximizes influence spread $\sigma(S)$, fair IM achieves trade-offs between spread and equity. Below are some common formulations:
\begin{enumerate}
    \item \textit{Constrained Optimization \cite{FairGraphMining_ASurvey}:} 
    Maximize influence spread while ensuring fairness as a hard constraint:
    \begin{equation*}
        \max_{S \subseteq V, |S|=k} \sigma(S) \quad \text{subject to} \quad \text{FairnessMetric}(S) \geq \tau,
    \end{equation*}
    where $\tau$ is a fairness threshold (e.g., minimum group coverage). This ensures fairness but may reduce overall spread.

    \item \textit{Multi-Objective Optimization \cite{MultipleFairness_IndvidualAndGroup,MultipleFairness_TradeOff1}:} 
    Considers fairness and influence as competing objectives, formulating Pareto-optimal solutions:
    \begin{equation*}
        \max_{S \subseteq V, |S|=k} \big(\sigma(S) -\text{Unfairness}(S)\big)
    \end{equation*}
    Methods like weighted-sum scalarization or evolutionary optimization methods \cite{10.1007/s00224-021-10066-5,LIU2024101466} balance these goals without guarantees on both fairness and influence spread.

    \item \textit{Regularized Optimization \cite{FairnessWithRegularisation,FairGraphMining_ASurvey}]:} 
    Defining the objective function by penalizing unfairness:
    \begin{equation*}
        \max_{S \subseteq V, |S|=k} \Big[ \sigma(S) - \lambda \cdot \text{Unfairness}(S) \Big],
    \end{equation*}
    where $\lambda$ balances the fairness and influence spread trade-off. Therefore, the objective function outcome is highly sensitive to $\lambda$ tuning.

    \item \textit{Max-Min Fairness \cite{G6.abs-1903-00967}:} 
    Prioritize the worst-off group by design:
    \begin{equation*}
        \max_{S \subseteq V, |S|=k} \min_{G_i \in {G}} \sigma(S, G_i),
    \end{equation*}
    where ${G_i}$ is the set of protected groups. This method can significantly limit the influence spread based on diverse group impact.

    \item \textit{Composite Objectives \cite{MultipleFairness_IndvidualAndGroup}:} 
    Combine multiple fairness criteria (e.g., group and individual fairness) using weighted sums or constraints.
\end{enumerate}

\begin{table*}[htbp]
\centering
\caption{Comparative Summary of Fairness Constraint in Influence Maximization.}
\resizebox{1.\columnwidth}{!}
{
\begin{tabular}{|p{2.cm}|p{2.8cm}|p{2.7cm}|p{2.8cm}|p{2.8cm}|p{2.4cm}|p{2.2cm}|p{2.5cm}|}
\hline
\textbf{Feature} & \textbf{Demographic \newline Parity} & \textbf{Cost-Aware \newline Fairness} & \textbf{Group Influence \newline Disparity} & \textbf{Max-Min \newline Fairness} & \textbf{Proportionality} & \textbf{Influence \newline Equity} & \textbf{Metric-Based \newline Fairness} \\
\hline
Fairness \newline Notion & Equal seed selection across groups & Equal seed impact relative to cost & Minimize disparity in group spread & Maximize influence to least-advantaged group & Proportional spread to group size or need & Individual-level equal treatment & Similar users treated similarly \\
\hline
Network Type & Static & Static/dynamic & Static & Static & Static/dynamic & Static/dynamic & Static, attributed \\
\hline
Topological Facets & Degree-aware, group distribution & Budget-weighted topology & Inter-group connection density & Group-wise minima & Group centrality alignment & Node-level distribution & Node similarity graphs \\
\hline
Theoretical Guarantees & Group parity bound & Budget-feasibility & Upper-bound on disparity & Guarantees under monotonicity & Relative fairness ratio & Topological assumptions & Metric-preserving  \\
\hline
Measurability & Quantitative & Quantitative & Quantitative & Quantitative & Quantitative & Quantitative & Quantitative or qualitative \\
\hline
Model \newline Dependence & Model-specific & Model-specific (budgeted IM) & Generalized  & Generalized & Needs group data & Model-specific & Depends on user similarity metric \\
\hline
Sample use cases & Affirmative action, equitable outreach & Fair access under budget limits & Community health, education outreach & Crisis response prioritization & Proportional vaccine/disaster aid & Personalized marketing & Privacy-aware targeting \\
\hline
References &\cite{G1.10.1145/3308560}\cite{G20.doi:10.1137/1.9781611976236.69}  &\cite{G6.abs-1903-00967}  & \cite{G2.10.1145/336642}\cite{G27.10.1145/3701737} & \cite{G6.abs-1903-00967}\cite{G16.10.1145/3437963.3441787}\cite{G27.10.1145/3701737} &\cite{G21.10.14778/3551793.3551824}  &\cite{G11.GONG2023119138}  &\cite{G9.10.5555/3491440}\cite{G20.doi:10.1137/1.9781611976236.69} \\
\hline
\end{tabular}
}
\label{tab:fairness_metrics}
\end{table*}

\subsection{Fairness Constraints}
In order to find out what makes an IM method fair, it is necessary to have clearly defined fairness constraints. Fairness in IM needs to consider \textit{network structure}, \textit{diffusion dynamics}, and \textit{disparate impacts on groups or individuals}. Table \ref{tab:fairness_metrics} presents a summary of fairness constraints for the IM problem.

\subsubsection{Seed Selection-based}
Selecting seed nodes is the first step towards fairness. Metrics \cite{G22.10.114424.3383555,FairGraphMining_ASurvey} here check to determine if the seed set $S$ has a fair number of individuals from protected groups, including race, gender, or geographic communities.
\begin{itemize}
    \item \textbf{Demographic Parity \cite{FairGraphMining_ASurvey}}: Ensures seeds reflect group demographics (e.g., if 30\% of the network belongs to group $G_i$, then $\frac{|S \cap G_i|}{|S|} \approx 0.3$).
    \item \textbf{Cost-Aware Fairness \cite{G13.10734665}}: Considers varying costs of selecting seeds from different groups (e.g., identifying influencers from marginalized communities may require higher incentives).
\end{itemize}

\subsubsection{Influence Outreach-based}
In order to measure disparities in influence spread to different groups, we need to aim at fair \textit{outcomes} \cite{G22.10.114424.3383555}, not only fair seeding.
\begin{itemize}
    \item \textbf{Group Influence Disparity \cite{FairGraphMining_ASurvey}}: Compares per-capita influence across groups, e.g., $\frac{\sigma(S, G_i)}{|G_i|} \approx \frac{\sigma(S, G_j)}{|G_j|}$.
    \item \textbf{Max-Min Fairness \cite{G6.abs-1903-00967,G22.10.114424.3383555}}: Maximizes the minimum influence received by any group ($\max_S \min_{G_i} \sigma (S, G_i)$), prioritizing the most underserved.
    \item \textbf{Proportionality \cite{G22.10.114424.3383555,G6.abs-1903-00967}}: Requires influence to mirror group size (e.g., if $G_i$ is 20\% of the network, it should receive $\approx$20\% of the influence).
\end{itemize}
These metrics nearly replicate real-world equity but are difficult to optimize due to the complexity of estimating $\sigma(S, G_i)$ under fairness constraints.

\subsubsection{Individual Characteristics-based}
Fairness can also apply to individuals, not only to groups. The challenge is defining similarity in a networked context.
\begin{itemize}
    \item \textbf{Influence Equity \cite{IndvidualFairness,FairGraphMining_ASurvey,MeasureIndvidualFairness}}: Nodes with similar influence (e.g., centrality, expertise) should have comparable probabilities of being influenced.
    \item \textbf{Metric-Based Fairness \cite{FairGraphMining_ASurvey}}: Adapts a framework, ensuring that nodes close in a fairness-aware metric space experience similar outcomes.
\end{itemize}

\subsection{Algorithmic Approach} 
Fairness-aware approaches are generally categorized into pre-processing, in-processing, and post-processing methods. Pre-processing methods modify the input dataset prior to model training in order to reduce or remove biases, with the goal that the resulting outcomes are fair. Post-processing methods, in contrast, first apply a standard algorithm and subsequently adjust the generated outputs to improve fairness. For instance, seed nodes are selected using different thresholds for different groups to achieve fairer outcomes. In-processing approaches incorporate fairness considerations directly into the learning or optimization procedure, enabling the algorithm to produce fair outputs by accounting for fairness constraints that mitigate the effects of structural biases present in the data. Most existing methods fall under the in-processing category. In the context of fairness-aware influence maximization (IM), these methods largely rely on optimization-based strategies, such as constraint-based formulations and multi-objective optimization, heuristics, regularization techniques, and approximation methods. In the next section, we discuss these approaches in detail.


\section{Algorithmic Classification of Fairness-aware IM Methods}
\label{sec:Fair-IM_Method}
This section reviews the existing literature on fairness-aware IM and categorizes it into six classes. The comparative summary of existing literature is illustrated in Table \ref{tab:MethosSummary} with respect to algorithmic classification, evaluation criteria, datasets, baseline method, research challenges, future directions, and fairness consideration. Table \ref{tab:Fairness_Function} presents a summary of the fairness functions used in the existing literature, which includes fairness functions, parameters, and evaluation metrics.

\subsection{Constrained Optimization Methods}
Constrained optimization methods ensure that every individual is treated fairly by setting boundaries on how many users can represent a group or how much power they can have during seed selection. These methods adopt greedy algorithms (like CELF, CELF++) or integer programming to satisfy fairness requirements while maximizing spread. These methods face scalability challenges for large networks.

\subsubsection{Demographic / Parity-Constrained IM}
The authors of \cite{G1.10.1145/3308560} aim to design fair algorithms for social IM that do not exacerbate demographic inequalities, such as those based on gender or race, in social networks. The authors propose adapting fairness definitions from classification settings to the context of social influence maximization and perform empirical analysis using Instagram data. They develop a mathematical model to compare strategic and non-strategic seeding, using the independent cascade model to quantify disparate outcomes among different populations. The study finds that feature-aware methods can lead to fairer and more efficient outcomes in social IM compared to feature-blind methods. The paper concludes that strategic seeding can achieve both diversity in information diffusion and better efficiency, providing a mathematical guarantee of optimality and fairness. The authors of \cite{G2.10.1145/336642} address how to maximize social influence in biased networks without exacerbating existing biases, particularly focusing on achieving both fairness and efficiency in the selection of early adopters or seed nodes. They introduce a theoretical model of biased networks using a biased preferential attachment model to study the effects of diversity on IM. It proposes diversity-enhancing seeding strategies, such as parity and diversity seeding, to improve both outreach and fairness, validated through experiments on the DBLP dataset. The study finds that diversity seeding requires fewer seeds (9,118) compared to parity seeding (11,286) to outperform agnostic seeding. They conclude that incorporating sensitive attributes in seed selection algorithms can lead to more equitable outcomes and suggest future research on more complex algorithms and diffusion processes. In another study  \cite{G3.inproceedings}, the authors address the problem of defining and understanding social capital within online social networks, as well as the methods to stimulate and measure it.

\subsubsection{Maximin / Group-Fairnes Constrained IM}
In \cite{G6.abs-1903-00967}, the authors discuss the problem of fair resource allocation in influence maximization, focusing on ensuring that all groups receive a fair share of influence, particularly marginalized groups. They introduce two fairness concepts: Maximin Fairness and Diversity Constraints. Maximin Fairness is a concept that aims to improve the conditions for the least well-off groups by maximizing the minimum influence received by any group. This is formalized in the utility function as  $U_{Maxmin}(A) = min_i \frac{IG_{C_i(A)}}{C_i}$,
where $IG_{C_i(A)}$ represents the expected number of activated vertices from group $C_i$ when seed nodes $A$ are activated. The goal is to maximize the overall influence while ensuring that the least advantaged group receives a fair share of influence. Diversity constraints ensure that each group receives influence proportional to its size and internal connectivity. They define diversity constraints as conditions that guarantee that the influence received by each group is at least equal to what they could achieve on their own with a fair allocation of resources. Specifically, for each group $C_i$, the expected influence $IG_{C_i(A)}$ from a seeding configuration $A$ must satisfy the condition $G_{C_i(A)} \geq IG$. An algorithmic framework is proposed to solve these fairness problems, reducing them to multi-objective submodular optimization. The proposed algorithm offers better approximation guarantees and runtime than existing methods. They find that standard influence maximization techniques often neglect smaller groups, causing fairness violations, which the proposed algorithm significantly reduces at a small cost to overall utility.

\subsubsection{LP / Approximation-based Constrained IM}
In \cite{G21.10.14778/3551793.3551824}, the authors address the FairCS problem, which aims to select a subset of candidate edges to maximize content spread while ensuring fairness. It is proven to be NP-hard and inapproximable in polynomial time unless P = NP. They propose a randomized approximation algorithm based on LP-relaxation and randomized rounding, achieving an approximation ratio of ($1 - e^{-1}$) for content spread and $\frac{(2e)}{(1 - e^{-1})}$ for fairness. They also introduce optimizations to enhance algorithm performance and a scalable algorithm that dynamically adds sampled nodes to the graph, allowing for easier problem-solving. Experiments conducted on real datasets demonstrate that their algorithms effectively reduce unfairness to less than 0.05\% while significantly increasing content spread. They emphasize the importance of fairness in content spread, particularly in social networks, where minority groups may be disproportionately affected by content dissemination. They also highlight that existing algorithms often fail to address fairness adequately, making their approach novel in the context of content spread optimization.

\subsection{Multi-Objective Optimization Methods}
Multi-objective methods investigate Pareto-optimal solutions that balance fairness and influence spread as competing goals. These methods evaluate trade-offs without rigid constraints by providing flexibility with right balance between fairness and effectiveness.

\subsubsection{Pareto-based Fairness–Influence Trade-off}
The authors of \cite{G16.10.1145/3437963.3441787} present a novel IM method focusing on optimizing the diffusion of content by selecting an initial set of users as seeds for spreading information by minimizing disparities in information exposure across categories. This is particularly relevant for applications like viral marketing, where reaching a broad audience is crucial. They emphasize the importance of balance and fairness in the seed set to ensure equitable information exposure among different user categories. They define homophily as the tendency of similar individuals to connect and influence each other, which complicates the influence maximization process. The results indicate that achieving balance in the seed set can enhance overall influence, especially in networks with high homophily. Overall, the paper contributes to the understanding of how to implement fair influence maximization strategies in social networks, addressing both theoretical and practical implications. Teng et al. \cite{G5.10.1145/34596} address the challenge of gender disparity in social media influence, specifically the glass ceiling effect that limits female visibility and influence. It aims to develop a seeding strategy that maximizes information spread while achieving a target gender ratio. They propose the Disparity Seeding framework, which uses gender-aware metrics like Target HI-index and Embedding index to rank influential users and select seeds. The Target HI-index is an extension of the traditional H-index that incorporates interaction intensity to quantify overall network visibility, while the Embedding index is derived via state-of-the-art graph neural networks. This framework employs simulations to compare with target-agnostic algorithms, showing effectiveness in achieving target gender ratios and maximizing spread. The core process involves selecting a suitable seed set through a simulation-based learning approach, ensuring that the target gender ratio is met effectively while maximizing information dissemination. 

In \cite{G20.doi:10.1137/1.9781611976236.69}, authors discuss the concept of information unfairness in social networks, which refers to the unequal access to information among different demographic groups. It highlights that individuals in networks often make significant life decisions based on the information they receive, and if this information is not distributed fairly, it can disadvantage certain groups, such as women and minorities. They propose a criterion to measure this unfairness and present an algorithm called MaxFair to improve information flow among groups. The study emphasizes the importance of understanding whether information flows equally to all groups, particularly those defined by sensitive attributes like race or gender. The paper also introduces an accessibility matrix to evaluate the flow of information between nodes in a network, aiming to identify disparities in information access. Furthermore, it discusses how a fair network should ensure that the flow of information between different groups is equitable, and proposes methods to quantify this unfairness. The authors of \cite{G28.xiao2025fairness} addresses the critical issue of fairness in the context of the influence blocking problem (IBM), which has been underexplored in existing research. It evaluates fairness using the maximin fairness concept, aiming to maximize the number of nodes protected from misinformation in the community with the lowest saving ratio. The main contributions of the study include: 1) the formal definition of the fairness-aware IBM (FIBM) problem, which is NP-hard for the Competitive Independent Cascade model; 2) the introduction of a fairness-aware blocking strategy that balances overall influence blocking effectiveness and protection fairness across communities; 3) the proposal of a sampling-based approximate algorithm called FIBMM, which is applicable for large networks. Extensive experiments on six real-world datasets demonstrate the effectiveness of FIBMM. The paper emphasizes the importance of incorporating fairness into the IBM framework to address disparities and develop balanced strategies that protect all communities, regardless of their size or influence. Fairness-blind approaches may leave marginalized groups more vulnerable, highlighting the need for fairness-aware strategies in socially sensitive decision-making scenarios.

\subsubsection{Metaheuristic Multi-objective}
In \cite{G11.GONG2023119138}, authors of introducing Fairness in Influence Maximization Problem (FIMP). The influence of a seed set S is defined as the fraction of influenced nodes in a network, represented mathematically as $\sigma(S) = \frac{\sum_{i=1}^{i=m}g_i}{|V|}$. The objective is to find a seed set $S$ of size $k$ that maximizes this influence within a time limit $T$. They highlight the importance of fairness in the context of IMP, noting that the fairness gap between different groups has garnered attention. It discusses various optimization objectives for FIMP, such as maximizing the minimum fraction of influenced nodes across all groups. The authors propose a multi-objective optimization framework called FIMMOGA, which incorporates prior knowledge to achieve a beneficial trade-off between influence and fairness. The framework aims to provide diverse solutions that balance both objectives effectively. The paper also emphasizes the need for accurate measurements of fairness, utilizing the Gini coefficient as a metric. Overall, the study aims to address the challenges of optimizing both influence and fairness simultaneously, which is a complex task due to the NP-hard nature of the IM.

\subsubsection{Welfare-based Optimization}
In \cite{G8.10.1613/jair.1.13367}, authors discuss the challenge of achieving fairness in IM, specifically focusing on the maximin criterion to ensure equitable information spread across communities. It explores the use of randomized strategies over deterministic ones to improve fairness in the selection of seed nodes. They introduce two probabilistic variants of the maximin problem: node-based and set-based approaches, allowing for approximation algorithms with a constant factor of $1 - \frac{1}{e}$ minus an additive error. For the node-based problem, a polynomial-sized linear program is used for approximation, while the set-based problem employs a multiplicative-weight routine. Experimental results show that the ex-ante fairness values of the proposed probabilistic strategies are significantly larger than those of previous methods, indicating the effectiveness of randomization in achieving fairness. The authors of \cite{G19.10.1145/3308558.3313680} emphasize a welfare approach to fairness, which focuses on optimizing access to information rather than ensuring equal access for all individuals. This approach is gaining traction as it allows for the modeling of utility received by individuals, rather than relying solely on sensitive features like race. The authors reference the difference principle by Rawls, which aims to benefit the least advantaged individuals by maximizing their probability of receiving information. The paper also connects to existing literature on algorithmic fairness, highlighting the need for decision-making systems that are non-discriminatory. Furthermore, it draws on Granovetter's work on the strength of weak ties, illustrating how network positions can create advantages or disadvantages in social contexts. Overall, the paper seeks to quantify aspects of social capital and proposes methods to mitigate the effects of inequality in information access. 

Rahmattalabi et al. \cite{G23.rahmattalabi2021fair} discuss various fairness notions in the context of influence maximization and evaluate their properties against established welfare principles. They highlight the following key concepts:
\begin{itemize}
    \item \textit{Maximin Fairness} aims to improve the minimum influence received by any community, but may degrade total influence due to its stringent requirements.
    \item \textit{Diversity Constraints} ensure that communities receive resources proportional to their size, but do not explicitly reduce inequality.
\end{itemize}
They emphasize the trade-off between efficiency and fairness, which can be controlled by an inequality aversion parameter. The Price of Fairness (PoF) is introduced as a measure of efficiency loss when fairness is enforced. They conclude that fairness is domain-dependent and requires careful consideration of community structures and decision-making processes.
In \cite{G15.9863801}, authors extend the notion of group fairness in IM by incorporating the speed of information propagation as a metric. It models group-fairness-aware IM as a multiobjective problem and proposes a discrete multiobjective optimizer called SetMOGWO to maximize both influence and fairness metrics. They emphasize the importance of maximin fairness, which aims to maximize the minimum influence received by any group. It also discusses group rationality, ensuring that no group gains more influence than its proportional allocation of resources. Experimental results examine the cost of ensuring group fairness and the relationships between fairness concepts and node attributes. 

\begin{table*}[htbp]
\centering
\caption{Illustration of the Fairness Function Used in Existing Literature. }
\resizebox{1.\columnwidth}{!}
{
\begin{tabular}{|p{1.cm}|p{6.4cm}|p{10.6cm}|p{2.5cm}|}
\hline
\textbf{Study} & \textbf{Fairness Function Used} &\textbf{Fairness Parameter} & \textbf{Fairness Metric} \\ 
\hline
\cite{G1.10.1145/3308560} &$\beta(B) = 1+\frac{1}{\frac{r \rho}{2\alpha+2(1-\alpha)\rho}+\frac{1-r}{2\alpha \rho+2(1-\alpha)}}$, \newline $\beta(B) = 1+\frac{1}{\frac{(1-r) \rho}{2\alpha \rho+2(1-\alpha)}+\frac{r}{2\alpha +2(1-\alpha) \rho}}$ &The fairness parameter is the proportion of communities represented in the seed set and the outreach achieved across communities. It ensures that the product or information being spread reaches the population. &Equality of proportions of communities\\
\hline
\cite{G2.10.1145/336642} &$S_{K^R(n)}^{R} \cup S_{K^B(n)}^{B} = \{ v \in R | deg(v) \geq K^R(n)\} \cup \{ v \in B | deg(v) \geq K^B(n)\}$ &The homophily parameter is described as a measure of the tendency of nodes to connect with others of the same group. &Parity, Diversity  \\
\hline
\cite{G5.10.1145/34596} &$H(v_i)= max_n min_{n \in I^{+}(N(v_i,n),n)}$ &Fairness parameter aims to ensure that the selected influential users align with the desired gender representation. &Target HI-index  \\
\hline
\cite{G6.abs-1903-00967} & $PoF_{Rational} = \frac{I_{OPT}^{G}}{I_{Rational}^{G}}$ &Fairness parameter involves ensuring that the influence received by each group is proportional to their population size. & Diversity Fairness \\
\hline
\cite{G7.becker2023cost} &$max_{S^* \in S}\{\sigma(S^*): \exists_{\gamma}: \sigma_{C_i}(S^*) = \gamma \text{ for all } C_i \in C\}$ &Fairness parameter indicates that it is a measure of how closely algorithm adheres to the ideal of equal coverage across different groups. & Fairness Violation \\
\hline
\cite{G8.10.1613/jair.1.13367} &$opt_p(G,C,k) = max_{p \in P} min_{G \in C} E_{S\sim P}[\sigma_G(S)]$ &Fairness parameter indicates that it ensures equitable distribution of resources among nodes, specifically in the context of maximizing spread. &Price of Fairness  \\
\hline
\cite{G9.10.5555/3491440} &$\frac{I_A}{|A|} \approx\frac{I_B}{|B|}$ &Fairness Parameter relates to ensuring that the spreading process does not favor one group over another. &Equal influence distribution\\
\hline
\cite{G10.MEENA2025127445} &$MF(S)=min_i \frac{I_{G_{C_i(S)}}}{|C_i|}$, $PoF=\frac{I^f}{I^{S^*}}$&Maximin fairness returns the minimum influenced proportion of the community, ensuring inclusivity by guaranteeing that all communities are influenced. &Price of Fairness, Maximin  \\
\hline
\cite{G11.GONG2023119138} &$F(S)=(1-gini_{G,S,p,k,T,M})$, $gini_{G,S,p,k,T,M}= (\int_0^1 xdx -\int_0^1 u(x)dx) / \int_0^1 xdx$ &The fairness parameter quantifies fairness by evaluating the influence gap between the maximum and minimum fractions of influenced nodes, aiming to minimize this gap for equitable distribution. & Gini coefficient \\
\hline
\cite{G12.MEENA2025108045} &$\varphi(S) =argmax_{s\subseteq V^t,|S|\leq k} \{(1-\lambda)\frac{|\sigma(S)|}{|V^t|}+\lambda \frac{\xi(\sigma(S))}{\xi(V^t)}\} $ &The parameter $\lambda \in [0, 1]$ is a tuning parameter that balances the trade-off between the influenced nodes and the diversity, i.e., influenced communities. &Diversify Fairness  \\
\hline
\cite{G13.10734665} &$\varphi(S) =argmax_{s\subseteq V^t,Cost(S)\leq B} \{(1-\lambda)\frac{|\sigma(S)|}{|V^t|}+\lambda {\xi(\sigma(S))}\} $ &Diversity parameter $\lambda \in [0, 1]$ is a tradeoff between diversity and influenced nodes, indicating a balance between these two aspects & Diversify Fairness \\
\hline
\cite{G14.10.1145/3664618} & $\varphi(S) =argmax_{s\subseteq V^t,|S|\leq k} \{(1-\lambda)\frac{|\sigma(S)|}{|V^t|}+\lambda \frac{\xi(\sigma(S))}{\xi(V^t)}\} $ &Parameter for controlling diversity $\lambda \in [0, 1]$ which aims to balance the influence and diversity of nodes in a dynamic social graph. &Diversify Fairness  \\
\hline
\cite{G15.9863801} &$GR(S) = \begin{cases}
        IG(S)    & \text{if } IG, C_i(S) > IG[C_i](k_i) \forall i\\
        0    & \text{Otherwise}
       \end{cases}$ &The fairness parameters are used to evaluate the fairness in resource allocation among different groups &Maximin, Group Rationality \\
\hline
\cite{G16.10.1145/3437963.3441787} &$F(S)=(1-\lambda)\frac{f(S)}{|V|}+\lambda \frac{\sqrt{\gamma M(S)} + \sqrt{(1 -\gamma) M(S)}}{\sqrt{\gamma p_M^n }+\sqrt{(1-\gamma)(1-p_M)^n }}$ &Fairness parameter $\lambda$ provide the trade-off between influence and categorical balance, with higher values promoting a more equitable distribution. &Categorical Balance \\
\hline
\cite{G17.9835234} &$extmax_{i,j}(\frac{f_t(S,V_i,G)}{|V_i|}-\frac{f_t(S,V_j,G)}{|V_j|})$ &The Fairness Parameter $c$, which indicates the maximum level of allowed disparity among the groups, constrained within the range [0, 1]. &Maximum Disparity \\
\hline
\cite{G18.87.article} &Fairness parameter ($\beta = \tau /2$) is used to estimate the number of meeting attempts &The fairness function is evaluated through MC simulations to ensure fair comparisons among different algorithms. &Influence Spread \\
\hline
\cite{G19.10.1145/3308558.3313680} &$\mu(S^*, V^{'}) - \mu(S^*, V) > \mu(S, V^{'}) - \mu(S, V) $  &It emphasizes a welfare approach to fairness, which is gaining popularity in the literature, as it seeks to provide benefits to the least-advantaged individuals in terms of access probability. &Welfare function \\
\hline
\cite{G20.doi:10.1137/1.9781611976236.69} &$IU_{G,p,k} = Max(Dist(D_{f1g1},D_{f2g2}): f_1,f_2,g_1,g_2 \in 1, \dots, t $ &Fairness parameter is characterizing the flow between nodes from each pair of groups using accessibility matrix and the distances. &Information Unfairness  \\
\hline
\cite{G21.10.14778/3551793.3551824} &$\frac{1}{g_i} \sum_{v_l \in g_i} F(v_l, E_p) = \frac{1}{g_i} \sum_{v_l \in g_j} F(v_l, E_p) $ &Fairness Parameter is defined as expected total content received by each group proportional to its size, ensuring that this is equal across all groups. &Ratio of content received to its size \\
\hline
\cite{G22.10.114424.3383555} & \textit{Equality}: $|S \cap V_C|/k \approx |C|/|V|$, \newline \textit{Diversity}: $ E[I(S,V_C)]\geq OPT_C$, \newline \textit{Equity}: $ E[I(S,V_C)] / E[I(S,V)] \approx |C|/|V|$  &The fairness parameter is represented by the price of fairness, which is defined as the ratio of optimal unconstrained total influence to the total influence subject to fairness constraints. &Equality, Equity, Maximin, Diversity \\
\hline
\cite{G23.rahmattalabi2021fair} &$extPoF = 1- \frac{OPT_{fair}}{OPT_{IM}}$  &The fairness parameter $\alpha$ directly influences the utility gap, which is the difference in utility between the most and least advantaged communities. &Price of Fairness (PoF) \\
\hline
\cite{G24.10.1145/3580305.3599847} &$f_s = \frac{2}{1+exp(CV)}$ &Fairness parameter serves as a penalty factor to down-sample influencers who initiated unfair information diffusion with respect to attribute $s$. &Proportion of influenced nodes \\
\hline
\cite{G25.10161600} &$GE = \frac{1}{k.w.(w-1)} \sum_{i=1}^{i=k} (\frac{R_i^t}{R^t})^{w-1}$ &The fairness parameter is represented by the generalized entropy (GE) metric. &Fairness Index (GE) \\
\hline
\cite{G26.zhang2025fairness} &Any utility vector $u \in R^m$, if $u_j - u_i > \epsilon > 0$, then $u + \epsilon e_i - \epsilon e_j$ s considered a fairer solution. &The fairness parameter involves ensuring that the solution satisfies criteria such as efficiency (Pareto optimality) and impartiality (equal treatment). &Utility Gap, Price of Fairness (PoF) \\
\hline
\cite{G27.10.1145/3701737} &$O^{PR}{S}=\sum_{C_i \in C} |V_{C_i}|. \varphi(u(C_i, S))$ &Fair utility $\varphi(.)$ subject to specific conditions ensuring that poorer groups receive larger increases in fair utility. &Concave Fairness (CFF) \\
\hline
\cite{G28.xiao2025fairness, saxena2025dq4fairim} &$\tau(S)= min_{C_i \in C} \prod_{C_i} (S)$ &Fairness parameter is defined by the maximin value $\tau(S)$, which serves as a basis for achieving maximin fairness. &Maximin Fairness \\
\hline
\cite{khajehnejad2022crosswalk} &$Disparity= Var(\{IG_{C_i(S)}\})$ &Fairness Computes the diaprity in influence spread amoung different groups. & Disparity Fairness \\
\hline
\end{tabular}
}
\label{tab:Fairness_Function}
\end{table*}

\subsection{Regularization Methods}
Regularization adds fairness directly to the objective function using a penalty, like closing the disparities in group influence. For a more fair spread, in-processing methods modify diffusion dynamics, including biased edge weights. Their performance depends on how well we select the hyperparameters, and may lack interpretability. 

\subsubsection{Demographic Parity Regularization}
The authors of \cite{G7.becker2023cost} present the influence maximization problem under the constraints of demographic parity. It analyzes the trade-offs between fairness and efficiency, demonstrating that a strict approach to fairness can lead to a substantial loss in overall spread and time complexity of algorithms. They prove that the price of fairness may be unbounded in this context, indicating that enforcing fairness can severely impact performance. They introduce three optimization problems aimed at maximizing overall spread while ensuring strict levels of demographic parity fairness, allowing users to choose their desired level of fairness. Demographic parity is defined as equality in the probability of being selected conditioned on group membership. It is discussed as a fairness criterion in machine learning, where the goal is to ensure that different groups receive similar treatment in terms of selection probabilities. For example, if two groups, $G_1$ and $G_2$, are considered, achieving demographic parity would mean that both groups have the same probability of being selected, regardless of their actual coverage levels. They also discuss the hardness of approximation for these problems, establishing that it is NP-hard to approximate the solutions. Furthermore, the authors develop an approximation algorithm for one of the problems, achieving a constant approximation factor while allowing for some fairness violation.  An extensive experimental study is conducted, showing that the proposed algorithms perform well in practice, achieving the best demographic parity fairness values while maintaining a spread comparable to the most efficient algorithms without fairness constraints.

\subsubsection{Surrogate / Relaxed Fairness Optimization}
The authors of \cite{G17.9835234} address the issue of time-critical influence maximization (TCIM) under two constraints: budget constraint (TCIM-BUDGET) and coverage constraint (TCIM-COVER). They highlight that existing algorithms aimed at maximizing total influence can create significant disparities in utility across different groups, particularly disadvantaging minority groups. To tackle this, the authors propose a notion of fairness and introduce two new formulations: FAIR-TCIM-BUDGET and FAIR-TCIM-COVER. By utilizing surrogate objective functions with submodular properties, they develop computationally efficient algorithms that ensure fair treatment of different groups during influence propagation. Their experiments demonstrate that these algorithms effectively reduce disparity in influence across groups, although enforcing fairness may lead to a slight reduction in overall performance. In \cite{G18.87.article}, authors introduce extensions to the IC model that incorporate deadline constraints and random meeting events, emphasizing the importance of timing in influence spread. The Maximum Influence In-Arborescence (MIIA) algorithm is presented as a method for maximizing influence spread, showing superior performance compared to standard heuristics. From a fairness perspective, the paper evaluates the quality of seed sets based on expected influence spread, ensuring fair comparisons through extensive Monte Carlo simulations. This approach aims to provide equitable opportunities for different seed sets to maximize their influence, which is crucial in marketing strategies that seek to engage diverse user groups effectively. 

\subsection{Pre-processing and Post-processing Methods}
Pre-processing debiases network data before seed selection, while post-hoc methods adjust seeds post-optimization to satisfy fairness criteria. These methods are computationally efficient but may suboptimally address core reasons of bias in influence diffusion. 

\subsubsection{Diversity Rebalancing}
In \cite{G13.10734665}, the authors address the lack of diversity and budget considerations in traditional IM algorithms, proposing a Diversified Budgeted Influence Maximization (DBIM) algorithm for dynamic social networks. They propose an objective function that maximizes diversity and active nodes, proving its monotone, submodular, and NP-hardness properties. A heuristic-based strategy is introduced to reduce the search space for seed nodes, minimizing overlap of active nodes. From a fairness perspective, they highlight the need for diversity in influence maximization strategies. Traditional methods often ignore the heterogeneous nature of social networks and treat all nodes equally, which can lead to unequal influence distribution. The proposed approach aims to balance influence across diverse communities while adhering to budgetary constraints, ensuring that influence is equitably distributed. The DBIM algorithm outperforms existing methods in terms of activated nodes and community influence across all eight datasets, demonstrating the effectiveness of integrating diversity and budget considerations.

In \cite{G12.MEENA2025108045}, authors highlight the emergence of diversified influence maximization, which aims to select nodes that not only maximize influence but also ensure diversity among the activated nodes. The DCDIM (Dynamic Community Diversified IM) algorithm is designed to identify seed nodes in dynamic social networks that maximize both the influence spread and the diversity of influenced nodes across communities. It dynamically detects communities using the TCD2 (Tree-based Community Detection in Dynamic networks) method \cite{MISHRA2021114493} and uses a proposed objective function that balances the number of influenced nodes and the diversity of these nodes across communities. The objective function is proven to be Monotonic, Submodular, and NP-hard under the linear threshold diffusion model. The algorithm iteratively adds edges to the dynamic network, updates community structures, and at checkpoints, selects a seed set of size $k$ that maximizes the objective function using a greedy approach. They also address the fairness aspect in IM, particularly through the diversification of seed across communities, which aims to achieve fairness alongside influence spread. This is crucial as it ensures that the influence spread does not disproportionately favor certain groups over others, promoting a more equitable distribution of influence across communities. Overall, the paper emphasizes the importance of balancing influence maximization with fairness to enhance the effectiveness and ethical implications of social network interventions.

In \cite{G14.10.1145/3664618}, authors introduce the Dynamic Community Diversified Influence Maximization using Bridge nodes (DCDIMB) algorithm. They emphasize the importance of diversification in activated nodes within dynamic social networks, utilizing bridge nodes to enhance this aspect. An objective function is proposed that maximizes the number of communities by leveraging bridge nodes, which serve as interfaces between different communities. The DCDIMB algorithm dynamically identifies a seed set that diversifies the communities of active nodes, enhancing the overall influence spread. A novel diffusion model is introduced that incorporates the influence of inactive nodes, expanding the understanding of the diffusion process. Authors prove the monotonicity and submodularity of the objective function under the proposed diffusion model, establishing a theoretical foundation for the algorithm. It also demonstrates the NP-hardness of the objective function, indicating the complexity of the problem addressed. Furthermore, they address the concept of group-fairness in IM, emphasizing the importance of equitable influence distribution among different groups within the network. The proposed work is validated through experiments on various real-world and synthetic datasets, showing that the objective function effectively maximizes the number of communities compared to benchmark algorithms.

\setlength\rotFPtop{0pt plus 1fil}
\begin{sidewaystable*}[!htbp]
\centering
\caption{Illustration of Comparative Analysis of Existing Fairness-aware Influence Maximization, where COM, MOO, RIP, PPP, and SAT denote Constrained Optimization Methods, Multi-Objective Optimization, Regularization and In-Processing Methods, Pre-processing and Post-processing Methods, Scalability and Approximation Techniques, respectively. The SS, IS, and IF denote Seed Set Fairness, Influence Spread Fairness, and Individual Fairness, respectively.}
\label{tab:MethosSummary}
\resizebox{1.\textwidth}{!}
{
\begin{tabular}{|p{0.8cm}|p{5.cm}|p{0.65cm}|p{0.65cm}|p{0.65cm}|p{0.65cm}|p{0.65cm}|p{0.65cm}|p{0.4cm}|p{0.4cm}|p{0.4cm}|p{1.4cm}|p{1.55cm}|p{1.5cm}|p{4.0cm}|p{4.0cm}|p{4.0cm}|}
\hline
 \textbf{Study} &\textbf{Contribution} & \multicolumn{6}{l|}{\textbf{Algorithmic Classification}} & \multicolumn{3}{l|}{\textbf{Evaluation}} &\textbf{Diffusion Models}  &\textbf{Dataset Used} &\textbf{Baseline Methods} &\textbf{Challenges} &\textbf{Future Directions}  &\textbf{Fairness Consideration} \\ 
\cline{3-11} 
& &\textbf{COM} &\textbf{MOO} &\textbf{RIP} &\textbf{PPP} &\textbf{SAT}&\textbf{AI} & \textbf{SS} & \textbf{IS}  & \textbf{IF} & & & & & &\\
\hline
\cite{G1.10.1145/3308560} & Contributes to the understanding of fairness in social influence maximization by analyzing algorithms and proposing definitions for fairness in diffusion processes. &\checkmark &\checkmark &- &- &-&- &\checkmark &\checkmark &\checkmark &IC &Instagram (115,796,284 photos and 539,023 users) &Greedy, Centrality-based methods &Addressing historical biases encoded in social networks and ensuring that algorithms do not reinforce these biases. &Future directions involve developing more feature-aware algorithms that can enhance fairness and efficiency in influence maximization. &Algorithms can reproduce and amplify biases present in the data, necessitating a focus on equitable outcomes across different demographic groups.\\
\hline
\cite{G2.10.1145/336642} &This incorporates sensitive features in seed selection algorithms, enhancing diversity without significantly sacrificing efficiency. &\checkmark &\checkmark -&- &- &-&- &\checkmark &\checkmark &\checkmark &IC &DBLP &Centrality, Diversity &A significant challenge is the trade-off between fairness and efficiency, particularly when the seed set is small. &Exploring complex methods and diffusion processes to understand the dynamics of fairness and efficiency. &Arguing that diversity-enhancing strategies can improve outreach while maintaining fairness.\\
\hline
\cite{G3.inproceedings} &The contribution is primarily in defining and analyzing the elements that characterize individual social capital and their impact on the network as a whole. &- &\checkmark &- &- &- &- &-&- &\checkmark &NA &NA &NA &Challenges include the lack of established definitions and classifications of social networks. &Apply social capital to multi-agent systems, and develop specialized recommendation systems to promote network growth. &Fairness considerations are implied through the discussion of social capital's impact on individuals within the network.\\
\hline
\cite{G4.10.1007/11523468_91} &Presents a greedy algorithm for maximizing the spread of influence in social networks using the decreasing cascade (DC) model. &- &\checkmark &- &- &-&- &- &\checkmark &- &DC &NA &NA &Computational hardness of evaluating influence and need for approximation guarantees. &Suggest investigating more general influence models and evaluating $\sigma(A)$ in polynomial time. &Fairness considerations focus on maximizing influence spread while ensuring equitable influence across nodes.\\
\hline
\cite{G5.10.1145/34596} &Presents a novel approach called Disparity Seeding aimed at optimizing information spread to achieve a target gender ratio in social networks. &- &\checkmark &- &- &-&- &- &\checkmark &- &IC &Instagram, Facebook &Diversity seeding, IM-balanced &Addresses the NP-hard Disparity IM problem, which makes seed group selection to meet the gender ratio challenging. &Include additional demographic information, and it can be adapted to enhance the influence of underrepresented minorities. &Selected seed groups reflect a specified gender ratio, thereby addressing gender disparity in information spread.\\
\hline
\cite{G6.abs-1903-00967} &Introduces a fair resource allocation problem in IM and proposes two fairness concepts: maximin and diversity constraints. &\checkmark &\checkmark &- &- &-&- &\checkmark &\checkmark &- &Submodu-lar based &HIV prevention for homeless youth &Greedy &Challenges include the price of group fairness, which can lead to a reduction in social welfare. &Future directions suggest improving algorithms to ensure fair treatment of marginalized groups. &Fairness considerations are crucial to avoid discrimination based on sensitive attributes like race or gender. \\
\hline
\cite{G7.becker2023cost} &Contributes to the IM by rigorously analyzing the trade-offs between fairness and efficiency in the context of demographic parity. &\checkmark &\checkmark &\checkmark &- &\checkmark&- &\checkmark &\checkmark &- &Triggering model &Random, Synthetic, Arenas, Irvine &greedy  &Challenges in balancing fairness and efficiency, noting that strict fairness can lead to a loss in overall spread and time complexity. &Future work may involve further exploration of approximation algorithms and their performance in various contexts of fairness. &Emphasizes that fairness should be treated as a requirement to be ensured by algorithms, rather than an objective to be optimized.\\
\hline
\cite{G8.10.1613/jair.1.13367} &Focuses on node-based methods that select a seed independently with a bounded expected size. &\checkmark &\checkmark &- &- &-&- &\checkmark &\checkmark &\checkmark &IC, LT &Arenas, ca-GrQc, Facebook &TIM, Greedy &Main challenge is the NP-hardness of the maximin problem and the difficulty in achieving good approximation ratios. &Improving approximation guarantees and exploring the use of randomization for other fairness criteria. & showing that traditional methods can lead to unfair coverage across communities.\\
\hline
\cite{G9.10.5555/3491440} &Present an IM method that incorporates fairness objectives, addressing the challenge of ensuring equitable information dissemination across different groups in a network. &\checkmark &\checkmark &- &- &-&- &\checkmark &\checkmark &\checkmark &Adversarial Graph Embeddings &Rice University Facebook dataset, Synthetic  &Greedy  &Balancing the trade-off between maximizing influence and ensuring fairness, especially with multiple sensitive attributes. &Explore enhancing fairness in larger networks and integrating more complex fairness constraints. &Emphasizes the importance of fairness in information dissemination, particularly in contexts like job opportunities and news spread, to prevent bias against minority groups.\\
\hline
\end{tabular}
}
\end{sidewaystable*}

\setlength\rotFPtop{0pt plus 1fil}
\begin{sidewaystable*}[!htbp]
\ContinuedFloat
\centering
\caption{Continued.}
\label{tab:MethosSummary}
\resizebox{1.\textwidth}{!}
{
\begin{tabular}{|p{0.8cm}|p{5.cm}|p{0.65cm}|p{0.65cm}|p{0.65cm}|p{0.65cm}|p{0.65cm}|p{0.65cm}|p{0.4cm}|p{0.4cm}|p{0.4cm}|p{1.4cm}|p{1.55cm}|p{1.5cm}|p{4.0cm}|p{4.0cm}|p{4.0cm}|}
\hline
 \textbf{Study} &\textbf{Contribution} & \multicolumn{6}{l|}{\textbf{Algorithmic Classification}} & \multicolumn{3}{l|}{\textbf{Evaluation}} &\textbf{Diffusion Models}  &\textbf{Dataset Used} &\textbf{Baseline Methods} &\textbf{Challenges} &\textbf{Future Directions}  &\textbf{Fairness Consideration} \\ 
\cline{3-11} 
& &\textbf{COM} &\textbf{MOO} &\textbf{RIP} &\textbf{PPP} &\textbf{SAT} &\textbf{AI} & \textbf{SS} & \textbf{IS}  & \textbf{IF} & & & & & &\\
\hline
\cite{G10.MEENA2025127445} &The proposed algorithm utilizes a Learning Automata-based Sine Cosine Algorithm with Levy Flights to optimize the influence spread while ensuring equitable distribution across different groups. &\checkmark &\checkmark &- &\checkmark &\checkmark &- &\checkmark &- &- &IC &CA, Amazon, Facebook &SCA, CELF, CELF++, GWIM, LAPSO-IM &The challenges identified include the need for enhanced fairness metrics and scalability in dynamic networks. &Future directions suggest exploring user feedback mechanisms to improve influence strategies. &Fairness considerations are integral to the algorithm, ensuring that influence is not disproportionately biased towards certain demographics or nodes.\\
\hline
\cite{G11.GONG2023119138} &Presents a method for IM that incorporates fairness considerations, proposing a multi-objective optimization model to balance influence and fairness in social networks. &- &\checkmark &- &- &-&- &\checkmark &\checkmark &\checkmark &IC &Facebook, Weibo, YouTube &Centrality, PageRank, K-core &Challenges include the NP-hard nature of the problem and the difficulty in simultaneously maximizing both influence and fairness. &Research can be extended to other contexts such as gender equality, religious equality, and racial equality. &Emphasizing the importance of equitable information access across different groups.\\
\hline
\cite{G12.MEENA2025108045} &Contributes to diversified IM in dynamic networks by proposing the DCDIM algorithm, which aims to maximize influence while ensuring diversity.  &- &\checkmark &\checkmark &- &\checkmark &-&\checkmark &\checkmark &- &LT &CondMAT, NetScience, Inploid &Greedy, CELF, CDIM &The overlapping of active nodes in dense networks and the need for dynamic updates in the network structure. &Focus on enhancing the fairness of influence maximization strategies and exploring additional community detection methods. &Emphasizes the importance of fairness in IM, particularly in ensuring that diverse communities are represented in the seed set.\\
\hline
\cite{G13.10734665} &Presents a heuristic-based strategy that reduces the search space of the seed set and minimizes the overlapping of active nodes. &\checkmark &\checkmark &- &\checkmark &-&- &\checkmark &\checkmark &- &IC &As-July, CondMAT, NetScience &Greedy, CELF, Budgeted Degree &Challenges addressed include the overlapping of active nodes and the need for diversification in IM. &Future directions suggest further exploration of community detection and IM strategies. &Fairness considerations involve ensuring a balanced distribution of influence across communities. \\
\hline
\cite{G14.10.1145/3664618} &Investigates the importance of bridge nodes in the seed set and their impact on the number of communities and activated nodes across different datasets. &- &- &\checkmark &\checkmark &-&- &\checkmark &\checkmark &- &ELTM &NetScience, Power, As-July, CondMAT, Email &CELF, PMIA, SIM-PATH, PageRank &Challenge of determining the optimal results based on the nature of the network and the influence capabilities of bridge nodes. &Exploring further enhancements in the diffusion model and its applications in various network types. &The focus on bridge nodes implies a consideration of equitable influence spread across communities.\\
\hline
\cite{G15.9863801} &The proposed method, SetMOGWO, is a binary multi-objective algorithm specifically designed for discrete optimization problems like IM, outperforming  continuous optimization methods. &\checkmark &\checkmark &- &- &\checkmark &- &\checkmark &\checkmark &- &Group Activation Speed &Obesity prevention dataset &SOGWO &The paper discusses the cost of ensuring group fairness and the complexity of defining fairness metrics. & Future research includes studying fairness in IM, parallelizing SetMOGWO, and designing a framework for group-fairness-aware IM. &Emphasizes the need for application-dependent fairness metrics and highlights the importance of ensuring smaller groups receive fair treatment.\\
\hline
\cite{G16.10.1145/3437963.3441787} &Presents an IM algorithm that balances influence spread and categorical balance among users, addressing disparities in information exposure across different categories. &\checkmark &\checkmark &- &- &- &- &\checkmark &\checkmark &\checkmark &IC with Homophily &Bofa\_careers, Upsjobs, Verizoncareers  &Diversity Seeding &Achieving a balance in outreach when applying classical IM algorithms, especially in the presence of homophily. &Explore measuring and mitigating imbalance concerning continuous node attributes and improving the scalability. &Emphasizes the importance of categorical balance in the active set to ensure fair representation of different user categories.\\
\hline
\cite{G17.9835234} &Introduces fairness in time-critical IM (TCIM), focusing on ensuring equal influence across different groups within a time deadline. &\checkmark &\checkmark &\checkmark &- &\checkmark &-&\checkmark &\checkmark &- &IC &Rice-Facebook Dataset &FAIRTCIM-COVER, TCIM &Ensuring fairness while minimizing the seed set size and addressing the disparity in influence across different groups. &Exploring different fairness and complex models of time-criticality in information propagation. &Emphasizes the importance of achieving parity of influence among different groups to consider minority groups.\\
\hline
\cite{G18.87.article} &Addressed IM by incorporating time-delayed influence diffusion and addressing the time-critical IM problem under the proposed IC-M model. &\checkmark &\checkmark &\checkmark &- &\checkmark &- &- &\checkmark &- &IC &NetHEPT, DBLP, Epinions &Greedy, Degree, MIA &Efficient computation of influence spread in tree structures and the incorporation of login probabilities into the IC model &Exploring more efficient computations for influence spread and considering time delays in various diffusion models. &Fair comparisons in evaluating seed set quality, ensuring that the influence spread is assessed accurately.\\
\hline
\cite{G19.10.1145/3308558.3313680} &Address the gaps in information access among individuals, particularly focusing on how to spread information while minimizing disparities in access. &\checkmark &\checkmark &- &- &- &- &\checkmark &- &\checkmark &IC, LT &Facebook, Irvine, ca-GrQc  &TIM+ &Highlight the disparity in information access among well-connected and poorly connected individuals in the network. &Explore more sophisticated models for fairness in information dissemination and the integration of additional sensitive features. &Emphasizes a welfare approach to fairness, focusing on maximizing the minimum probability of receiving information.\\
\hline
\cite{G20.doi:10.1137/1.9781611976236.69} &Presents an approach to measuring information unfairness and proposes the MaxFair algorithm to reduce unfairness by adding edges to the network.  &- &\checkmark &- &- &- &- &\checkmark  &- &\checkmark  &NA &Parallel-Full, Graphics-Full, DBLP  &Degree, Centrality &Difficulty in estimating the change in flow when adding edges to the network. &Explore further enhancements to the MaxFair algorithm and its application to other types of networks and fairness criteria. & Highlights the need to prevent differential treatment based on protected group membership. \\ 
\hline
\end{tabular}
}
\end{sidewaystable*}

\setlength\rotFPtop{0pt plus 1fil}
\begin{sidewaystable*}[!htbp]
\ContinuedFloat
\centering
\caption{Continued.}
\label{tab:MethosSummary}
\resizebox{1.\textwidth}{!}
{
\begin{tabular}{|p{0.8cm}|p{5.cm}|p{0.65cm}|p{0.65cm}|p{0.65cm}|p{0.65cm}|p{0.65cm}|p{0.65cm}|p{0.4cm}|p{0.4cm}|p{0.4cm}|p{1.4cm}|p{1.55cm}|p{1.5cm}|p{4.0cm}|p{4.0cm}|p{4.0cm}|}
\hline
 \textbf{Study} &\textbf{Contribution} & \multicolumn{6}{l|}{\textbf{Algorithmic Classification}} & \multicolumn{3}{l|}{\textbf{Evaluation}} &\textbf{Diffusion Models}  &\textbf{Dataset Used} &\textbf{Baseline Methods} &\textbf{Challenges} &\textbf{Future Directions}  &\textbf{Fairness Consideration} \\ 
\cline{3-11} 
& &\textbf{COM} &\textbf{MOO} &\textbf{RIP} &\textbf{PPP} &\textbf{SAT} &\textbf{AI} & \textbf{SS} & \textbf{IS}  & \textbf{IF} & & & & & &\\
\hline
\cite{G21.10.14778/3551793.3551824} &Introduces the FairCS problem, which aims to maximize content spread while ensuring fairness in the selection of candidate edges. &\checkmark &\checkmark &- &- &- &- &\checkmark &\checkmark &\checkmark &IC &Antelope Valley dataset &Continuous Greedy, IRFA, SpGreedy &Challenges faced include the NP-hardness of the problem and the need for scalable solutions to handle large networks. &Exploring edge suggestion without considering recommendation acceptance probabilities \& determining fair content spread across topics. &Fairness considerations are addressed by inequity in information spread across social networks, particularly affecting minority groups. \\
\hline
\cite{G22.10.114424.3383555} &Presents a flexible framework for modeling and solving fairness-aware IM, achieving optimal solutions unlike many existing methods that are sub-optimal. &\checkmark &\checkmark &\checkmark &- &\checkmark &- &\checkmark &\checkmark &\checkmark &IC &Antelope
Valley dataset &Maximin, Diversity, MIP &Include scalability and the need for robustness guarantees in sampling techniques, as small changes in graph structure can drastically alter solutions. &Designing more effective sampling approaches and exploring new diffusion models to better capture the dynamic nature of networks. &Emphasizes that fairness is subjective and context-dependent, proposing multiple definitions of fairness and discussing the trade-offs involved.\\
\hline
\cite{G23.rahmattalabi2021fair} &Contributes a unifying framework for comparing different fairness measures in algorithmic decision-making, allowing for the evaluation of trade-offs between fairness and utility. &\checkmark &\checkmark &- &- &- &- &- &- &\checkmark &LT &Suicide prevention dataset &Diversity Constraints, Maximin Fairness &Introduce undesirable biases, and that the extreme of inequality aversion can complicate fairness considerations. &Suggests that further study is needed on the implications of different fairness definitions and their trade-offs. &Highlight that fairness is domain-dependent and that different fairness definitions can conflict with each other.\\
\hline
\cite{G24.10.1145/3580305.3599847}&This is the first policy-driven solution that can be applied at scale to maximize influence while ensuring fairness across different groups. &\checkmark &\checkmark &\checkmark &- &- &-&\checkmark &\checkmark &\checkmark &IC &Sina Weibo, Digg &Set-based, Node-based, FIM, MIP &A significant challenge is the scalability of existing methods, which often fail to handle large datasets effectively. &Future work could explore more sophisticated models that further enhance the balance between fairness and influence spread. &Emphasizes the importance of fairness in information diffusion, particularly in contexts where biased information can lead to unequal influence.\\
\hline
\cite{G25.10161600} &The paper presents a novel Competitive Bidding Influence Maximization (CBIM) problem, which is proven to be NP-hard. &\checkmark &- &- &- &\checkmark &- &\checkmark &\checkmark &\checkmark &CIC, CLT &Facebook, YouTube &TD3, SAC, DDPG &Simulating interaction of competitors and design effective method to optimize  policies to ensure fairness. &Future work is to explore multitime activation situations and further refine the bidding optimization process. &Emphasizes the importance of fairness among competitors in the context of influence spread and cost distribution.\\
\hline
\cite{G26.zhang2025fairness} &Incorporates a multi-objective reward function and utilizes a generalized Gini welfare function to balance fairness and utility. &- &\checkmark &\checkmark &- &- &- &\checkmark &- &\checkmark &IC &HIV prevention dataset &Maximin, AGE, Greedy, ABRIS &Challenges include designing efficient algorithms for large-scale social networks and addressing the privacy of demographic data. &To develop a method that can operate effectively without demographic data and explore fairness-oriented social welfare functions. &Ensuring that the proportions of influenced nodes are approximately equal across different demographic groups.\\
\hline
\cite{G27.10.1145/3701737} &Presents a novel fairness notion called Poverty Reward (PR), which balances fairness and total influence spread in IM problems.  &\checkmark &\checkmark &- &- &\checkmark &- &\checkmark &\checkmark &- &IC &Obesity prevention dataset &DC, MMF &Challenges include balancing fairness with total influence spread and developing scalable algorithms. &Analyzing the influence gap and exploring relationships between fairness notions and the proposed framework. &Addressed through the group-level fairness framework, ensuring that different demographic groups receive equitable influence.\\
\hline
\cite{G28.xiao2025fairness} &Contributes to the fairness in influence maximization by proposing a method called FIBMM that effectively balances fairness and effectiveness in social networks. &- &\checkmark &- &- &- &- &\checkmark &\checkmark &\checkmark &CIC &Hamsterster, Facebook, Government &Degree, PageRank &A significant challenge highlighted is the trade-off between maximizing influence and ensuring fairness across different communities. &The paper suggests exploring further enhancements in algorithmic fairness and community structures in future research. &The proposed method, FIBMM, aims to achieve the best performance in terms of fairness by effectively protecting all communities involved.\\
\hline
\cite{khajehnejad2022crosswalk} & Proposes a fair embedding method that biases random walks toward cross-group interactions to reduce disparity &- &- &- &- &- &\checkmark &- &\checkmark &\- & IC & Synthetic and Rice-Facebook & Greedy, DeepWalk, Node2Vec, FairWalk & Scalability is a concern as they tested only on small networks. & The authors did not mention any, but scalable fairness-aware embedding methods for large dynamic networks will beenefit the field. & Considers Disparity fairness constraint for evealution.\\
\hline
\cite{saxena2025dq4fairim} & Proposes a deep RL framework that jointly optimizes influence spread and fairness while selecting seed nodes &- &- &- &- &- &\checkmark & - &\checkmark &- & IC & Synthetic network (Homophily-BA and Obesity), Facebook, Twitter & CELF, Degree, PageRank, Parity, Fair PageRank, CrossWalk, CEA & Tested on small size evolving networks & Explore RL-based solutions for different diffusion models and context-specific conditions & Uses Maximin fairness constraint.\\ 
\hline
\end{tabular}
}
\end{sidewaystable*}

\begin{figure*}[!ht]
        \centering \includegraphics[height=2.8in,width=.9\textwidth]{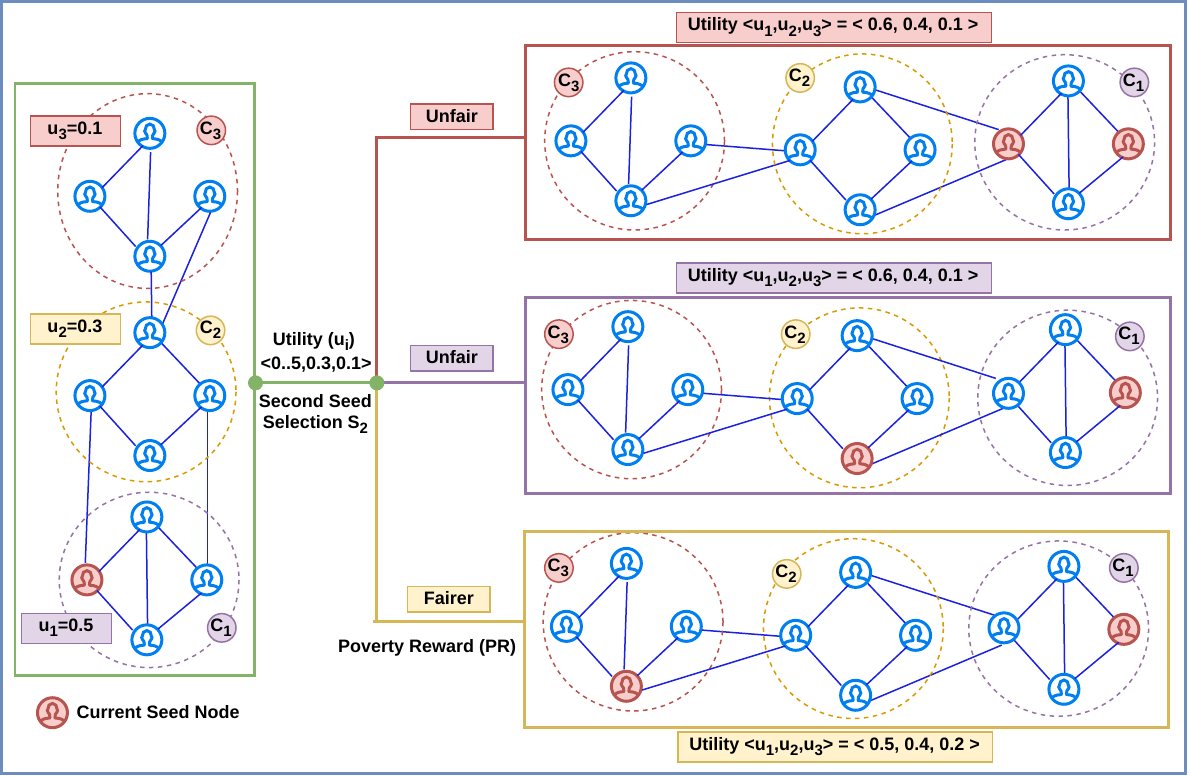}
        \caption{Illustration of the Three Choices for Second Seed Selection with their Respective Utility Distribution using Poverty Reward.}
        \label{fig:Poverty}
\end{figure*}

\subsection{Scalable Approximation Techniques}
To deal with fairness requirements in large-scale social networks, scalable methods use sampling (like Fair-RIS) or heuristic pruning. Community-aware approximations use the network structure to reduce complexity. These methods compromise theoretical guarantees in favor of practicality in the real world. In \cite{G27.10.1145/3701737}, authors introduce a novel fairness notion called poverty reward (PR), which considers both fairness and total influence spread in the context of IM. Unlike traditional fairness concepts such as Max-Min fairness, demographic parity, and diversity constraints, which focus on a single aspect of fairness, PR aims to balance these two factors effectively. The concave fairness framework (CFF) is derived from PR, allowing the use of any concave function to mitigate unfairness at the group level. This framework emphasizes the relationship between the characteristics of concave functions and fairness levels, providing a quantitative analysis of fairness strategies. For instance, Fig. \ref{fig:Poverty} shows that communities $C_1$, $C_2$, and $C_3$ have utility 0.5, 0.3, and 0.1, respectively, with the current seed set $S_1$. The second seed selection $S_2$ has three choices, those can make utility increased to $<0.6, 0.4, 0.1>$, $<0.6, 0.4, 0.1>$, and $<0.5, 0.4, 0.2>$, respectively. The total utility increase from $S_1$ to $S_2$ is 0.2 in all three choices. Therefore, all the choices are the same for classical seed selection methods, even the third choice $<0.5, 0.4, 0.2>$ offers a fairer seed selection using poverty reward. The paper also highlights limitations of existing methods; while they attempt to address unfairness, they often overlook the population size of groups, which can lead to inequities. Overall, the proposed approach seeks to enhance fairness by rewarding poorer groups while ensuring that richer groups maintain a significant influence spread.

The authors of \cite{G22.10.114424.3383555} present a unifying framework for fairness-aware IM problems from an optimization perspective. It emphasizes that fairness is subjective and varies across different real-world applications, indicating that a single definition of fairness is insufficient for all contexts. The authors propose multiple definitions of fairness, highlighting their relative advantages and differences based on context. They also explore the trade-offs between enforcing fairness and the potential loss of total influence, suggesting that addressing discrimination incurs costs regardless of the chosen group-level fairness notion. The framework allows users to combine various definitions of fairness and even encode new definitions, thus providing flexibility. The paper also discusses the limitations of fairness-aware IM and identifies open questions for future research. Furthermore, it positions the work within the existing literature on IM and fairness in AI, noting the lack of comprehensive studies on fairness-aware IM problems. The authors argue that while many fairness metrics exist, it is often impractical to satisfy all definitions simultaneously, as highlighted by the impossibility theorem. They focus primarily on group-level definitions of fairness but note that the framework can be extended to individual fairness measures, which prioritize the welfare of individuals regardless of group affiliation. Overall, the paper contributes to the understanding of fairness in IM by providing a flexible and comprehensive framework that can adapt to various fairness definitions and contexts.

The authors of \cite{G25.10161600} propose a fairness-aware multiagent competitive bidding influence maximization (FMCBIM) framework, which consists of two main modules: 1) a competition module that develops a multiagent bidding particle environment (MBE) for competitors to bid dynamically, and 2) a training module that introduces a practical multiagent competitive bidding influence maximization (MCBIM) algorithm for optimizing policies. This framework aims to create a win-win situation for both the platform and competitors while ensuring fairness in the bidding process. The experiments conducted on five real-world datasets demonstrate that the proposed algorithm effectively addresses the competitive bidding influence maximization (CBIM) problem. The results indicate that cooperation with the platform fosters a healthier competitive atmosphere compared to pure competition. The fairness index is defined using generalized entropy, which measures fairness among competitors. A lower value of this index indicates a fairer competition. The paper also highlights the importance of considering fairness and budget allocation in the bidding scenario, as disparities in promotional effects can lead to unfair advantages.

\subsection{Network Embedding and  Learning-based Approaches}

\subsubsection{Network Embedding-based Approaches} 
Network embedding methods represent nodes in a graph as low-dimensional vectors that preserve structural and relational properties of the network, enabling efficient downstream tasks such as node classification, link prediction, and influential nodes identification \cite{cui2018survey}. 
Building on this, CrossWalk \cite{khajehnejad2022crosswalk} introduces a fairness-aware embedding by biasing random walks to cross group boundaries, thereby ensuring that nodes from different groups are better connected in the embedding space and reducing structural bias. After generating the network embeddings, CrossWalk selects the top-$k$ seed nodes using a $k$-medoids approach, which chooses nodes that are well separated in the embedding space, thereby promoting fairer outcomes in influence maximization. Extending this idea, MoonWalk \cite{moens2022reproducibility} further improves fairness by refining how proximity to group boundaries is handled, enabling better generalization to networks with multiple groups while maintaining influence spread performance. On the other hand, methods such as IMNE \cite{hu2024imne} leverage deep learning–based node embeddings combined with structural features and diffusion models to efficiently identify influential nodes while reducing computational complexity through search space reduction. Similarly, SGNN \cite{kumar2022influence} integrates graph embedding (e.g., struc2vec) with graph neural networks to model influence maximization as a regression problem, where embeddings serve as node features and learned models predict influence scores to select optimal seed nodes. Together, these approaches demonstrate how embedding techniques, combined with fairness constraints and deep learning, provide scalable and effective solutions for influence maximization in complex networks.

The authors of \cite{G24.10.1145/3580305.3599847}  address the problem of influence maximization with fairness, focusing on selecting k influential nodes to maximize information spread in a network while ensuring that sensitive user attributes (e.g., gender, location, origin, race) are fairly represented among the influenced users. The authors propose two deep-learning-based solutions: (a) fairness-based participant sampling (FPS) and (b) fairness as context (FAC). FPS derives node embeddings and diffusion probabilities independently, while FAC adopts a joint learning approach to consider both fairness and information cascades together. The paper emphasizes the need for scalability in these solutions, as previous studies were limited to small networks. The authors introduce a new formulation for fair influence maximization, aiming for equity in the influence spread across different groups. They utilize a fair-greedy algorithm to select diffusion seeds based on node embeddings and diffusion probabilities. The paper also discusses the coefficient of variation (CV) as a measure of fairness, where perfect fairness is achieved when the influenced nodes' proportions across categories are equal. The ideal scenario is to maintain the same proportions of influenced users as in the original population, ensuring that the spread does not disproportionately favor any group. Overall, the paper contributes to the field by providing scalable methods for achieving fairness in influence maximization, which is crucial for applications like social advertising campaigns that require unbiased information dissemination.

\subsubsection{Learning-based Approaches} Reinforcement learning (RL) solves influence maximization problem by modeling seed selection as a sequential decision process, that is formulated as a Markov Decision Process (MDP), where the state encodes the current graph and selected seed set, actions correspond to choosing the seed node, and rewards capture the marginal spread achieved under diffusion models like IC and LT \cite{khalil2017learning,saxena2025dq4fairim}. Khalil et al. \cite{khalil2017learning} proposed S2V-DQN, which integrates Structure2Vec embeddings with Deep Q-Learning to learn greedy solutions for network analysis tasks. The method first maps the network to low-dimensional node representations using Structure2Vec and then trains a Deep Q-Network (DQN) using deep reinforcement learning. The top-$k$ seed nodes are subsequently selected based on the DQN's scores. 
Li et al. \cite{li2022piano} proposed PIANO, a method that combines network embedding with reinforcement learning to estimate the expected influence of nodes. They further demonstrated that models trained on similar network types generalize effectively, allowing a model trained on one network to be directly applied to another with similar characteristics. 
Saxena et al. proposed DQ4FairIM \cite{saxena2025dq4fairim}, a fairness-aware RL-based method that incorporates additional fairness-aware reward components (based on max-min fairness) to balance overall spread with equitable outreach, enabling dynamic trade-offs via reward shaping. RL-based methods have also been designed for influence blocking maximization, where nodes are sequentially selected for immunization or counter-campaigning to minimize the spread of harmful influence, often under adversarial or competing diffusion settings.  
In influence blocking maximization (IBM), the problem is formulated as an MDP, where the state captures the current diffusion status and protected nodes, actions correspond to selecting the next node for intervention, and the reward is defined as the reduction in expected spread of the adversarial cascade \cite{he2022reinforcement}. RL method ranging from value-based approaches like Deep Q-Networks to policy-gradient and actor–critic methods learn adaptive blocking strategies that can generalize across different graph structures and adversarial scenarios.

In \cite{G9.10.5555/3491440}, authors address the challenge of achieving fair IM, where the goal is to ensure that the set of influenced nodes is diverse with respect to sensitive attributes, such as age, gender, and race, while maintaining competitive IM results. They propose an adversarial network embedding, using an autoencoder and a discriminator in an adversarial setting to create embeddings that are similarly distributed across sensitive attributes. The discriminator is designed to distinguish between the embeddings of two communities, thereby guiding the embedder to generate embeddings that are as similar as possible in distribution. The method involves pre-training the embedder and discriminator, followed by adversarial training to ensure fair distribution of embeddings, and selecting initial seeds using a fair selection algorithm. The approach is competitive with existing methods in maximizing the total number of influenced people and reducing disparity, demonstrating potential for application in other fairness tasks over social networks. Most of the existing IBM algorithms are highly sensitive to the diffusion process, and the influence spread is coupled with the diffusion model. Therefore, the authors of \cite{XIAO2026131192} proposed a centrality-based graph embedding method to address this limitation in the fairness-aware IBM problem. The proposed method chooses a truth seed set that balances blocking effectiveness and equal protection across communities without diffusion models by integrating original-space centrality measurements with embedding-space proximity. They have validated the performance of the proposed method on four real-world networks against five state-of-the-art baselines for blocking efficacy and maximum fairness difference. The proposed method outperforms state-of-the-art baselines in blocking and reduces community fairness discrepancies.

In \cite{G26.zhang2025fairness}, authors discuss the concept of equity in resource allocation, emphasizing the Pigou–Dalton principle, which states that transferring rewards from a better-off group to a worse-off group results in a fairer solution. This principle is mathematically represented as: for any utility vector $u \in R^m$, if $u_{j} - u_{i} > \epsilon > 0$, then $u + \epsilon e_{i} - \epsilon e_{j}$ is considered fairer, where $e_{i}$ and $e_{j}$ are null vectors with specific components equal to 1. The paper also highlights the importance of social welfare functions in implementing fairness, ensuring that the selected functions align with the principles of equity. Furthermore, it explores the maximin combined utility method, which excels in distributing opportunities equitably among various groups, particularly benefiting the most disadvantaged. The classical IM problem is critiqued for potentially neglecting underrepresented demographics, leading to inequitable outcomes in campaigns, such as public health initiatives.  Recent research has introduced fairness methods like maximin fairness and adversarial graph embeddings, each with its strengths and limitations. The paper proposes a fairness optimization method based on reinforcement learning, utilizing a generalized Gini welfare function to balance fairness and utility, validated through experiments. Overall, the paper emphasizes the need for a comprehensive approach to influence maximization that incorporates fairness while maintaining interpretability in decision-making.

\section{Datasets and Evaluation Metrics}
\label{sec:Datasets}

\subsection{Datasets}
Despite the growing interest in fairness-aware influence maximization, most of the commonly used datasets\footnote{https://snap.stanford.edu/data/}, such as Twitter, Facebook, Ego-Nets, or citation networks, do not provide the protected attributes information to evaluate fairness \cite{snapnets}. As a result, researchers need to simulate or infer group information, which may introduce noise or bias into the evaluation process. The evolution of real-world networks is influenced by homophily \cite{saxena2025homophily}, and some efforts have been made to generate synthetic homophilic networks with major and minor communities using homophilic models like Barabási-Albert (HICH-BA) model \cite{193.0489-023-04928-3}, diversified homophily BA model \cite{232.wang2021infoetwork}, homophily BA model \cite{232.wang2021infoetwork}, directed homophily model \cite{G16.10.1145/3437963.3441787}, and organic growth model \cite{211.10.1145/3178876.3186140}. Similarly, some of the works have been generated real-world social network datasets listed in Table \ref{tab:datasets}. There is still a necessity of generating selective social network datasets that are ethically sourced and provide a wide range of user attributes, ground-truth group labels, and true impact propagation traces.

\begin{table*}[htbp]
\centering
\caption{Dataset Information, including Protected Attributes. }
\resizebox{1.\columnwidth}{!}
{
\begin{tabular}{|p{1.4cm}|p{3.6cm}|p{1.4cm}|p{1.4cm}|p{7.2cm}|p{3.0cm}|}
\hline
\textbf{Reference} & \textbf{Dataset} &\textbf{Nodes} & \textbf{Edges} & \textbf{Description} &\textbf{Protected Attributes}  \\
\hline
\cite{G1.10.1145/3308560} &Instagram &539,023 &640,211  & 45.57\% males and 54.43\% females &Gender \\
\cline{2-6}
&DBLP & 53,307 &288,864  &81\% males and 19\% females &Gender \\
\cline{2-6}
&DBLP-continent & 3,980 & 6,965  &Labels have 5 continent values &Continent \\
\hline
\cite{10.5555/3367471.3367498} &Instagram-London & 53,902 &165,184  &Races: African, Caucasian, Asian & Gender, Race \\
\cline{2-6}
&Instagram-Los Angeles &82,607 &482,305  &Races: African, Caucasian, Asian & Gender, Race \\
\hline
\cite{148.2999134.2999195} &Google+ Network &4,938 &547,923  &Anonymized gender data &Gender \\
\cline{2-6}
&Facebook Network & 4,039 &88,234  &Anonymized gender data &Gender \\
\hline
\cite{10.1145/3485447.3512249} &DBLP Data Mining &16,501 &66,613  &
Publications data from 2011 to 2020 &Gender \\
\hline
\cite{182.10.1371/journal.pcbi.1001109} &Brazilian network & 16,730 &39,044  & 10,106 sex workers \& 6,624 sex
buyers; 40\% minority nodes &Gender \\
\hline
\cite{203.Shetty2004TheEE} &Enron & 144 &1,344  &76\% men and 24\% women &Gender \\
\hline
\cite{202.SEIERSTAD201144} &Norwegian Interlocking
Directorate Network &1,421 &3,855  & 63\% men and 37\% women &Gender \\
\hline
\cite{116.10.1145/5} &Github Follower network &89,630 &167,359  &5.7\% females &Gender \\
\hline
\cite{104.HOLME2004155} &POK Swedish Dating &29,341 &–  &Minority group has 44\% nodes &Gender \\
\hline
\cite{205.SNIJDERS201044} &Dutch School Friendship &26 &221  &17 Girls and 9 Boys &Gender \\
\hline
\cite{G6.abs-1903-00967} &HIV Prevention for homeless youth &60–70 &- &4 datasets each having 60-70 nodes &Birth sex, gender, race, and sexual orientation \\
\hline
\cite{macedo2026gender} &Twitter Soccer-Pt & 194,795 & 2,117,105 & Soccer communication temporal network in English from X & Gender \\
\hline
\cite{macedo2026gender} &Twitter Soccer-En & 782,550 & 4,308,009 & Soccer communication temporal network in English from X &Gender \\
\hline
\cite{kumar2018community} &Reddit Hyperlink Network &55,863  &858,490  &Network consists hyperlinks between subreddits on Reddit  &Gender \\
\hline
\cite{kumar2016edge} &Bitcoin OTC &5,881  &35,592  &Bitcoin OTC web of trust network  &Users Information \\
\hline
\end{tabular}
}
\label{tab:datasets}
\end{table*}

\begin{table*}
\centering
\caption{Description of various fairness evaluation metrics for IM.}
\label{tab:Evaluationfairness_metrics}
\resizebox{\textwidth}{!}{
\begin{tabular}{|p{2.3cm}|p{2.1cm}|p{4.0cm}|p{2.8cm}|p{2.7cm}|p{2.1cm}|p{2.5cm}|}
\hline
\textbf{Metric name} & \textbf{Formula} & \textbf{Description} & \textbf{Advantage} & \textbf{Disadvantage} & \textbf{Behavior (low value)} & \textbf{Behavior (high value)} \\
\hline
Influence Spread \cite{10.1145/956750.956769}
& $|I|$ 
& Total number of activated (influenced) nodes in the network. 
& Direct measure of effectiveness of the model. 
& Ignores fairness; may favor only certain communities. 
& Low influence propagation. 
& High overall influence in the network. \\
\hline
Diversity (community count) \cite{G14.10.1145/3664618}
& $|C|$ 
& Number of communities considered in the network. 
& Captures structural diversity of the network. 
& Does not reflect fairness or influence distribution. 
& Few communities (low diversity). 
& High structural diversity. \\
\hline
Equity (influence coverage) \cite{G22.10.114424.3383555}
& $\sum_{c \in C} \frac{|I \cap c|}{|c|}$ 
& Measures the proportion of influenced nodes in each community. 
& Encourages fair influence spread. 
& Can be high even with an imbalance across communities. 
& Uneven or poor influence coverage. 
& Strong influence coverage across communities. \\
\hline
Equality (seed distribution) \cite{G22.10.114424.3383555}
& $\sum_{c \in C} \frac{|S \cap c|}{|c|}$ 
& Measures the distribution of seed nodes across communities. 
& Ensures fair initial allocation of seeds. 
& Does not ensure actual influence spread. 
& Biased seed placement. 
& Balanced seed allocation. \\
\hline
Min-Max fairness \cite{G22.10.114424.3383555}
& $\frac{\min\limits_{c \in C} \left( \frac{|I \cap c|}{|c|} \right)}{\max\limits_{c \in C} \left( \frac{|I \cap c|}{|c|} \right)}$ 
& Ratio of the least to the most influenced community. 
& Captures disparity explicitly. 
& Sensitive to outliers. 
& High inequality across communities. 
& Uniform influence distribution. \\
\hline
Disparity ($Log$ penalty) \cite{2025_TNSE_MineMultiFairness_PDTF_IM}
& $-\sum_{c \in C} \log(r_c)$, \quad $r_c = \frac{|I \cap c|}{|c|}$ 
& Penalizes communities with low influence ratios using a logarithmic penalty. 
& Strongly penalizes underrepresented communities. 
& Sensitive to very small values; requires smoothing. 
& High disparity (large penalty). 
& Low disparity (small or zero penalty). \\
\hline
Gini index \cite{G11.GONG2023119138}
& $\frac{\sum_{i=1}^{n} \sum_{j=1}^{n} |r_i - r_j|}{2n \sum_{i=1}^{n} r_i}$ 
& Measures inequality in influence distribution across communities. 
& Standard measure of inequality; widely interpretable. 
& Computationally expensive for large $n$. 
& High inequality (unfair distribution). 
& Low inequality (fair distribution). \\
\hline
\end{tabular}
}
\end{table*}

\subsection{Evaluation Metrics}
We provide the analysis of the evaluation metric in Table \ref{tab:Evaluationfairness_metrics}. It provides a comprehensive overview of various key fairness metrics used in analysis of fair IM. Table \ref{tab:Evaluationfairness_metrics} contains its mathematical formula, core description, advantages, disadvantages, etc. However, in the experiments of this survey, we utilize the most widely use metric for the result analysis.

\section{Empirical Evaluation and Reproducibility}
\label{sec:Datasets1}

\subsection{Experimental Set-Up}
\begin{figure}[!ht]
    \centering
    \includegraphics[width=0.38\linewidth]{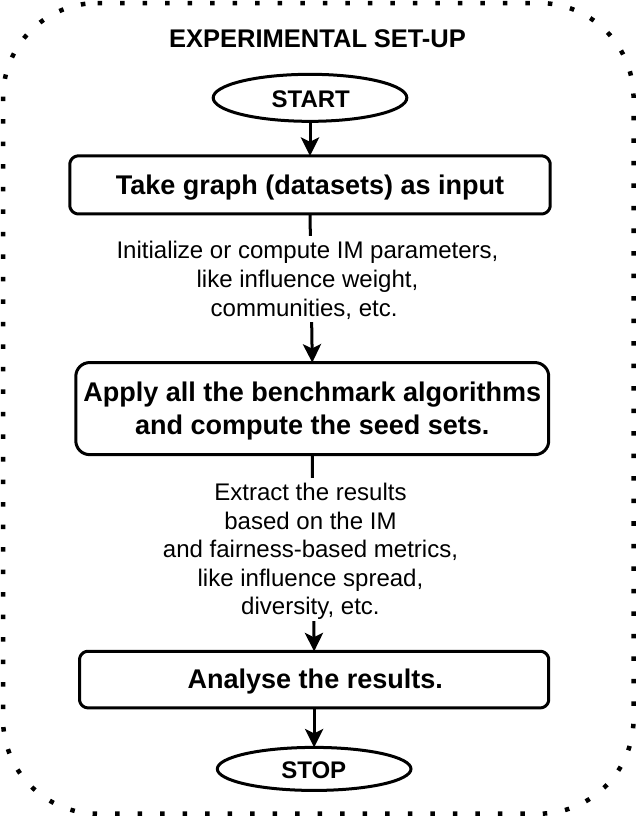}
    \caption{Flowchart of the experimental execution process.}
    \label{fig:EXPERIMENTAL}
\end{figure}

The framework of the experimental setup is shown in Fig. \ref{fig:EXPERIMENTAL}. In this figure, the experimental procedure for IM begins with taking the graph (datasets) as input. After taking the graph as input, we configure the graph by assigning the IM parameters and other attributes. The proposed work initialized or computed the attribute values. For example, the experiments initialize the influence weight and compute the communities using the Louvian method \cite{2008_CommunityDetection_FirstBaseLouvain}. In the next phase, we apply various benchmark algorithms to the input graph to identify and compute respective seed sets. Following the algorithmic execution, results are extracted and evaluated. We evaluate results using IM metrics, such as total influence spread, alongside fairness-based metrics. We also analyze the results based on the other fairness-based metric  to ensure a comprehensive assessment of the algorithms. The final stage of the workflow involves a detailed analysis of these results to draw meaningful conclusions based on the experimental outcomes. 

\subsection{Compared Algorithms}
This survey paper compares several existing IM algorithms. To analyze them effectively, this work classifies them as fairness-based and non-fairness-based algorithms. We have taken the following algorithms for the evaluation. 
\begin{itemize}
    \item \textbf{Fairness-based algorithms}: DCDIM \cite{G12.MEENA2025108045}, CDIM \cite{CDIM_CommunityDiversity_StaticGraph_JianxinLi}, SSR-PEA \cite{2023_Zhang_SSRPEA_Diversity_SpaceReduction_EvolutionaryAlgo}, FIMMOGA \cite{G11.GONG2023119138}, Bridge \cite{G14.10.1145/3664618}, DIM \cite{2014_Tang_IM_FirstDiversified_GeneralFreamework}, DBIM \cite{G13.10734665}, DALO \cite{2025_Mine_AntLionOptimixation_DALO}, PDTF-IM \cite{2025_TNSE_MineMultiFairness_PDTF_IM}.  
    \item \textbf{Non-fairness based algorithms}:  Greedy \cite{10.1145/956750.956769}, GWIM \cite{2020_Xareie_IM_NatureInsp_Wolf}, CELF \cite{CELF_Leskovec_2007}, CELF++ \cite{CELF_PP_2011_Goyal}, LMP \cite{2022_LMP_Bouyer_IM_Bridge}, PAGERANK \cite{PAGERANK_1999}, DEGREE \cite{centrality_Freeman_1978}, HUB \cite{2021_SocialNetworkAnalysis_Book_Tanmoy}, AUTHORITY \cite{2021_SocialNetworkAnalysis_Book_Tanmoy}, neighborhood coreness (N-Core) \cite{2021_SocialNetworkAnalysis_Book_Tanmoy}, RR \cite{RIS2014Borgs}, DGA \cite{2022_DynamicDGA_GeneticAlgorithmForDynamicNetwork}. 
\end{itemize}

\begin{table}[!ht]
\caption{Illustration of the Comparison of State-of-the-Art Algorithms in terms of Influence Spread.}
\label{tab:InfluencedSpread}
\resizebox{\textwidth}{!}{
\begin{tabular}{|lr|rrr|rrr|rrr|rrr|rrr|rrr|rrr|}
\toprule
\multicolumn{1}{|l}{\textbf{Datasets}} & \multicolumn{1}{l}{\textbf{\shortstack{Seed\\set\\size}}} & \multicolumn{1}{c}{\textbf{SSRPEA}} & \multicolumn{1}{c}{\textbf{Greedy}} & \multicolumn{1}{c}{\textbf{GWIM}} & \multicolumn{1}{c}{\textbf{CELF}} & \multicolumn{1}{c}{\textbf{CELFPP}} & \multicolumn{1}{c}{\textbf{PR}} & \multicolumn{1}{c}{\textbf{Degree}} & \multicolumn{1}{c}{\textbf{Hub}} & \multicolumn{1}{c}{\textbf{Auth. }} & \multicolumn{1}{c}{\textbf{\shortstack{Neigh.}}} & \multicolumn{1}{c}{\textbf{\shortstack{FIMM-\\OGA}}} & \multicolumn{1}{c}{\textbf{LMP}} & \multicolumn{1}{c}{\textbf{Bridge}} & \multicolumn{1}{c}{\textbf{DCDIM}} & \multicolumn{1}{c}{\textbf{CDIM}} & \multicolumn{1}{c}{\textbf{RR}} & \multicolumn{1}{c}{\textbf{IM\_C}} & \multicolumn{1}{c}{\textbf{DIM}} & \multicolumn{1}{c}{\textbf{DGA}} & \multicolumn{1}{c}{\textbf{PDTF}} & \multicolumn{1}{c|}{\textbf{DBIM}} \\
\midrule
\multirow{5}{*}{Facebook} & \textbf{2} & 295.0 & 766.0 & 398.0 & 573.0 & 425.0 & 2.0 & 187.0 & 563.0 & 12.0 & 563.0 & 271.0 & 177.0 & 760.0 & 766.0 & 766.0 & 113.0 & 381.0 & 329.0 & 295.0 & 316.0 & 766.0 \\
 & \textbf{4} & 563.0 & 990.0 & 622.0 & 577.0 & 445.0 & 13.0 & 207.0 & 564.0 & 21.0 & 563.0 & 333.0 & 207.0 & 768.0 & 990.0 & 990.0 & 472.0 & 897.0 & 579.0 & 633.0 & 316.0 & 882.0 \\
 & \textbf{8} & 420.0 & 1150.0 & 771.0 & 580.0 & 704.0 & 16.0 & 371.0 & 564.0 & 34.0 & 567.0 & 213.0 & 480.0 & 935.0 & 1150.0 & 1145.0 & 175.0 & 1127.0 & 683.0 & 509.0 & 316.0 & 1150.0 \\
 & \textbf{16} & 563.0 & 1282.0 & 989.0 & 585.0 & 850.0 & 77.0 & 679.0 & 564.0 & 58.0 & 567.0 & 561.0 & 522.0 & 1074.0 & 1282.0 & 1275.0 & 763.0 & 1272.0 & 1001.0 & 782.0 & 316.0 & 1246.0 \\
 & \textbf{32} & 565.0 & 1394.0 & 1003.0 & 809.0 & 1032.0 & 89.0 & 776.0 & 569.0 & 68.0 & 766.0 & 844.0 & 583.0 & 1177.0 & 1394.0 & 1388.0 & 665.0 & 1389.0 & 1079.0 & 862.0 & 316.0 & 1302.0 \\
 \midrule
\multirow{5}{*}{Congress} & \textbf{2} & 215.0 & 395.0 & 279.0 & 381.0 & 334.0 & 2.0 & 215.0 & 247.0 & 126.0 & 247.0 & 315.0 & 215.0 & 381.0 & 395.0 & 395.0 & 9.0 & 381.0 & 334.0 & 163.0 & 255.0 & 395.0 \\
 & \textbf{4} & 166.0 & 411.0 & 296.0 & 389.0 & 335.0 & 4.0 & 247.0 & 247.0 & 129.0 & 274.0 & 329.0 & 247.0 & 395.0 & 411.0 & 411.0 & 7.0 & 389.0 & 382.0 & 237.0 & 255.0 & 406.0 \\
 & \textbf{8} & 272.0 & 428.0 & 347.0 & 395.0 & 341.0 & 8.0 & 282.0 & 291.0 & 147.0 & 310.0 & 266.0 & 321.0 & 411.0 & 428.0 & 428.0 & 375.0 & 389.0 & 389.0 & 297.0 & 255.0 & 413.0 \\
 & \textbf{16} & 366.0 & 439.0 & 385.0 & 415.0 & 347.0 & 16.0 & 302.0 & 336.0 & 228.0 & 367.0 & 281.0 & 333.0 & 430.0 & 439.0 & 439.0 & 290.0 & 389.0 & 391.0 & 291.0 & 255.0 & 414.0 \\
 & \textbf{32} & 370.0 & 455.0 & 402.0 & 417.0 & 374.0 & 33.0 & 359.0 & 344.0 & 232.0 & 384.0 & 378.0 & 376.0 & 444.0 & 455.0 & 455.0 & 327.0 & 389.0 & 402.0 & 286.0 & 255.0 & 414.0 \\
 \midrule
\multirow{5}{*}{Deezer} & \textbf{2} & 38.0 & 41.0 & 32.0 & 41.0 & 35.0 & 2.0 & 32.0 & 20.0 & 2.0 & 35.0 & 2.0 & 32.0 & 38.0 & 31.0 & 41.0 & 24.0 & 23.0 & 19.0 & 9.0 & 754.0 & 22.0 \\
 & \textbf{4} & 50.0 & 76.0 & 65.0 & 76.0 & 55.0 & 4.0 & 62.0 & 23.0 & 4.0 & 65.0 & 6.0 & 62.0 & 70.0 & 70.0 & 76.0 & 26.0 & 61.0 & 49.0 & 39.0 & 754.0 & 27.0 \\
 & \textbf{8} & 92.0 & 142.0 & 105.0 & 142.0 & 90.0 & 8.0 & 105.0 & 34.0 & 8.0 & 97.0 & 10.0 & 105.0 & 122.0 & 138.0 & 138.0 & 37.0 & 129.0 & 96.0 & 14.0 & 754.0 & 54.0 \\
 & \textbf{16} & 196.0 & 249.0 & 187.0 & 225.0 & 145.0 & 17.0 & 190.0 & 44.0 & 16.0 & 179.0 & 21.0 & 190.0 & 235.0 & 248.0 & 248.0 & 73.0 & 239.0 & 162.0 & 23.0 & 754.0 & 81.0 \\
 & \textbf{32} & 325.0 & 424.0 & 318.0 & 384.0 & 303.0 & 40.0 & 344.0 & 218.0 & 33.0 & 334.0 & 61.0 & 346.0 & 387.0 & 424.0 & 424.0 & 48.0 & 416.0 & 292.0 & 57.0 & 754.0 & 133.0 \\
 \midrule
\multirow{5}{*}{LastFM} & \textbf{2} & 135.0 & 523.0 & 305.0 & 523.0 & 212.0 & 2.0 & 419.0 & 348.0 & 10.0 & 419.0 & 3.0 & 408.0 & 419.0 & 519.0 & 481.0 & 17.0 & 37.0 & 15.0 & 104.0 & 382.0 & 513.0 \\
 & \textbf{4} & 382.0 & 856.0 & 437.0 & 531.0 & 90.0 & 4.0 & 607.0 & 348.0 & 56.0 & 423.0 & 55.0 & 509.0 & 712.0 & 792.0 & 796.0 & 92.0 & 557.0 & 344.0 & 485.0 & 382.0 & 851.0 \\
 & \textbf{8} & 654.0 & 1228.0 & 681.0 & 694.0 & 356.0 & 8.0 & 832.0 & 356.0 & 102.0 & 610.0 & 11.0 & 812.0 & 1089.0 & 801.0 & 1198.0 & 149.0 & 1124.0 & 523.0 & 417.0 & 382.0 & 991.0 \\
 & \textbf{16} & 812.0 & 1497.0 & 812.0 & 1167.0 & 599.0 & 18.0 & 1146.0 & 375.0 & 141.0 & 888.0 & 128.0 & 1103.0 & 1235.0 & 811.0 & 1467.0 & 334.0 & 1477.0 & 641.0 & 316.0 & 382.0 & 1140.0 \\
 & \textbf{32} & 862.0 & 1745.0 & 999.0 & 1252.0 & 1068.0 & 34.0 & 1266.0 & 388.0 & 179.0 & 1205.0 & 388.0 & 1258.0 & 1481.0 & 1450.0 & 1615.0 & 385.0 & 1742.0 & 667.0 & 550.0 & 382.0 & 1430.0 \\
 \midrule
\multirow{5}{*}{Twitter} & \textbf{2} & 2156.0 & 3072.0 & 1563.0 & 2010.0 & 928.0 & 3.0 & 1813.0 & 894.0 & 4.0 & 2156.0 & 2.0 & 1813.0 & 3072.0 & 3072.0 & 3072.0 & 264.0 & 949.0 & 2266.0 & 966.0 & 3314.0 & 3072.0 \\
 & \textbf{4} & 2808.0 & 3662.0 & 2427.0 & 3072.0 & 1910.0 & 6.0 & 2898.0 & 959.0 & 158.0 & 2470.0 & 5.0 & 2898.0 & 3659.0 & 3659.0 & 3659.0 & 1709.0 & 3131.0 & 2643.0 & 1815.0 & 3314.0 & 3077.0 \\
 & \textbf{8} & 3679.0 & 4346.0 & 3063.0 & 3256.0 & 2469.0 & 10.0 & 3178.0 & 1585.0 & 163.0 & 3624.0 & 104.0 & 3178.0 & 3897.0 & 4336.0 & 4301.0 & 2647.0 & 4053.0 & 3235.0 & 766.0 & 3314.0 & 3087.0 \\
 & \textbf{16} & 3945.0 & 4913.0 & 3325.0 & 4077.0 & 3082.0 & 28.0 & 4383.0 & 1654.0 & 191.0 & 4241.0 & 22.0 & 4334.0 & 4666.0 & 4903.0 & 4882.0 & 336.0 & 4828.0 & 3817.0 & 1156.0 & 3314.0 & 3470.0 \\
 & \textbf{32} & 3815.0 & 5359.0 & 3000.0 & 4573.0 & 4358.0 & 47.0 & 4836.0 & 2546.0 & 215.0 & 4708.0 & 95.0 & 4836.0 & 5015.0 & 5011.0 & 5108.0 & 1456.0 & 5329.0 & 4109.0 & 1796.0 & 3314.0 & 3499.0\\
 \bottomrule
\end{tabular}
}
\end{table}

\begin{table}[!ht]
\caption{Illustration of the Comparison of State-of-the-Art Algorithms in terms of Influence Diversity (at least one node is influenced in the community). A high value indicates a high number of communities are influenced.}
\label{tab:diversity}
\resizebox{\textwidth}{!}{
\begin{tabular}{|lr|rrr|rrr|rrr|rrr|rrr|rrr|rrr|}
\toprule
\multicolumn{1}{|l}{\textbf{Datasets}} & \multicolumn{1}{l}{\textbf{\shortstack{Seed\\set\\size}}} & \multicolumn{1}{c}{\textbf{SSRPEA}} & \multicolumn{1}{c}{\textbf{Greedy}} & \multicolumn{1}{c}{\textbf{GWIM}} & \multicolumn{1}{c}{\textbf{CELF}} & \multicolumn{1}{c}{\textbf{CELFPP}} & \multicolumn{1}{c}{\textbf{PR}} & \multicolumn{1}{c}{\textbf{Degree}} & \multicolumn{1}{c}{\textbf{Hub}} & \multicolumn{1}{c}{\textbf{Auth. }} & \multicolumn{1}{c}{\textbf{\shortstack{Neigh.}}} & \multicolumn{1}{c}{\textbf{\shortstack{FIMM-\\OGA}}} & \multicolumn{1}{c}{\textbf{LMP}} & \multicolumn{1}{c}{\textbf{Bridge}} & \multicolumn{1}{c}{\textbf{DCDIM}} & \multicolumn{1}{c}{\textbf{CDIM}} & \multicolumn{1}{c}{\textbf{RR}} & \multicolumn{1}{c}{\textbf{IM\_C}} & \multicolumn{1}{c}{\textbf{DIM}} & \multicolumn{1}{c}{\textbf{DGA}} & \multicolumn{1}{c}{\textbf{PDTF}} & \multicolumn{1}{c|}{\textbf{DBIM}} \\
\midrule
\multirow{5}{*}{Facebook} & \textbf{2} & 7.0 & 11.0 & 11.0 & 11.0 & 5.0 & 2.0 & 10.0 & 11.0 & 3.0 & 11.0 & 7.0 & 7.0 & 11.0 & 11.0 & 11.0 & 3.0 & 7.0 & 7.0 & 7.0 & 11.0 & 11.0 \\
 & \textbf{4} & 11.0 & 11.0 & 10.0 & 11.0 & 6.0 & 4.0 & 10.0 & 11.0 & 3.0 & 11.0 & 11.0 & 10.0 & 11.0 & 11.0 & 11.0 & 10.0 & 11.0 & 8.0 & 11.0 & 11.0 & 11.0 \\
 & \textbf{8} & 5.0 & 11.0 & 10.0 & 11.0 & 11.0 & 5.0 & 10.0 & 11.0 & 3.0 & 11.0 & 9.0 & 10.0 & 11.0 & 11.0 & 11.0 & 6.0 & 11.0 & 10.0 & 11.0 & 11.0 & 11.0 \\
 & \textbf{16} & 11.0 & 11.0 & 11.0 & 11.0 & 11.0 & 9.0 & 10.0 & 11.0 & 3.0 & 11.0 & 11.0 & 11.0 & 11.0 & 11.0 & 11.0 & 11.0 & 11.0 & 11.0 & 11.0 & 11.0 & 11.0 \\
 & \textbf{32} & 11.0 & 11.0 & 11.0 & 11.0 & 11.0 & 9.0 & 11.0 & 11.0 & 3.0 & 11.0 & 11.0 & 11.0 & 11.0 & 11.0 & 11.0 & 10.0 & 11.0 & 11.0 & 11.0 & 11.0 & 11.0 \\
 \midrule
\multirow{5}{*}{Congress} & \textbf{2} & 3.0 & 3.0 & 3.0 & 3.0 & 3.0 & 1.0 & 3.0 & 3.0 & 3.0 & 3.0 & 3.0 & 3.0 & 3.0 & 3.0 & 3.0 & 3.0 & 3.0 & 3.0 & 3.0 & 3.0 & 3.0 \\
 & \textbf{4} & 3.0 & 3.0 & 3.0 & 3.0 & 3.0 & 3.0 & 3.0 & 3.0 & 3.0 & 3.0 & 3.0 & 3.0 & 3.0 & 3.0 & 3.0 & 3.0 & 3.0 & 3.0 & 3.0 & 3.0 & 3.0 \\
 & \textbf{8} & 3.0 & 3.0 & 3.0 & 3.0 & 3.0 & 3.0 & 3.0 & 3.0 & 3.0 & 3.0 & 3.0 & 3.0 & 3.0 & 3.0 & 3.0 & 3.0 & 3.0 & 3.0 & 3.0 & 3.0 & 3.0 \\
 & \textbf{16} & 3.0 & 3.0 & 3.0 & 3.0 & 3.0 & 3.0 & 3.0 & 3.0 & 3.0 & 3.0 & 3.0 & 3.0 & 3.0 & 3.0 & 3.0 & 3.0 & 3.0 & 3.0 & 3.0 & 3.0 & 3.0 \\
 & \textbf{32} & 3.0 & 3.0 & 3.0 & 3.0 & 3.0 & 3.0 & 3.0 & 3.0 & 3.0 & 3.0 & 3.0 & 3.0 & 3.0 & 3.0 & 3.0 & 3.0 & 3.0 & 3.0 & 3.0 & 3.0 & 3.0 \\
 \midrule
\multirow{5}{*}{Deezer} & \textbf{2} & 2.0 & 3.0 & 2.0 & 3.0 & 2.0 & 2.0 & 2.0 & 1.0 & 1.0 & 2.0 & 2.0 & 2.0 & 3.0 & 4.0 & 3.0 & 2.0 & 2.0 & 2.0 & 3.0 & 292.0 & 3.0 \\
 & \textbf{4} & 4.0 & 5.0 & 4.0 & 5.0 & 4.0 & 4.0 & 4.0 & 1.0 & 1.0 & 4.0 & 4.0 & 4.0 & 6.0 & 6.0 & 5.0 & 4.0 & 5.0 & 4.0 & 5.0 & 292.0 & 5.0 \\
 & \textbf{8} & 8.0 & 9.0 & 8.0 & 9.0 & 8.0 & 7.0 & 8.0 & 2.0 & 1.0 & 8.0 & 8.0 & 8.0 & 10.0 & 10.0 & 10.0 & 8.0 & 9.0 & 8.0 & 7.0 & 292.0 & 10.0 \\
 & \textbf{16} & 16.0 & 17.0 & 15.0 & 15.0 & 16.0 & 13.0 & 16.0 & 5.0 & 1.0 & 16.0 & 15.0 & 16.0 & 18.0 & 18.0 & 18.0 & 15.0 & 17.0 & 16.0 & 11.0 & 292.0 & 18.0 \\
 & \textbf{32} & 31.0 & 34.0 & 31.0 & 30.0 & 32.0 & 28.0 & 31.0 & 21.0 & 1.0 & 31.0 & 28.0 & 31.0 & 33.0 & 34.0 & 34.0 & 28.0 & 34.0 & 32.0 & 20.0 & 292.0 & 34.0 \\
 \midrule
\multirow{5}{*}{LastFM} & \textbf{2} & 5.0 & 12.0 & 15.0 & 12.0 & 7.0 & 2.0 & 12.0 & 10.0 & 1.0 & 12.0 & 2.0 & 11.0 & 12.0 & 16.0 & 17.0 & 3.0 & 2.0 & 4.0 & 4.0 & 41.0 & 16.0 \\
 & \textbf{4} & 10.0 & 17.0 & 16.0 & 12.0 & 7.0 & 3.0 & 16.0 & 10.0 & 1.0 & 12.0 & 4.0 & 13.0 & 19.0 & 22.0 & 20.0 & 6.0 & 12.0 & 13.0 & 16.0 & 41.0 & 21.0 \\
 & \textbf{8} & 14.0 & 22.0 & 14.0 & 13.0 & 10.0 & 4.0 & 17.0 & 10.0 & 3.0 & 17.0 & 5.0 & 16.0 & 21.0 & 26.0 & 22.0 & 8.0 & 22.0 & 14.0 & 12.0 & 41.0 & 25.0 \\
 & \textbf{16} & 18.0 & 22.0 & 20.0 & 22.0 & 15.0 & 9.0 & 20.0 & 12.0 & 3.0 & 20.0 & 8.0 & 20.0 & 25.0 & 34.0 & 22.0 & 13.0 & 22.0 & 19.0 & 9.0 & 41.0 & 30.0 \\
 & \textbf{32} & 18.0 & 22.0 & 22.0 & 22.0 & 19.0 & 13.0 & 21.0 & 12.0 & 5.0 & 21.0 & 15.0 & 20.0 & 25.0 & 41.0 & 28.0 & 18.0 & 22.0 & 35.0 & 15.0 & 41.0 & 30.0 \\
 \midrule
\multirow{5}{*}{Twitter} & \textbf{2} & 20.0 & 24.0 & 19.0 & 20.0 & 16.0 & 1.0 & 21.0 & 13.0 & 1.0 & 20.0 & 2.0 & 21.0 & 24.0 & 24.0 & 24.0 & 12.0 & 15.0 & 20.0 & 18.0 & 669.0 & 24.0 \\
 & \textbf{4} & 22.0 & 25.0 & 23.0 & 24.0 & 21.0 & 1.0 & 23.0 & 13.0 & 3.0 & 22.0 & 4.0 & 23.0 & 26.0 & 26.0 & 26.0 & 20.0 & 24.0 & 24.0 & 21.0 & 669.0 & 26.0 \\
 & \textbf{8} & 25.0 & 26.0 & 23.0 & 24.0 & 25.0 & 3.0 & 24.0 & 21.0 & 3.0 & 25.0 & 7.0 & 24.0 & 26.0 & 27.0 & 27.0 & 25.0 & 26.0 & 25.0 & 10.0 & 669.0 & 30.0 \\
 & \textbf{16} & 25.0 & 27.0 & 21.0 & 26.0 & 25.0 & 7.0 & 26.0 & 21.0 & 3.0 & 27.0 & 12.0 & 26.0 & 27.0 & 28.0 & 28.0 & 13.0 & 27.0 & 26.0 & 19.0 & 669.0 & 38.0 \\
 & \textbf{32} & 26.0 & 28.0 & 25.0 & 27.0 & 28.0 & 11.0 & 27.0 & 24.0 & 3.0 & 27.0 & 17.0 & 27.0 & 28.0 & 43.0 & 39.0 & 26.0 & 28.0 & 35.0 & 22.0 & 669.0 & 54.0\\
 \bottomrule
\end{tabular}
}
\end{table}

\begin{table}[!ht]
\caption{Illustration of the Comparison of State-of-the-Art Algorithms in terms of Equality.}
\label{tab:Equality}
\resizebox{\textwidth}{!}{
\begin{tabular}{|lr|rrr|rrr|rrr|rrr|rrr|rrr|rrr|}
\toprule
\multicolumn{1}{|l}{\textbf{Datasets}} & \multicolumn{1}{l}{\textbf{\shortstack{Seed\\set\\size}}} & \multicolumn{1}{c}{\textbf{SSRPEA}} & \multicolumn{1}{c}{\textbf{Greedy}} & \multicolumn{1}{c}{\textbf{GWIM}} & \multicolumn{1}{c}{\textbf{CELF}} & \multicolumn{1}{c}{\textbf{CELFPP}} & \multicolumn{1}{c}{\textbf{PR}} & \multicolumn{1}{c}{\textbf{Degree}} & \multicolumn{1}{c}{\textbf{Hub}} & \multicolumn{1}{c}{\textbf{Auth. }} & \multicolumn{1}{c}{\textbf{\shortstack{Neigh.}}} & \multicolumn{1}{c}{\textbf{\shortstack{FIMM-\\OGA}}} & \multicolumn{1}{c}{\textbf{LMP}} & \multicolumn{1}{c}{\textbf{Bridge}} & \multicolumn{1}{c}{\textbf{DCDIM}} & \multicolumn{1}{c}{\textbf{CDIM}} & \multicolumn{1}{c}{\textbf{RR}} & \multicolumn{1}{c}{\textbf{IM\_C}} & \multicolumn{1}{c}{\textbf{DIM}} & \multicolumn{1}{c}{\textbf{DGA}} & \multicolumn{1}{c}{\textbf{PDTF}} & \multicolumn{1}{c|}{\textbf{DBIM}} \\
\midrule
\multirow{5}{*}{Facebook} & \textbf{2} & 0.0 & 0.0 & 0.0 & 0.0 & 0.0 & 0.0 & 0.0 & 0.0 & 0.0 & 0.0 & 0.0 & 0.0 & 0.0 & 0.0 & 0.0 & 0.0 & 0.0 & 0.0 & 0.0 & 0.1 & 0.0 \\
 & \textbf{4} & 0.0 & 0.0 & 0.0 & 0.0 & 0.0 & 0.0 & 0.0 & 0.0 & 0.0 & 0.0 & 0.0 & 0.0 & 0.0 & 0.0 & 0.0 & 0.0 & 0.0 & 0.0 & 0.0 & 0.1 & 0.0 \\
 & \textbf{8} & 0.0 & 0.1 & 0.0 & 0.0 & 0.0 & 0.0 & 0.0 & 0.0 & 0.0 & 0.0 & 0.0 & 0.0 & 0.0 & 0.1 & 0.1 & 0.1 & 0.1 & 0.1 & 0.0 & 0.1 & 0.1 \\
 & \textbf{16} & 0.1 & 0.1 & 0.1 & 0.1 & 0.1 & 0.1 & 0.1 & 0.1 & 0.1 & 0.1 & 0.1 & 0.1 & 0.1 & 0.1 & 0.1 & 0.1 & 0.1 & 0.1 & 0.0 & 0.1 & 0.1 \\
 & \textbf{32} & 0.1 & 0.2 & 0.2 & 0.1 & 0.2 & 0.1 & 0.1 & 0.1 & 0.1 & 0.1 & 0.2 & 0.1 & 0.2 & 0.2 & 0.2 & 0.2 & 0.2 & 0.2 & 0.1 & 0.1 & 0.2 \\
 \midrule
\multirow{5}{*}{Congress} & \textbf{2} & 0.0 & 0.0 & 0.0 & 0.0 & 0.0 & 0.0 & 0.0 & 0.0 & 0.0 & 0.0 & 0.0 & 0.0 & 0.0 & 0.0 & 0.0 & 0.0 & 0.0 & 0.0 & 0.0 & 0.0 & 0.0 \\
 & \textbf{4} & 0.0 & 0.0 & 0.0 & 0.0 & 0.0 & 0.0 & 0.0 & 0.0 & 0.0 & 0.0 & 0.0 & 0.0 & 0.0 & 0.0 & 0.0 & 0.0 & 0.0 & 0.0 & 0.0 & 0.0 & 0.0 \\
 & \textbf{8} & 0.0 & 0.1 & 0.1 & 0.1 & 0.1 & 0.1 & 0.0 & 0.0 & 0.0 & 0.0 & 0.1 & 0.1 & 0.1 & 0.1 & 0.1 & 0.1 & 0.0 & 0.1 & 0.0 & 0.0 & 0.1 \\
 & \textbf{16} & 0.1 & 0.1 & 0.1 & 0.2 & 0.1 & 0.1 & 0.1 & 0.1 & 0.1 & 0.1 & 0.1 & 0.1 & 0.1 & 0.1 & 0.1 & 0.1 & 0.0 & 0.1 & 0.1 & 0.0 & 0.1 \\
 & \textbf{32} & 0.2 & 0.2 & 0.2 & 0.3 & 0.2 & 0.2 & 0.2 & 0.2 & 0.2 & 0.2 & 0.2 & 0.2 & 0.2 & 0.2 & 0.2 & 0.2 & 0.0 & 0.2 & 0.1 & 0.0 & 0.1 \\
 \midrule
\multirow{5}{*}{Deezer} & \textbf{2} & 0.0 & 0.0 & 0.0 & 0.0 & 0.0 & 0.0 & 0.0 & 0.0 & 0.0 & 0.0 & 0.0 & 0.0 & 0.0 & 0.0 & 0.0 & 0.0 & 0.0 & 0.0 & 0.0 & 9.1 & 0.3 \\
 & \textbf{4} & 0.0 & 0.0 & 0.0 & 0.0 & 0.0 & 0.0 & 0.0 & 0.0 & 0.0 & 0.0 & 0.0 & 0.0 & 0.0 & 0.0 & 0.0 & 0.0 & 0.0 & 0.0 & 0.0 & 9.1 & 0.7 \\
 & \textbf{8} & 0.1 & 0.0 & 0.0 & 0.0 & 0.0 & 0.1 & 0.0 & 0.0 & 0.0 & 0.0 & 0.1 & 0.0 & 0.0 & 0.0 & 0.0 & 0.1 & 0.1 & 0.0 & 0.1 & 9.1 & 1.0 \\
 & \textbf{16} & 0.1 & 0.1 & 0.1 & 0.1 & 0.1 & 0.2 & 0.1 & 0.1 & 0.1 & 0.1 & 0.2 & 0.1 & 0.1 & 0.1 & 0.1 & 0.2 & 0.1 & 0.1 & 0.1 & 9.1 & 2.0 \\
 & \textbf{32} & 0.2 & 0.2 & 0.2 & 0.2 & 0.2 & 0.4 & 0.2 & 0.2 & 0.1 & 0.2 & 0.4 & 0.2 & 0.2 & 0.2 & 0.2 & 0.3 & 0.2 & 0.2 & 0.3 & 9.1 & 3.5 \\
 \midrule
\multirow{5}{*}{LastFM} & \textbf{2} & 0.0 & 0.0 & 0.0 & 0.0 & 0.0 & 0.0 & 0.0 & 0.0 & 0.0 & 0.0 & 0.0 & 0.0 & 0.0 & 0.0 & 0.0 & 0.0 & 0.0 & 0.0 & 0.0 & 5.5 & 0.0 \\
 & \textbf{4} & 0.0 & 0.0 & 0.0 & 0.0 & 0.0 & 0.0 & 0.0 & 0.0 & 0.0 & 0.0 & 0.0 & 0.0 & 0.0 & 0.0 & 0.0 & 0.0 & 0.0 & 0.0 & 0.0 & 5.5 & 0.0 \\
 & \textbf{8} & 0.0 & 0.0 & 0.0 & 0.0 & 0.0 & 0.0 & 0.0 & 0.0 & 0.0 & 0.0 & 0.0 & 0.0 & 0.0 & 1.1 & 0.0 & 0.0 & 0.0 & 0.0 & 0.0 & 5.5 & 0.7 \\
 & \textbf{16} & 0.0 & 0.0 & 0.0 & 0.0 & 0.0 & 0.1 & 0.0 & 0.0 & 0.0 & 0.0 & 0.1 & 0.0 & 1.1 & 2.7 & 0.1 & 0.1 & 0.0 & 0.2 & 0.0 & 5.5 & 1.4 \\
 & \textbf{32} & 0.1 & 0.1 & 0.1 & 0.1 & 0.1 & 0.1 & 0.1 & 0.0 & 0.0 & 0.1 & 0.1 & 0.1 & 1.1 & 5.3 & 1.7 & 0.1 & 0.1 & 4.9 & 0.1 & 5.5 & 2.0 \\
 \midrule
\multirow{5}{*}{Twitter} & \textbf{2} & 0.0 & 0.0 & 0.0 & 0.0 & 0.0 & 0.0 & 0.0 & 0.0 & 0.0 & 0.0 & 0.0 & 0.0 & 0.0 & 0.0 & 0.0 & 0.0 & 0.0 & 0.0 & 0.0 & 278.4 & 0.0 \\
 & \textbf{4} & 0.0 & 0.0 & 0.0 & 0.0 & 0.0 & 0.0 & 0.0 & 0.0 & 0.0 & 0.0 & 0.1 & 0.0 & 0.0 & 0.0 & 0.0 & 0.0 & 0.0 & 0.0 & 0.0 & 278.4 & 0.8 \\
 & \textbf{8} & 0.0 & 0.0 & 0.0 & 0.0 & 0.0 & 0.0 & 0.0 & 0.0 & 0.0 & 0.0 & 0.3 & 0.0 & 0.0 & 0.0 & 0.0 & 0.0 & 0.0 & 0.0 & 0.5 & 278.4 & 1.9 \\
 & \textbf{16} & 0.0 & 0.0 & 0.0 & 0.0 & 0.0 & 0.0 & 0.0 & 0.0 & 0.0 & 0.0 & 0.5 & 0.0 & 0.0 & 0.0 & 0.0 & 0.0 & 0.0 & 0.0 & 0.0 & 278.4 & 3.6 \\
 & \textbf{32} & 0.0 & 0.0 & 0.0 & 0.0 & 0.0 & 0.1 & 0.0 & 0.0 & 0.0 & 0.0 & 1.0 & 0.0 & 0.0 & 1.8 & 1.4 & 1.0 & 0.0 & 0.6 & 0.2 & 278.4 & 8.1\\
 \bottomrule
\end{tabular}
}
\end{table}

\begin{table}[!ht]
\caption{Illustration of the Comparison of State-of-the-Art Algorithms in terms of Equity. A high Equity sum across communities suggests that the influence is effectively spread across different groups.}
\label{tab:Equity}
\resizebox{\textwidth}{!}{
\begin{tabular}{|lr|rrr|rrr|rrr|rrr|rrr|rrr|rrr|}
\toprule
\multicolumn{1}{|l}{\textbf{Datasets}} & \multicolumn{1}{l}{\textbf{\shortstack{Seed\\set\\size}}} & \multicolumn{1}{c}{\textbf{SSRPEA}} & \multicolumn{1}{c}{\textbf{Greedy}} & \multicolumn{1}{c}{\textbf{GWIM}} & \multicolumn{1}{c}{\textbf{CELF}} & \multicolumn{1}{c}{\textbf{CELFPP}} & \multicolumn{1}{c}{\textbf{PR}} & \multicolumn{1}{c}{\textbf{Degree}} & \multicolumn{1}{c}{\textbf{Hub}} & \multicolumn{1}{c}{\textbf{Auth. }} & \multicolumn{1}{c}{\textbf{\shortstack{Neigh.}}} & \multicolumn{1}{c}{\textbf{\shortstack{FIMM-\\OGA}}} & \multicolumn{1}{c}{\textbf{LMP}} & \multicolumn{1}{c}{\textbf{Bridge}} & \multicolumn{1}{c}{\textbf{DCDIM}} & \multicolumn{1}{c}{\textbf{CDIM}} & \multicolumn{1}{c}{\textbf{RR}} & \multicolumn{1}{c}{\textbf{IM\_C}} & \multicolumn{1}{c}{\textbf{DIM}} & \multicolumn{1}{c}{\textbf{DGA}} & \multicolumn{1}{c}{\textbf{PDTF}} & \multicolumn{1}{c|}{\textbf{DBIM}} \\
\midrule
\multirow{5}{*}{Facebook} & \textbf{2} & 1.5 & 4.0 & 2.2 & 3.1 & 2.1 & 0.0 & 1.1 & 3.0 & 0.1 & 3.0 & 1.4 & 1.0 & 3.9 & 4.0 & 4.0 & 0.6 & 2.0 & 1.8 & 1.5 & 2.0 & 4.0 \\
 & \textbf{4} & 3.0 & 4.7 & 3.2 & 3.1 & 2.4 & 0.1 & 1.1 & 3.0 & 0.1 & 3.0 & 2.0 & 1.1 & 4.0 & 4.7 & 4.7 & 2.4 & 4.4 & 2.8 & 3.4 & 2.0 & 5.0 \\
 & \textbf{8} & 2.1 & 6.3 & 3.7 & 3.1 & 3.8 & 0.1 & 1.8 & 3.0 & 0.2 & 3.0 & 1.1 & 2.4 & 4.8 & 6.3 & 6.4 & 0.7 & 6.1 & 3.5 & 2.8 & 2.0 & 6.3 \\
 & \textbf{16} & 3.0 & 7.0 & 4.9 & 3.1 & 4.2 & 0.3 & 3.3 & 3.0 & 0.3 & 3.0 & 2.8 & 2.8 & 6.0 & 7.0 & 7.0 & 3.9 & 6.9 & 5.3 & 3.9 & 2.0 & 6.9 \\
 & \textbf{32} & 3.0 & 7.6 & 5.0 & 4.2 & 5.2 & 0.4 & 4.0 & 3.0 & 0.3 & 4.0 & 4.1 & 3.1 & 6.6 & 7.6 & 7.7 & 3.4 & 7.5 & 5.7 & 4.2 & 2.0 & 7.1 \\
 \midrule
\multirow{5}{*}{Congress} & \textbf{2} & 1.2 & 2.5 & 1.7 & 2.4 & 2.0 & 0.0 & 1.2 & 1.4 & 0.7 & 1.4 & 1.8 & 1.2 & 2.4 & 2.5 & 2.5 & 0.1 & 2.4 & 2.0 & 0.9 & 1.5 & 2.5 \\
 & \textbf{4} & 1.0 & 2.6 & 1.8 & 2.4 & 2.0 & 0.0 & 1.4 & 1.4 & 0.8 & 1.6 & 2.0 & 1.4 & 2.5 & 2.6 & 2.6 & 0.1 & 2.4 & 2.4 & 1.3 & 1.5 & 2.5 \\
 & \textbf{8} & 1.6 & 2.7 & 2.1 & 2.5 & 2.1 & 0.1 & 1.6 & 1.7 & 0.8 & 1.8 & 1.5 & 1.9 & 2.6 & 2.7 & 2.7 & 2.3 & 2.4 & 2.4 & 1.7 & 1.5 & 2.6 \\
 & \textbf{16} & 2.2 & 2.8 & 2.4 & 2.6 & 2.1 & 0.1 & 1.7 & 1.9 & 1.3 & 2.2 & 1.7 & 2.0 & 2.7 & 2.8 & 2.8 & 1.7 & 2.4 & 2.4 & 1.7 & 1.5 & 2.6 \\
 & \textbf{32} & 2.3 & 2.8 & 2.5 & 2.6 & 2.3 & 0.2 & 2.1 & 2.0 & 1.3 & 2.4 & 2.3 & 2.3 & 2.8 & 2.8 & 2.9 & 2.0 & 2.4 & 2.5 & 1.6 & 1.5 & 2.6 \\
 \midrule
\multirow{5}{*}{Deezer} & \textbf{2} & 0.2 & 0.2 & 0.1 & 0.2 & 0.1 & 0.0 & 0.1 & 0.1 & 0.0 & 0.1 & 0.0 & 0.1 & 0.2 & 0.1 & 0.2 & 0.1 & 0.1 & 0.1 & 0.1 & 18.2 & 0.6 \\
 & \textbf{4} & 0.2 & 0.3 & 0.2 & 0.3 & 0.2 & 0.0 & 0.2 & 0.1 & 0.0 & 0.2 & 0.0 & 0.2 & 0.3 & 0.3 & 0.3 & 0.1 & 0.3 & 0.2 & 0.2 & 18.2 & 1.5 \\
 & \textbf{8} & 0.5 & 0.7 & 0.4 & 0.6 & 0.4 & 0.1 & 0.5 & 0.2 & 0.0 & 0.3 & 0.1 & 0.5 & 0.6 & 0.6 & 0.6 & 0.3 & 0.6 & 0.4 & 0.1 & 18.2 & 2.7 \\
 & \textbf{16} & 1.1 & 1.2 & 0.9 & 1.1 & 0.8 & 0.2 & 1.1 & 0.3 & 0.1 & 0.8 & 0.2 & 1.1 & 1.3 & 1.2 & 1.2 & 0.6 & 1.3 & 0.8 & 0.1 & 18.2 & 5.2 \\
 & \textbf{32} & 1.9 & 2.9 & 1.9 & 2.6 & 1.8 & 0.5 & 2.2 & 1.2 & 0.1 & 1.9 & 0.7 & 2.3 & 2.4 & 2.9 & 2.9 & 0.6 & 2.8 & 1.6 & 0.4 & 18.2 & 8.7 \\
 \midrule
\multirow{5}{*}{LastFM} & \textbf{2} & 0.2 & 0.9 & 0.6 & 0.9 & 0.3 & 0.0 & 0.6 & 0.6 & 0.0 & 0.6 & 0.0 & 0.7 & 0.6 & 0.9 & 1.1 & 0.0 & 0.1 & 0.0 & 0.1 & 7.5 & 1.2 \\
 & \textbf{4} & 0.6 & 1.6 & 0.8 & 0.9 & 0.2 & 0.0 & 1.2 & 0.6 & 0.1 & 0.7 & 0.1 & 0.8 & 1.5 & 1.8 & 1.8 & 0.2 & 1.0 & 0.6 & 1.1 & 7.5 & 1.9 \\
 & \textbf{8} & 1.1 & 2.6 & 1.1 & 1.2 & 0.7 & 0.0 & 1.7 & 0.6 & 0.1 & 1.3 & 0.0 & 1.6 & 2.3 & 4.2 & 2.8 & 0.3 & 2.5 & 1.0 & 0.7 & 7.5 & 3.0 \\
 & \textbf{16} & 1.7 & 3.3 & 1.7 & 2.6 & 1.1 & 0.1 & 2.3 & 0.6 & 0.1 & 1.9 & 0.2 & 2.2 & 3.7 & 6.2 & 3.6 & 0.7 & 3.2 & 1.5 & 0.6 & 7.5 & 4.5 \\
 & \textbf{32} & 1.8 & 4.3 & 2.3 & 2.8 & 2.2 & 0.1 & 2.6 & 0.6 & 0.3 & 2.5 & 0.8 & 2.5 & 4.3 & 10.2 & 7.7 & 0.8 & 4.2 & 7.8 & 1.0 & 7.5 & 6.3 \\
 \midrule
\multirow{5}{*}{Twitter} & \textbf{2} & 1.3 & 1.9 & 1.1 & 1.1 & 0.7 & 0.0 & 1.2 & 0.7 & 0.0 & 1.3 & 0.0 & 1.2 & 1.9 & 1.9 & 1.9 & 0.2 & 0.5 & 1.4 & 0.8 & 335.1 & 1.9 \\
 & \textbf{4} & 1.7 & 2.2 & 1.6 & 1.9 & 1.3 & 0.0 & 1.8 & 0.7 & 0.1 & 1.5 & 0.1 & 1.8 & 2.3 & 2.3 & 2.3 & 1.1 & 1.9 & 1.6 & 1.2 & 335.1 & 3.9 \\
 & \textbf{8} & 2.3 & 2.7 & 1.9 & 2.0 & 1.6 & 0.0 & 2.0 & 1.1 & 0.1 & 2.3 & 0.4 & 2.0 & 2.4 & 2.7 & 2.7 & 1.6 & 2.5 & 2.0 & 1.5 & 335.1 & 6.5 \\
 & \textbf{16} & 2.4 & 3.1 & 2.0 & 2.5 & 1.9 & 0.0 & 2.7 & 1.1 & 0.1 & 2.7 & 0.5 & 2.6 & 2.8 & 3.1 & 3.1 & 0.2 & 3.0 & 2.5 & 0.7 & 335.1 & 10.3 \\
 & \textbf{32} & 2.4 & 3.4 & 1.9 & 2.9 & 2.8 & 0.1 & 3.0 & 1.6 & 0.1 & 3.0 & 1.0 & 3.0 & 3.1 & 8.7 & 7.8 & 2.2 & 3.4 & 3.9 & 1.7 & 335.1 & 17.1\\
 \bottomrule
\end{tabular}
}
\end{table}

\begin{table}[!ht]
\caption{Illustration of the Comparison of State-of-the-Art Algorithms in terms of Disparity. Minimizing the disparity (penalty) shows the equitable distribution of
influence.}
\label{tab:Disparity}
\resizebox{\textwidth}{!}{
\begin{tabular}{|lr|rrr|rrr|rrr|rrr|rrr|rrr|rrr|}
\toprule
\multicolumn{1}{|l}{\textbf{Datasets}} & \multicolumn{1}{l}{\textbf{\shortstack{Seed\\set\\size}}} & \multicolumn{1}{c}{\textbf{SSRPEA}} & \multicolumn{1}{c}{\textbf{Greedy}} & \multicolumn{1}{c}{\textbf{GWIM}} & \multicolumn{1}{c}{\textbf{CELF}} & \multicolumn{1}{c}{\textbf{CELFPP}} & \multicolumn{1}{c}{\textbf{PR}} & \multicolumn{1}{c}{\textbf{Degree}} & \multicolumn{1}{c}{\textbf{Hub}} & \multicolumn{1}{c}{\textbf{Auth. }} & \multicolumn{1}{c}{\textbf{\shortstack{Neigh.}}} & \multicolumn{1}{c}{\textbf{\shortstack{FIMM-\\OGA}}} & \multicolumn{1}{c}{\textbf{LMP}} & \multicolumn{1}{c}{\textbf{Bridge}} & \multicolumn{1}{c}{\textbf{DCDIM}} & \multicolumn{1}{c}{\textbf{CDIM}} & \multicolumn{1}{c}{\textbf{RR}} & \multicolumn{1}{c}{\textbf{IM\_C}} & \multicolumn{1}{c}{\textbf{DIM}} & \multicolumn{1}{c}{\textbf{DGA}} & \multicolumn{1}{c}{\textbf{PDTF}} & \multicolumn{1}{c|}{\textbf{DBIM}} \\
\midrule
\multirow{5}{*}{Facebook} & \textbf{2} & 43.8 & 17.8 & 23.0 & 21.1 & 49.7 & 75.6 & 35.6 & 21.1 & 69.9 & 21.1 & 41.4 & 45.4 & 17.8 & 17.8 & 17.8 & 64.1 & 42.1 & 40.9 & 43.8 & 21.9 & 17.8 \\
 & \textbf{4} & 21.1 & 11.5 & 22.5 & 21.1 & 45.0 & 67.1 & 33.2 & 21.1 & 68.9 & 21.1 & 23.9 & 33.2 & 17.8 & 11.5 & 11.5 & 26.8 & 15.4 & 32.8 & 16.9 & 21.9 & 14.0 \\
 & \textbf{8} & 49.7 & 5.8 & 18.3 & 21.1 & 13.8 & 64.4 & 28.8 & 21.1 & 68.2 & 21.1 & 36.2 & 23.6 & 13.9 & 5.8 & 5.0 & 48.7 & 6.8 & 19.6 & 20.9 & 21.9 & 5.8 \\
 & \textbf{16} & 21.1 & 2.0 & 10.0 & 21.1 & 11.6 & 46.2 & 20.4 & 21.1 & 67.6 & 21.1 & 18.3 & 18.7 & 7.3 & 2.0 & 1.8 & 13.9 & 2.0 & 8.2 & 14.1 & 21.9 & 2.8 \\
 & \textbf{32} & 21.1 & 1.6 & 8.3 & 17.6 & 7.5 & 43.0 & 15.4 & 21.1 & 67.4 & 17.8 & 11.1 & 18.7 & 4.2 & 1.6 & 0.7 & 19.3 & 1.6 & 7.0 & 13.8 & 21.9 & 2.6 \\
 \midrule
\multirow{5}{*}{Congress} & \textbf{2} & 2.9 & 0.0 & 1.5 & 0.0 & 0.0 & 19.0 & 2.9 & 2.0 & 4.7 & 2.0 & 1.1 & 2.9 & 0.0 & 0.0 & 0.0 & 12.3 & 0.0 & 0.0 & 4.3 & 1.4 & 0.0 \\
 & \textbf{4} & 3.8 & 0.0 & 0.8 & 0.0 & 0.0 & 14.4 & 2.0 & 2.0 & 4.5 & 1.5 & 0.0 & 2.0 & 0.0 & 0.0 & 0.0 & 13.0 & 0.0 & 0.0 & 2.0 & 1.4 & 0.0 \\
 & \textbf{8} & 1.6 & 0.0 & 0.0 & 0.0 & 0.0 & 12.8 & 1.6 & 1.4 & 4.1 & 1.2 & 1.3 & 0.8 & 0.0 & 0.0 & 0.0 & 0.0 & 0.0 & 0.0 & 1.6 & 1.4 & 0.0 \\
 & \textbf{16} & 0.0 & 0.0 & 0.0 & 0.0 & 0.0 & 10.6 & 1.3 & 1.1 & 2.7 & 0.0 & 0.8 & 0.8 & 0.0 & 0.0 & 0.0 & 1.4 & 0.0 & 0.0 & 1.4 & 1.4 & 0.0 \\
 & \textbf{32} & 0.0 & 0.0 & 0.0 & 0.0 & 0.0 & 8.6 & 0.7 & 1.1 & 2.0 & 0.0 & 0.0 & 0.0 & 0.0 & 0.0 & 0.0 & 0.8 & 0.0 & 0.0 & 1.5 & 1.4 & 0.0 \\
 \midrule
\multirow{5}{*}{Deezer} & \textbf{2} & 1861.1 & 1857.1 & 1861.5 & 1857.1 & 1861.3 & 1867.0 & 1861.5 & 1866.3 & 1868.6 & 1861.3 & 1867.0 & 1861.5 & 1857.2 & 1855.5 & 1857.1 & 1862.9 & 1863.3 & 1862.5 & 1862.8 & 1000.6 & 1858.8 \\
 & \textbf{4} & 1852.6 & 1846.8 & 1851.3 & 1846.8 & 1852.4 & 1862.4 & 1851.6 & 1866.2 & 1867.9 & 1851.3 & 1861.3 & 1851.6 & 1845.0 & 1845.0 & 1846.8 & 1857.2 & 1848.9 & 1852.7 & 1852.8 & 1000.6 & 1851.7 \\
 & \textbf{8} & 1834.5 & 1826.3 & 1833.1 & 1826.3 & 1836.1 & 1854.8 & 1833.2 & 1862.0 & 1867.2 & 1834.1 & 1852.1 & 1833.2 & 1825.6 & 1824.5 & 1824.5 & 1845.0 & 1828.4 & 1833.9 & 1852.6 & 1000.6 & 1833.7 \\
 & \textbf{16} & 1795.5 & 1787.2 & 1800.1 & 1796.8 & 1803.8 & 1838.9 & 1796.2 & 1853.1 & 1866.5 & 1797.9 & 1833.2 & 1796.2 & 1786.2 & 1785.1 & 1785.1 & 1820.3 & 1789.0 & 1801.5 & 1841.1 & 1000.6 & 1806.7 \\
 & \textbf{32} & 1731.7 & 1709.9 & 1731.5 & 1728.3 & 1731.7 & 1801.6 & 1727.4 & 1779.0 & 1865.8 & 1730.3 & 1795.8 & 1727.2 & 1717.2 & 1709.9 & 1709.9 & 1795.9 & 1711.5 & 1734.2 & 1817.8 & 1000.6 & 1754.0 \\
 \midrule
\multirow{5}{*}{LastFM} & \textbf{2} & 218.1 & 180.3 & 177.1 & 180.3 & 205.3 & 232.9 & 188.6 & 196.9 & 232.9 & 188.6 & 232.2 & 189.0 & 188.6 & 172.2 & 159.4 & 227.3 & 227.6 & 224.2 & 221.3 & 97.9 & 167.7 \\
 & \textbf{4} & 191.8 & 154.1 & 166.3 & 180.3 & 208.3 & 229.9 & 161.6 & 196.9 & 231.2 & 188.6 & 222.6 & 179.8 & 151.3 & 141.3 & 139.1 & 214.9 & 178.6 & 180.4 & 168.1 & 97.9 & 137.1 \\
 & \textbf{8} & 173.0 & 126.4 & 167.1 & 172.0 & 185.7 & 226.2 & 151.9 & 196.9 & 225.3 & 160.2 & 222.4 & 154.1 & 130.9 & 128.2 & 124.6 & 202.3 & 128.8 & 167.1 & 185.5 & 97.9 & 124.4 \\
 & \textbf{16} & 149.2 & 119.6 & 138.2 & 127.9 & 161.6 & 211.9 & 134.2 & 192.2 & 225.0 & 142.5 & 208.6 & 136.8 & 119.6 & 108.0 & 116.5 & 181.0 & 120.4 & 148.7 & 192.6 & 97.9 & 106.1 \\
 & \textbf{32} & 151.5 & 111.7 & 126.8 & 126.8 & 139.3 & 198.8 & 130.4 & 192.2 & 218.8 & 132.7 & 174.4 & 134.7 & 113.4 & 69.9 & 88.7 & 162.6 & 112.2 & 103.2 & 170.6 & 97.9 & 92.7 \\
 \midrule
\multirow{5}{*}{Twitter} & \textbf{2} & 2184.5 & 2168.1 & 2192.7 & 2187.8 & 2216.4 & 2288.5 & 2188.3 & 2238.6 & 2288.2 & 2184.5 & 2287.3 & 2188.3 & 2168.1 & 2168.1 & 2168.1 & 2234.9 & 2211.5 & 2182.8 & 2212.7 & 295.6 & 2168.1 \\
 & \textbf{4} & 2171.1 & 2159.1 & 2170.3 & 2168.1 & 2183.8 & 2287.8 & 2167.3 & 2238.6 & 2279.2 & 2177.1 & 2282.0 & 2167.3 & 2154.3 & 2154.3 & 2154.3 & 2187.2 & 2167.5 & 2167.0 & 2187.8 & 295.6 & 2161.7 \\
 & \textbf{8} & 2157.8 & 2151.9 & 2166.6 & 2166.9 & 2166.9 & 2282.9 & 2163.8 & 2195.1 & 2279.2 & 2157.2 & 2266.7 & 2163.8 & 2153.5 & 2147.9 & 2147.3 & 2165.4 & 2153.2 & 2158.5 & 2252.5 & 295.6 & 2147.3 \\
 & \textbf{16} & 2156.6 & 2144.9 & 2170.1 & 2150.4 & 2161.8 & 2269.1 & 2151.8 & 2195.0 & 2278.3 & 2147.8 & 2259.1 & 2151.9 & 2146.8 & 2142.5 & 2141.3 & 2233.8 & 2145.3 & 2149.8 & 2197.2 & 295.6 & 2112.3 \\
 & \textbf{32} & 2157.1 & 2138.7 & 2158.4 & 2145.9 & 2147.0 & 2254.6 & 2145.2 & 2173.4 & 2278.2 & 2144.9 & 2237.9 & 2145.2 & 2142.4 & 2088.2 & 2097.2 & 2167.3 & 2138.7 & 2120.2 & 2179.8 & 295.6 & 2062.9\\
 \bottomrule
\end{tabular}
}
\end{table}

\subsection{Evaluation}

We conduct experiments on five datasets: Facebook, Congress, Deezer, LastFM, and Twitter. We next discuss the empirical insights from these experiments. For influence propagation we use the ICM model \cite{10.1145/956750.956769} where the influence probability is 0.85, and the number of seed nodes is $\{2,4,8,16,32\}$.

\subsubsection{Influence spread}
The analysis of the influence spread on all the datasets is shown in Table \ref{tab:InfluencedSpread}. As the seed set size ($k$) increases, there is a consistent upward trend in the total number of nodes reached. Though the scale of this spread varies significantly on different networks. In the Facebook and Twitter datasets, traditional greedy-based algorithms like Greedy and CELF perform exceptionally well. But, diversity-based algorithms (Bridge, DCDIM, CDIM), which also work on a greedy approach, showed competitiveness with respect to Greedy and CELF algorithms. This concludes that, regardless of objective functions, greedy approach-based algorithms showed equivalent influence spread. Across all datasets, heuristic-based measures such as PageRank, Hub, and Authority consistently underperformed compared to influence maximization and fairness-aware algorithms. PageRank showed the lowest influence spread across the board. Overall, fairness-based algorithms achieve equivalent influence spread as compared to other algorithms; however, fairness-based algorithms excel in terms of diversity. 

\subsubsection{Diversity (number of communities)} 
The performance results across the five datasets reveal a distinct hierarchy in algorithm effectiveness as shown in Table \ref{tab:diversity}. On the Facebook and Congress datasets, most algorithms reach a quick saturation point, i.e., with many achieving a maximum of 11.0 and 3.0 communities, respectively. However, in larger and more complex Twitter and Deezer datasets, the disparity between methods becomes stark. The PDTF algorithm emerges as the clear outlier, consistently maximizing community coverage far beyond the capabilities of traditional heuristics. On the Twitter dataset, the PDTF algorithm is giving the maximum communities, and classical centralities algorithms (Degree or PageRank) perform significantly lower. This concludes that the PDTF ensures information propagates across diverse communities.
The results conclude that the fairness-oriented algorithms DIM, DBIM, and DCDIM provide more robust scalability compared to Greedy baselines like CELF or SSRPEA. Because the Greedy approaches perform well in low-complexity environments. It failed to capture the breadth of communities in datasets like LastFM (the number of communities reached by DBIM (up to 30.0) or DGA (up to 41.0)). This is significantly surpasses classical IM techniques. Conversely, Hub and Authority metrics show very low community diversity. The results highlight that fairness- and diversity-based algorithms ensure a more equitable distribution of influence across the entire social topology, whereas classical algorithms merely maximize the total number of influenced nodes.

\subsubsection{Equality}
Equality (seed set equality) measures the fairness of the initial allocation of seed nodes in the communities. It is calculated as the ratio of seed nodes selected from a specific community relative to that community's total size. A high equality score in all communities suggests that the algorithm is distributing seed nodes proportionately. The results are shown in Table \ref{tab:Equality}. The experimental data showed disparity in equality across the various algorithms as the seed set size increases. The traditional algorithms (Greedy, CELF, PageRank) maintain extremely low and stable scores. This suggests these heuristic and greedy-based models behave conservatively and similarly regarding influence spread. Furthermore, the relationship between the seed set size and the equality shows that as $k$ increases from $2$ to $32$, the equality in most algorithms becomes smaller. However,  DBIM and CDIM variants show more sensitivity to the underlying graph structure. In the Deezer and LastFM datasets, we see the equality scores begin to diverge more significantly as the seed set grows. This shows that as more seeds are added, the community density of the networks becomes more influential than the algorithm's selection logic itself. The Twitter dataset shows the highest volatility; for example, the Bridge and DCDIM algorithms show a sharp increase in values only at $k=32$. Overall, the data suggest that most of the algorithms aim to maximize the influence and fail to choose a seed set with respect to community size. The diversity-based algorithms like CDIM, DCDIM, Bridge, etc., performed slightly better than others. All the values in equality are close to zero due to the disparity in the size of seed nodes and community size. 

\subsubsection{Equity}
Equity in IM measures the fairness of the final impact in different communities, and the results are shown in Table \ref{tab:Equity}. Equity is defined as the sum of ratios of influenced nodes within a community to the total size of that community. A high Equity sum across communities suggests that the influence is effectively spread across different groups. The experimental results across the five datasets are shown in Table \ref{tab:Equity}. In the Facebook dataset, we observe that DCDIM and CDIM reach relatively high equity among algorithms. This suggests that the Facebook graph is highly interconnected, and the influence is distributed in a higher proportion in the communities. Conversely, the Congress dataset shows significantly lower equity values. This indicates the influence struggle to spread the influence in the communities. As $k$ scales, centrality-based algorithms showed a very low equity, which proves that centrality measures are insufficient for ensuring broad outcome equity compared to community-aware or greedy optimization strategies. Overall, the algorithms that consider fairness (Bridge, DCDIM, CDIM, DIM, DBIM, PDTF) showed the highest equity values. These algorithms surpassed the classical algorithms like Greedy and CELF algorithms.

\subsubsection{Disparity}
In this analysis, disparity (as a penalty) is calculated using a logarithmic transformation of the ratio of the influenced proportion of communities. This analysis is a sensitive form of equity analysis.  The penalty is non-linear and aggressive. As the influence in a community approaches zero, the penalty value increases exponentially. This creates a "soft constraint" that heavily discourages the exclusion of any single group. The analysis of disparity (as penalty is shown in Table \ref{tab:Disparity}). To achieve equitable distribution of influence, algorithms aim to minimize the disparity or penalty. The experimental results of disparity in the form of penalty \cite{2025_TNSE_MineMultiFairness_PDTF_IM} is shown in Table \ref{tab:Disparity}. Minimum penalty shows the higher equatable distribution. The analysis of the experimental data reveals a significant disparity in different algorithms. In dense networks like Facebook, the DCDIM and CDIM algorithms minimize their penalty scores as the seed set size increases. This indicates that in well-connected graphs, high influence naturally leads to broader coverage. However, in Deezer and Twitter datasets, the disparity is higher. For instance, in the Twitter dataset, traditional heuristics like PageRank and Authority maintain massive penalty scores exceeding $2250.0$. It shows that the algorithm failed to spread influence equitably. The PDTF stands out as a high-performing outlier in the Twitter and Deezer datasets, while other algorithms suffer from penalties. The PDTF maintains a constant and significantly lower penalty on Twitter dataset. This behavior indicates that PDTF is mathematically structured to achieve equitable distribution of influence. Furthermore, the high penalties observed for Hub and Authority on the LastFM dataset. This concludes that high-degree nodes often exacerbate disparity by concentrating influence within a dominant core.

\subsubsection{Key empirical observation on fairness-based algorithms} This section highlights the key observation of fairness-based algorithms based on the experimental results. 
\begin{enumerate}
    \item DCDIM and CDIM consistently match or nearly match the influence spread of the Greedy algorithm. These algorithms optimize diversity without compromising the spread of influence. 
    \item DCDIM, DBIM, and Bridge reach significantly more distinct communities than classical algorithms. Furthermore, DCDIM, CDIM, and DBIM consistently yield lower disparity compared to Greedy and CELF.  
    \item Among the fairness-based algorithms, SSR-PEA and FIMMOGA perform worse in terms of influence spread.
    \item PDTF achieves high (and often constant) diversity scores but flat influence spread across seed sizes. Thus, it suggests it might be selecting seeds solely from the maximum communities, which leads to a sacrifice in the influence spread. 
\end{enumerate}

\subsection{Reproducibility Challenges}
Reproducibility remains a challenge in fairness-aware influence maximization. While classical IM methods are sensitive to network structure, diffusion parameters, and stochastic simulations, extending them to incorporate fairness constraints further amplifies these challenges. One major challenge is the limited availability of real-world social network datasets with reliable protected attributes. As discussed in Section \ref{sec:Datasets1}, most widely used benchmarks do not provide demographic or group labels, forcing researchers to either assume attributes or depend on small, domain-specific datasets. This might lead to noise, hidden biases, and inconsistencies, making it challenging to reproduce fairness results across studies. Another challenge arises due to Monte Carlo simulations for influence estimation \cite{10.1145/956750.956769}. Fairness metrics such as equity, disparity, and max–min coverage are highly sensitive to random seeds, the number of simulation runs, and diffusion probability settings. Small variations in these experimental settings can yield highly varied results across influence spread and fairness, yet these parameters are often ignored or inconsistently defined in the existing studies \cite{CDIM_CommunityDiversity_StaticGraph_JianxinLi,DiversifiedEffiecient}.
As a result, even when methods are publicly available, direct comparison and faithful replication remain challenging without standardized benchmarks, metrics, and reporting experimental settings. Further, scalability-driven heuristics, approximation techniques, and implementation-level optimizations can significantly affect results. Addressing these challenges requires open-source frameworks, shared datasets with protected attributes, fixed evaluation metrics, and transparent reporting of diffusion, fairness, and simulation settings.


\section{Emerging Trends, Open Challenges, and Future Directions}
\label{sec:Challenges}
\subsection{Emerging Trends and Open Challenges}
\subsubsection{Dynamic/Adaptive Fair IM:}
Social networks in the real world are not static; they change over time as people create and dissolve ties and their interests change. 
In recent years, fair IM methods have started focusing on applying dynamic and adaptive policies in place of fixed fairness policies for continuous monitoring of influence spread. These fair IM methods adjust the seed selection strategy in real-time to overcome disparities of location, gender, etc., and ensure fairness in the diffusion process. This is vital in long-term campaigns where the initial network structure might change significantly. In \cite{10.1109/TNET.2016.2563397}, authors have highlighted that most of the existing IM methods have limitations in fully capturing the characteristics of real-world social networks. These social networks also have many aspects of uncertainty due to a large population of users and high-speed data transmission.

\subsubsection{Causal Fairness in IM} 
In the context of IM, causal fairness moves beyond simple statistical correlations to address the underlying mechanisms of bias. Traditional fairness metrics (as defined in Table \ref{tab:Fairness_Function}) often focus on "equality of representation", i.e., ensuring different demographic groups receive similar levels of information. Existing studies \cite{2025_TNSE_MineMultiFairness_PDTF_IM,Meena.10.1145/3779058} ensure fair representation in influence spread as well as in the seed set. However, these metrics are superficial because they ignore the causal relationships between sensitive attributes (like race or gender), the network structure, and the ultimate outcome of the diffusion process. Causal fairness seeks to ensure that a node's probability of being influenced is not unfairly dictated by their protected characteristics through path-specific effects. This approach typically utilizes structural causal models (SCMs) \cite{StructuralCasualModelsSCMsForCasualFairness} to identify and mitigate unfair causal dependencies. The challenge lies in distinguishing between "fair" pathways (where differences in influence are justified by legitimate network roles) and "unfair" pathways (where the sensitive attribute directly or indirectly leads to systematic exclusion). By modeling the diffusion process as a causal graph, future work may design algorithms that intervene to break these biased links.

\subsubsection{Multi-sided Fairness}
Information receivers receive the influence in the network. Information providers are the nodes that are competing to spread the message. Information sources are the seed nodes or influencers selected to trigger the process. Multi-sided fairness refers to the simultaneous protection of diverse stakeholders involved in or affected by a diffusion process. While traditional fair IM \cite{Meena.10.1145/3779058} typically focuses on the information receivers. This ensures a balanced reach across demographic groups. The multi-sided fairness recognize that unfairness also impact the information providers and the information sources. For instance, in a job recruitment campaign on a social network, a multi-sided approach ensures that the information is equitably distributed among job seekers and employers to get fair access to the influential hubs of the network. It also addresses that specific influencers are not over-exploited or systematically excluded based on their own protected attributes. Balancing these competing objectives is a major open challenge in IM problem. Optimizing for one side, such as maximizing diversity among receivers, leads to reduced utility on the provider side. It require complex constrained optimization techniques to reach a stable equilibrium.

\subsubsection{Combining Fairness with Other Objectives } 
Existing work treats fairness as a standalone constraint that often incurs a \textit{price of fairness} \cite{CDIM_CommunityDiversity_StaticGraph_JianxinLi,G11.GONG2023119138}. This constraint leads to a significant drop in the total number of influenced nodes. This challenge represents a move from theoretical isolation to real-world deployment.  Emerging trends \cite{2025_TNSE_MineMultiFairness_PDTF_IM} seeking for a fairness that is integrated with practicality constraints like robustness, cost-efficiency, utility, etc., simultaneously. Every constraint addresses different practicality challenges. For example, robustness in fair IM addresses the uncertainty of social ties. Real-world network data is often stochastic and incomplete \cite{2021_SocialNetworkAnalysis_Book_Tanmoy}; therefore, a robustly fair algorithm ensures that the influence spread remains equitable even if the network topology changes slightly. When combined with cost-efficiency, the challenge becomes selecting a diverse set of seed nodes that satisfy fairness requirements while staying within a strict financial budget \cite{G13.10734665}. For example, in a public health campaign in a location, an algorithm must find influencers across different groups (addressing fairness) who are affordable to recruit (cost-efficiency) and whose impact won't collapse if a few lose interest (robustness). Thus, multi-objective optimization is important because a fair IM is useless if it is highly sensitive to noise in the network graph or if the cost of fair IM seeds exceeds a practical budget. This objective coupling often utilizes advanced mathematical frameworks like multi-objective submodular maximization \cite{2021_MultiObjectiveSubmodularOptimization_IM}. The goal is to identify a "Pareto optimal" \cite{G10.MEENA2025127445} frontier where no single objective, be it fairness, budget, or reach, are improved without compromising another. Research in this area is currently tackling the stability-fairness trade-off \cite{2025_TNSE_MineMultiFairness_PDTF_IM}. Ultimately, this emerging trend ensures that fair algorithms are not just socially responsible but also incorporate various challenges to deploy them for practical solutions.

\subsubsection{Defining Context-Aware Fairness}
Traditional fairness definitions \cite{Meena.10.1145/3779058} often assume that the social cost of missing an individual is uniform across all groups. However, context-aware fairness recognizes that the definition of what is \textit{fair} must change based on the nature of the information and specific socio-technical environment of the network. For instance, achieving gender parity in a viral marketing campaign for a soft drink may be less critical than achieving it in a campaign for life-saving medical information. Defining context-aware fairness involves mathematically encoding these situational priorities into the objective function. This ensures that the algorithm prioritizes fairness where the real-world impact of exclusion is highest. This fairness also addresses the structural context of the network, which acknowledges that different communities have varying levels of organic reach. A context-aware model accounts for legitimate underlying differences in node characteristics and still protects against discrimination based on protected attributes. The challenge lies in creating a formal contextual mapping that translates qualitative social norms into quantitative constraints. This often requires the use of context-dependent utility functions. Integrating these contexts into an optimization framework is a significant open challenge. It requires algorithms to be adaptive, which must be able to address different fairness depending on the input data. Researchers are currently exploring reinforcement learning (RL) approaches \cite{G25.10161600} where the agent learns the appropriate fairness context through interaction with the network environment. 

\subsubsection{Mitigating Underlying Network Bias}
Mitigating underlying network bias refers to the active intervention strategies used to counteract disparities \cite{2014_Tang_IM_FirstDiversified_GeneralFreamework} rooted in the structural properties of social networks. Unlike algorithmic bias, which arises from the optimization of mathematical logic, network bias is inherent in the graph itself. These features lead to saturation of information around the key influential nodes, and they deprive the marginalized groups, which are naturally isolated from the network’s main influential hubs. Mitigating this means the algorithm must recognize when the network’s architecture is stacked against certain groups. The algorithm adjust its selection to bridge these structural divides to spread influence in marginalized clusters.
Strategies for mitigation typically involve structural reweighting. In structural reweighting, the algorithm modifies the perceived strength of edges. This artificially boosts the importance of links that connect different demographic groups to prevent the cascade from getting trapped in privileged communities. In more advanced pre-processing approaches, researchers might use fair graph embeddings that map the social network into a latent space where structural biases are neutralized before the influence maximization process begins. This is an open challenge because correcting the network structure often comes at the cost of total influence spread (utility). The goal is to develop structure-aware fair algorithms \cite{G14.10.1145/3664618} that can identify and leverage bridge nodes to achieve equitable reach.

\subsubsection{Handling Uncertainty in Group Membership and Network Structure}
Handling uncertainty in group membership and network structure is still an open challenge that addresses the gap between idealized mathematical models and messy real-world data. Most fairness-aware IM algorithms \cite{Meena.10.1145/3779058} assume static mapping in which the demographic group/communities every user belongs to, and the precise strength of the connections between them. In practice, this data is often missing, noisy, or protected by privacy laws. Group membership uncertainty occurs when an individual’s sensitive attributes are unknown. This makes it challenging to calculate standard fairness metrics accurately. Meanwhile, network structure uncertainty arises from incomplete knowledge of the graph topology or the stochastic nature of influence probabilities. If an algorithm optimizes for fairness based on a probability that turns out to be wrong, it may inadvertently worsen disparities. This leads to an inaccurate estimation of fairness, where the social impact collapses under even minor data perturbations.
To mitigate these risks, emerging research is moving toward robust fair optimization. This involves designing algorithms that guarantee fairness across a range of scenarios. Instead of maximizing influence for a single perceived network, these models seek a seed set that performs reasonably well across an uncertainty set of potential graph structures and group assignments. Techniques such as distributionally robust optimization (DRO) \cite{RIS2014Borgs} and interval-based RIS (reverse influence sampling) are being explored. These techniques ensure that even if the group labels are only 70\% accurate, the resulting influence spread remains equitable. 


\subsubsection{Scalability to Very Large Networks with Complex Constraints}
The challenge of scalability to very large networks with complex constraints refers to the need for algorithms that handle massive, million-scale social graphs. Further, with this size, this challenge addresses multi-dimensional fairness and operational requirements. Traditionally, IM is an NP-hard problem, and adding fairness constraints such as ensuring demographic parity often breaks the sub-modularity property that simple greedy algorithms rely on for efficiency \cite{2025_TNSE_MineMultiFairness_PDTF_IM}. As a result, many early fair IM methods were computationally expensive \cite{CDIM_CommunityDiversity_StaticGraph_JianxinLi,G14.10.1145/3664618}. Therefore, most of the fairness-based IM algorithms are applicable to small datasets. Scalability in this context means developing techniques that approximate the optimal fair seed set on realistic networks with billions of edges without a prohibitive increase in runtime. 
The complex constraints aspect further complicates this by requiring the algorithm to balance fairness alongside other real-world limitations like heterogeneous node costs, time-critical deadlines, multi-item budgets, etc. For example, a public health agency might need to spread information across a diverse population where recruiting certain community leaders is more expensive than others. They also ensure that the message reaches everyone before a specific outbreak peak. Though a few studies \cite{EfficientFairAlgo2,G24.10.1145/3580305.3599847,EfficientFairAlgo1} have been developed recently that find the seed set efficiently for fair IM. But, solving this requires moving beyond simple heuristics toward distributed computing frameworks and unbiased estimators (to quickly evaluate the fairness-utility trade-off across millions of potential seed combinations). The open challenge lies in maintaining strong theoretical approximation guarantees (like $(1 - 1/e - \epsilon)$) while ensuring the algorithm remains computationally inexpensive and robust enough for real-world networks.

\subsubsection{Longitudinal Fairness and Long-Term Impacts}
The concept of longitudinal fairness and long-term impacts addresses the reality that social influence is rarely a one-time event. Traditional fairness focuses on a single snapshot of information spread in a static network \cite{2024_MineCSODynamicNetwork}. This indicate that a specific campaign reaches different demographic groups equitably at one point in time. However, longitudinal fairness considers the cumulative effects of repeated influence interventions over a time period. For example, if certain influencers are repeatedly selected as seed nodes because they are easy to reach, they may gain excessive social capital, status, and economic advantages over time. This results in other groups becoming chronically influence-deprived. Therefore, this creates a feedback loop in which existing social biases are not only reflected but also actively amplified by the algorithm's long-term behavior.
To address these long-term impacts, the emerging trends require moving beyond simple parity metrics toward dynamic fairness models \cite{2024_MineCSODynamicNetwork,2025_Mine_AntLionOptimixation_DALO,DynamicAndMultiNWIM_TONE_DMNIMP} and Markov decision processes (MDPs) \cite{205.SNIJDERS201044}. Researchers are exploring how to prevent digital stratification, where a community's structural position in the network degrades because it is consistently overlooked by high-value information cascades. A key challenge here is the long-term utility trade-off: an intervention that looks fair today might lead to a more segregated or less resilient network tomorrow. Emerging trends in this area may involve developing fairness-aware policies that optimize for equitable outcomes over an entire time horizon.

\subsubsection{Bridging Algorithmic Fairness and Real-World Impact Assessment}
Bridging algorithmic fairness and real-world impact assessment refers to the transition from purely mathematical definitions of equity to evaluating the actual social outcomes of an intervention. Traditionally, fairness in IM is measured by static, internal metrics like having equal probability of two demographic groups for being influenced \cite{2014_Tang_IM_FirstDiversified_GeneralFreamework,CDIM_CommunityDiversity_StaticGraph_JianxinLi}. However, bridging this gap means asking whether achieving parity in the algorithm actually leads to a reduction in real-world inequality. For example, in a public health campaign for HIV vaccination, an algorithm might be mathematically fair by reaching an equal percentage of two neighborhoods. But if one neighborhood has significantly lower access to clinics or higher baseline infection rates, the real-world impact remains unequal. This trend emphasizes the need for impact-oriented metrics that include socio-economic conditions. This challenge involves integrating domain-specific constraints and qualitative field data into the optimization process. It requires researchers to collaborate with social scientists and policymakers to define what success looks like beyond simple reach numbers. A major open challenge here is the measurement gap, like measuring long-term behavioral changes, which is challenging. This fairness aims to ensure that fair influence strategies are not just technically sound but are also ethically effective in showing the impact.

\subsection{Future Research Directions}
\subsubsection{Developing Theoretical Frameworks for Fair IM}
Developing theoretical frameworks for fair IM refers to the quest for a rigorous, mathematically unified foundation that formalizes what fairness means across various social and computational contexts. Currently, the field is fragmented, with different researchers employing isolated metrics such as maximin fairness (improving the worst-off group), demographic parity (equalizing influence ratios), diversity constraints, etc. A robust theoretical framework aims to provide an axiomatic characterization of fairness. This involves identifying a core set of desirable properties (axioms) such as monotonicity, submodularity, and scale invariance, that any fair algorithm should satisfy \cite{G13.10734665,CDIM_CommunityDiversity_StaticGraph_JianxinLi}. By establishing these ground rules, researchers prove general theorems about the inherent trade-offs between fairness and total influence spread. 
Furthermore, these frameworks are essential for addressing the non-submodularity of many common fairness objectives. In traditional IM \cite{Meena.10.1145/3779058}, the objective function is submodular, allowing for a simple greedy approach with a $(1 - 1/e)$ approximation guarantee. However, introducing fairness constraints often breaks this property, making the optimization problem significantly more complex. Future theoretical research focuses on developing surrogate functions or multi-objective optimization frameworks that restore these mathematical guarantees for fair outcomes. Ultimately, a unified theoretical framework aims to provide a fairness-by-design blueprint that selects and tunes fairness parameters with a clear understanding of the resulting algorithmic behavior and social consequences.

\subsubsection{Exploring Novel Fairness Notions suited for Network Dynamics}
This research direction aims to develop the equity definitions that account for how social networks change over time. Traditional fairness metrics assume the network structure and user attributes are fixed. However, real-world networks are time-evolving graphs where users join or leave, and relationships (edges) constantly form or dissolve. Novel notions for dynamics move away from measuring fairness as a single end-state and instead evaluate it as a continuous process. This includes temporal fairness, which ensures that the delay in receiving information is balanced across groups. The recurrent fairness prevents the same influential nodes from being repeatedly targeted as seeds. Together, both fairness avoids the long-term exhaustion or over-saturation of specific communities under dynamic settings. These novel notions also address the stochastic nature of influence. Future research aims to define fairness not just in terms of expected outreach, but through the lens of risk-aware fairness or mutual fairness. These concepts indicate that even in the worst-case scenario of a probabilistic diffusion process, no group is systematically neglected. For instance, in a dynamic setting, a fair algorithm might need to adapt its strategy in real-time as the network topology shifts. By moving beyond static snapshots, this research direction may ensure that influence remains equitable even as social trends evolve.

\subsubsection{Designing Highly Scalable Algorithms with Guarantees}
Designing highly scalable algorithms with guarantees refers to the development of computational frameworks that can handle massive, real-world social networks while providing rigorous mathematical guarantees on both performance and fairness. Traditionally, as constraints are added to the IM problem, the mathematical property of submodularity often breaks down. Without submodularity, simple greedy algorithms no longer guarantee an optimal or near-optimal solution. Research in this direction focuses on creating approximation algorithms that ensure the final seed set is at least a specific fraction, i.e., $1 - 1/e - \epsilon$ of the theoretically perfect fair solution, regardless of the network's complexity. Achieving this requires a dual focus on computational efficiency and theoretical rigor. High scalability is typically achieved through techniques like reverse influence sampling (RIS) or unbiased estimators. These techniques allow the algorithm to approximate influence spread and fairness metrics using a small, representative sample of the network rather than performing exhaustive, time-consuming simulations. The guarantees serve as a safety net with less complexity. But, it will also produce a solution that consistently meets a predefined threshold of fairness and utility. This is a vital frontier because, for fair IM, the underlying algorithms must be able to process global-scale data in near-real-time without sacrificing the ethical standards.

\subsubsection{Improved Evaluation Methodologies and Benchmarks}
Improved evaluation methodologies and benchmarks refers to the development of standardized, rigorous, and multifaceted systems for assessing how well fair algorithms perform. Historically, research in this area has been fragmented, with different studies using inconsistent datasets, incompatible fairness metrics, and varied diffusion models. This makes it challenging to objectively determine which algorithms are superior or how they truly trade off total influence spread against social equity. Improved methodologies seek to create a common ground by establishing a shared library of large-scale, real-world social network datasets that include verified demographic labels. Also, there is need for a unified suite of performance indicators that capture not just mathematical fairness, but also computational efficiency and robustness.
A core component of this direction is the creation of context-specific benchmarks that reflect the diverse domains where IM is applied. For example, a benchmark for public health outreach would include metrics for reaching vulnerable sub-populations. Other viral marketing benchmarks might focus more on consumer demographic parity. These methodologies also aim to address the price of fairness (the loss in total influence incurred to satisfy fairness constraints) by providing standardized ways to visualize and compare the \textit{Pareto frontier} \cite{G10.MEENA2025127445} of different models. By standardizing how we measure success, the field can move away from isolated experiments toward a more collaborative and cumulative scientific understanding of fair social influence.

\subsubsection{Fairness in Complex Diffusion Models (e.g., competitive IM, topic-aware IM)} Most existing methods assume simple diffusion models such as the Independent Cascade Model (ICM) or the Linear Threshold Model (LTM) to simulate information propagation. However, real-world diffusion processes are often governed by more complex dynamics. These include edge-diversity-based spreading models \cite{saxena2015understanding}, compartmental epidemic models such as SIR (Susceptible–Infected–Recovered) \cite{anderson1992infectious}, SI (Susceptible–Infected) \cite{allen1994some}, and SIS (Susceptible–Infected–Susceptible) \cite{hethcote1995sis}, as well as opinion dynamics models \cite{evans2018opinion, saxena2020mitigating}. Future work should examine fairness in influence propagation under these more complex diffusion settings and develop fairness-aware methods that account for diverse information propagation dynamics.
Additionally, Diffusion models in the IM problem describe how influence propagates in the network. In most of the IM studies \cite{Meena.10.1145/3779058}, the diffusion model assumes that one message spreads in isolation. However, in competitive IM \cite{2020_RNVGA_CompetitiveIM_Follower_based}, multiple entities, such as rival political campaigns, competing brands, are considered for the same audience. In this context, fairness isn't just about the reach of one message; it is about ensuring that the competitive equilibrium does not systematically marginalize specific demographic groups. For example, if two competing health providers only target affluent neighborhoods, their diffusion model must account for the information gap created by their combined strategic behavior. This ensures that marginalized communities are not left to the influence spread.
In this context, topic-aware IM \cite{DynamicIM_TTG_TTH_TopicBased_Min_2020,DynamicTopic_TCIC_CFDI_FeatureBased_2022_Qin} adds a layer of content sensitivity to the fairness equation. All information is not perceived equally by all users; a message about computer science internships may spread differently across a network than one about Public Policy due to varying user interests and modal topic-receptive weights. Fair topic-aware models must ensure that influence is not just balanced in aggregate, but balanced across specific domains of knowledge. This prevents digital homophily, where certain groups only receive information related to specific topics. The open challenge here involves modeling these complex interactions using hypergraphs or multi-layer networks, where an edge contains a multidimensional vector of topic-specific strengths. This challenge is also addressed using the game theory \cite{2021_CompetitiveIM_GCW_GameTheoryBased_AwarenessModel}. Thus, this direction aim to address how to prevent fairness manipulation where one competitor might exploit another's fairness constraints to gain an unfair share of the total market. 


\subsubsection{Interdisciplinary Research: Integrating social science perspectives}
This research direction refers to the necessity of merging computational algorithms with the deep qualitative insights of sociology, psychology, and behavioral economics. Computer science has treated fairness as a static mathematical constraint (like parity in influence numbers). However, future research recognizes that social networks are not just graphs; they are complex human ecosystems shaped by historical biases, power dynamics, and different cultural norms. By integrating social science, researchers can move beyond blind algorithmic fairness to understand why certain groups are marginalized. This includes using social welfare theory to define equity or applying sociological theories of homophily to model how information actually flows through fragmented communities. This approach is essential for shifting from numerical fairness to substantive fairness. Social science perspectives provide the context needed to determine which fairness metrics are most appropriate for a given real-world application. Future research aims to build \textit{human-in-the-Loop} systems where qualitative feedback from affected communities informs the reward functions of the AI. This direction of IM addresses the systemic inequalities and lived experiences of the people within the network.

\section{Concluding Remarks}
\label{sec:Concluding Remarks}
As the IM problem continues to shape critical decisions in digital society, addressing fairness becomes essential. This survey emphasizes the need to design IM methods that balance performance with equitable impact, particularly in contexts where social, economic, or demographic disparities exist. We discuss existing fairness-aware IM models, fairness metrics, and optimization methods that balance maximum influence spread and fair distribution of spread. 
Empirical analysis highlights the need for parity seeding and diversity constraints in multi-objective fairness settings in this rapidly evolving domain. Experimental results show that the fairness-based DCDIM and CDIM emerge as the most effective algorithms in terms of maximizing community coverage and minimizing disparity. In contrast, other fairness-based approaches like SSR-PEA, FIMMOGA, and PDTF fail to achieve this balance. These algorithms sacrifice influence to spread to satisfy their specific fairness constraints. Therefore, this work empirically underscores the need for a framework that balances fairness and influence spread across applications. Other significant gaps remain in terms of scalability, contextual fairness, handling uncertainty, real-world deployment, and interdisciplinary validation. Future research needs to make fairness-conscious models that are not only robust but also transparent, interpretable, and sensitive to the real-world users and communities. 

\section*{Acknowledgments}
The authors gratefully acknowledge that this work was supported by {Anusandhan National Research Foundation} (ANRF) under Grant No. {ANRF/ECRG/2025/000150/ENS} for providing financial support for the study. The authors would also like to acknowledge that language editing tools were used to assist with grammar and spelling refinement only during the preparation of this manuscript. 

\section*{Conflict of Interest}
None of the authors has a conflict of interest to disclose.


\bibliographystyle{unsrt}
\bibliography{SNAB}

%% file: SNAB.bib
@inproceedings{G1.10.1145/3308560,
author = {Stoica, Ana-Andreea and Chaintreau, Augustin},
title = {Fairness in Social Influence Maximization},
year = {2019},
isbn = {9781450366755},
publisher = {Association for Computing Machinery},
address = {New York, NY, USA},
url = {https://doi.org/10.1145/3308560.3317588},
doi = {10.1145/3308560.3317588},
booktitle = {Companion Proceedings of The 2019 World Wide Web Conference},
pages = {569–574},
numpages = {6},
location = {San Francisco, USA},
series = {WWW '19}
}

@inproceedings{10.5555/3367471.3367498,
author = {Rahman, Tahleen and Surma, Bartlomiej and Backes, Michael and Zhang, Yang},
title = {Fairwalk: towards fair graph embedding},
year = {2019},
isbn = {9780999241141},
publisher = {AAAI Press},
booktitle = {Proceedings of the 28th International Joint Conference on Artificial Intelligence},
pages = {3289–3295},
numpages = {7},
location = {Macao, China},
series = {IJCAI'19}
}

@article{macedo2026gender,
  title={Gender biases in online communication: A case study of soccer},
  author={Macedo, Mariana and Saxena, Akrati},
  journal={Applied Intelligence},
  volume={56},
  number={1},
  pages={33},
  year={2026},
  publisher={Springer}
}

@inproceedings{10.1145/3485447.3512249,
author = {Tsioutsiouliklis, Sotiris and Pitoura, Evaggelia and Semertzidis, Konstantinos and Tsaparas, Panayiotis},
title = {Link Recommendations for PageRank Fairness},
year = {2022},
isbn = {9781450390965},
publisher = {Association for Computing Machinery},
address = {New York, NY, USA},
url = {https://doi.org/10.1145/3485447.3512249},
doi = {10.1145/3485447.3512249},
booktitle = {Proceedings of the ACM Web Conference 2022},
pages = {3541–3551},
numpages = {11},
location = {Virtual Event, Lyon, France},
series = {WWW '22}
}

@inproceedings{G2.10.1145/336642,
author = {Stoica, Ana-Andreea and Han, Jessy Xinyi and Chaintreau, Augustin},
title = {Seeding Network Influence in Biased Networks and the Benefits of Diversity},
year = {2020},
isbn = {9781450370233},
publisher = {Association for Computing Machinery},
address = {New York, NY, USA},
url = {https://doi.org/10.1145/3366423.3380275},
doi = {10.1145/3366423.3380275},
booktitle = {Proceedings of The Web Conference 2020},
pages = {2089–2098},
numpages = {10},
location = {Taipei, Taiwan},
series = {WWW '20}
}

@inproceedings{G3.inproceedings,
author = {Kazienko, Przemysław and Musial, Katarzyna},
year = {2006},
month = {10},
pages = {417-424},
title = {Social Capital in Online Social Networks},
booktitle = {Knowledge-Based Intelligent Information and Engineering Systems},
isbn = {978-3-540-46537-9},
doi = {10.1007/11893004_54}
}

@inproceedings{G4.10.1007/11523468_91,
author = {Kempe, David and Kleinberg, Jon and Tardos, \'{E}va},
title = {Influential nodes in a diffusion model for social networks},
year = {2005},
isbn = {3540275800},
publisher = {Springer-Verlag},
address = {Berlin, Heidelberg},
url = {https://doi.org/10.1007/11523468_91},
doi = {10.1007/11523468_91},
booktitle = {Proceedings of the 32nd International Conference on Automata, Languages and Programming},
pages = {1127–1138},
numpages = {12},
location = {Lisbon, Portugal},
series = {ICALP'05}
}

@inproceedings{G5.10.1145/34596,
author = {Teng, Ya-Wen and Chen, Hsi-Wen and Yang, De-Nian and Pignolet, Yvonne-Anne and Li, Ting-Wei and Chen, Lydia},
title = {On Influencing the Influential: Disparity Seeding},
year = {2021},
isbn = {9781450384469},
publisher = {Association for Computing Machinery},
address = {New York, NY, USA},
url = {https://doi.org/10.1145/3459637.3482375},
doi = {10.1145/3459637.3482375},
booktitle = {Proceedings of the 30th ACM International Conference on Information \& Knowledge Management},
pages = {1804–1813},
numpages = {10},
location = {Virtual Event, Queensland, Australia},
series = {CIKM '21}
}

@article{G6.abs-1903-00967,
  author       = {Alan Tsang and
                  Bryan Wilder and
                  Eric Rice and
                  Milind Tambe and
                  Yair Zick},
  title        = {Group-Fairness in Influence Maximization},
  journal      = {CoRR},
  volume       = {abs/1903.00967},
  year         = {2019},
  url          = {http://arxiv.org/abs/1903.00967},
  eprinttype    = {arXiv},
  eprint       = {1903.00967},
  bibsource    = {dblp computer science bibliography, https://dblp.org}
}

@misc{G7.becker2023cost,
      title={On the Cost of Demographic Parity in Influence Maximization}, 
      author={Ruben Becker and Gianlorenzo D'Angelo and Sajjad Ghobadi},
      year={2023},
      eprint={2302.13110},
      archivePrefix={arXiv},
      primaryClass={cs.SI},
      url={https://arxiv.org/abs/2302.13110}, 
}

@article{G8.10.1613/jair.1.13367,
author = {Becker, Ruben and D’Angelo, Gianlorenzo and Ghobadi, Sajjad and Gilbert, Hugo},
title = {Fairness in Influence Maximization through Randomization},
year = {2022},
issue_date = {May 2022},
publisher = {AI Access Foundation},
address = {El Segundo, CA, USA},
volume = {73},
issn = {1076-9757},
url = {https://doi.org/10.1613/jair.1.13367},
doi = {10.1613/jair.1.13367},
journal = {J. Artif. Int. Res.},
month = may,
numpages = {33}
}

@inproceedings{G9.10.5555/3491440,
author = {Khajehnejad, Moein and Rezaei, Ahmad Asgharian and Babaei, Mahmoudreza and Hoffmann, Jessica and Jalili, Mahdi and Weller, Adrian},
title = {Adversarial graph embeddings for fair influence maximization over social networks},
year = {2021},
isbn = {9780999241165},
booktitle = {Proceedings of the Twenty-Ninth International Joint Conference on Artificial Intelligence},
articleno = {594},
numpages = {7},
location = {Yokohama, Yokohama, Japan},
series = {IJCAI'20}
}

@article{G10.MEENA2025127445,
title = {Fairness-aware influence maximization: A novel Learning Automata-based approach},
journal = {Expert Systems with Applications},
volume = {280},
pages = {127445},
year = {2025},
issn = {0957-4174},
doi = {https://doi.org/10.1016/j.eswa.2025.127445},
author = {Sunil Kumar Meena and Kuldeep Singh and Bhaskar Biswas}
}

@article{G11.GONG2023119138,
title = {Influence maximization considering fairness: A multi-objective optimization approach with prior knowledge},
journal = {Expert Systems with Applications},
volume = {214},
pages = {119138},
year = {2023},
issn = {0957-4174},
doi = {https://doi.org/10.1016/j.eswa.2022.119138},
author = {Hao Gong and Chunxiang Guo}
}

@article{G12.MEENA2025108045,
title = {DCDIM: Diversified influence maximization on dynamic social networks},
journal = {Computer Communications},
volume = {232},
pages = {108045},
year = {2025},
issn = {0140-3664},
doi = {https://doi.org/10.1016/j.comcom.2025.108045},
author = {Sunil Kumar Meena and Shashank Sheshar Singh and Kuldeep Singh}
}

@article{CDIM_CommunityDiversity_StaticGraph_JianxinLi,
title = {Community-diversified influence maximization in social networks},
journal = {Information Systems},
volume = {92},
pages = {101522},
year = {2020},
issn = {0306-4379},
doi = {https://doi.org/10.1016/j.is.2020.101522},
url = {https://www.sciencedirect.com/science/article/pii/S0306437920300326},
author = {Jianxin Li and Taotao Cai and Ke Deng and Xinjue Wang and Timos Sellis and Feng Xia}
}

@INPROCEEDINGS{2014_Tang_IM_FirstDiversified_GeneralFreamework,
  author={Tang, Fangshuang and Liu, Qi and Zhu, Hengshu and Chen, Enhong and Zhu, Feida},
  booktitle={2014 IEEE/ACM International Conference on Advances in Social Networks Analysis and Mining (ASONAM 2014)}, 
  title={Diversified social influence maximization}, 
  year={2014},
  volume={},
  number={},
  pages={455-459},
  doi={10.1109/ASONAM.2014.6921625}
}

@ARTICLE{2025_TNSE_MineMultiFairness_PDTF_IM,
  author={Meena, Sunil Kumar and Singh, Shashank Sheshar and Singh, Kuldeep},
  journal={IEEE Transactions on Network Science and Engineering}, 
  title={Achieving Fair Influence Maximization: Integrating Privacy, Diversity, and Temporal Fairness}, 
  year={2025},
  volume={},
  number={},
  pages={1-13},
  doi={10.1109/TNSE.2025.3598819}}

@article{2025_Mine_AntLionOptimixation_DALO,
title = {Pruning-enabled dynamic influence maximization using antlion optimization},
journal = {Knowledge-Based Systems},
volume = {317},
pages = {113406},
year = {2025},
issn = {0950-7051},
doi = {https://doi.org/10.1016/j.knosys.2025.113406},
url = {https://www.sciencedirect.com/science/article/pii/S0950705125004538},
author = {Sunil Kumar Meena and Shashank {Sheshar Singh} and Kuldeep Singh}
}

@ARTICLE{G13.10734665,
  author={Meena, Sunil Kumar and Sheshar Singh, Shashank and Singh, Kuldeep},
  journal={IEEE Transactions on Computational Social Systems}, 
  title={Diversified Budgeted Influence Maximization in Dynamic Social Networks}, 
  year={2024},
  volume={},
  number={},
  pages={1-13},
  doi={10.1109/TCSS.2024.3473948}
}

@ARTICLE{2023_Zhang_SSRPEA_Diversity_SpaceReduction_EvolutionaryAlgo,
  author={Zhang, Lei and Ma, Kaicong and Yang, Haipeng and Zhang, Cheng and Ma, Haiping and Liu, Qi},
  journal={IEEE Transactions on Computational Social Systems}, 
  title={A Search Space Reduction-Based Progressive Evolutionary Algorithm for Influence Maximization in Social Networks}, 
  year={2023},
  volume={10},
  number={5},
  pages={2385-2399},
  doi={10.1109/TCSS.2022.3189036}
}

@article{G14.10.1145/3664618,
author = {Meena, Sunil Kumar and Singh, Shashank Sheshar and Singh, Kuldeep},
title = {DCDIMB: Dynamic Community-based Diversified Influence Maximization using Bridge Nodes},
year = {2024},
issue_date = {November 2024},
publisher = {Association for Computing Machinery},
address = {New York, NY, USA},
volume = {18},
number = {4},
issn = {1559-1131},
url = {https://doi.org/10.1145/3664618},
doi = {10.1145/3664618},
journal = {ACM Trans. Web},
month = oct,
articleno = {47},
numpages = {32}
}

@article{MISHRA2021114493,
title = {TCD2: Tree-based community detection in dynamic social networks},
journal = {Expert Systems with Applications},
volume = {169},
pages = {114493},
year = {2021},
issn = {0957-4174},
doi = {https://doi.org/10.1016/j.eswa.2020.114493},
author = {Sneha Mishra and Shashank Sheshar Singh and Shivansh Mishra and Bhaskar Biswas}
}

@ARTICLE{G15.9863801,
  author={Razaghi, Behnam and Roayaei, Mehdy and Charkari, Nasrollah Moghadam},
  journal={IEEE Transactions on Computational Social Systems}, 
  title={On the Group-Fairness-Aware Influence Maximization in Social Networks}, 
  year={2023},
  volume={10},
  number={6},
  pages={3406-3414},
  doi={10.1109/TCSS.2022.3198096}}

@inproceedings{G16.10.1145/3437963.3441787,
author = {Anwar, Md Sanzeed and Saveski, Martin and Roy, Deb},
title = {Balanced Influence Maximization in the Presence of Homophily},
year = {2021},
isbn = {9781450382977},
publisher = {Association for Computing Machinery},
address = {New York, NY, USA},
url = {https://doi.org/10.1145/3437963.3441787},
doi = {10.1145/3437963.3441787},
booktitle = {Proceedings of the 14th ACM International Conference on Web Search and Data Mining},
pages = {175–183},
numpages = {9},
location = {Virtual Event, Israel},
series = {WSDM '21}
}

@INPROCEEDINGS{G17.9835234,
  author={Ali, Junaid and Babaei, Mahmoudreza and Chakraborty, Abhijnan and Mirzasoleiman, Baharan and Gummadi, Krishna P. and Singla, Adish},
  booktitle={2022 IEEE 38th International Conference on Data Engineering (ICDE)}, 
  title={On the Fairness of Time-Critical Influence Maximization in Social Networks (Extended Abstract)}, 
  year={2022},
  volume={},
  number={},
  pages={1541-1542},
  doi={10.1109/ICDE53745.2022.00147}}

@article{G18.87.article,
author = {Chen, Wei and Lu, Wei and Zhang, Ning},
year = {2012},
month = {04},
pages = {},
title = {Time-Critical Influence Maximization in Social Networks with
Time-Delayed Diffusion Process},
volume = {26},
journal = {Proceedings of the AAAI Conference on Artificial Intelligence},
doi = {10.1609/aaai.v26i1.8204}
}

@inproceedings{G19.10.1145/3308558.3313680,
author = {Fish, Benjamin and Bashardoust, Ashkan and Boyd, Danah and Friedler, Sorelle and Scheidegger, Carlos and Venkatasubramanian, Suresh},
title = {Gaps in Information Access in Social Networks?},
year = {2019},
isbn = {9781450366748},
publisher = {Association for Computing Machinery},
address = {New York, NY, USA},
url = {https://doi.org/10.1145/3308558.3313680},
doi = {10.1145/3308558.3313680},
booktitle = {The World Wide Web Conference},
pages = {480–490},
numpages = {11},
location = {San Francisco, CA, USA},
series = {WWW '19}
}

@inbook{G20.doi:10.1137/1.9781611976236.69,
author = {Zeinab S. Jalali and Weixiang Wang and Myunghwan Kim and Hema Raghavan and Sucheta Soundarajan},
title = {On the Information Unfairness of Social Networks},
booktitle = {Proceedings of the 2020 SIAM International Conference on Data Mining (SDM)},
chapter = {},
pages = {613-521},
doi = {10.1137/1.9781611976236.69}
}

@article{G21.10.14778/3551793.3551824,
author = {Swift, Ian P. and Ebrahimi, Sana and Nova, Azade and Asudeh, Abolfazl},
title = {Maximizing fair content spread via edge suggestion in social networks},
year = {2022},
issue_date = {July 2022},
publisher = {VLDB Endowment},
volume = {15},
number = {11},
issn = {2150-8097},
url = {https://doi.org/10.14778/3551793.3551824},
doi = {10.14778/3551793.3551824},
journal = {Proc. VLDB Endow.},
month = jul,
pages = {2692–2705},
numpages = {14}
}

@inproceedings{G22.10.114424.3383555,
author = {Farnad, Golnoosh and Babaki, Behrouz and Gendreau, Michel},
title = {A Unifying Framework for Fairness-Aware Influence Maximization},
year = {2020},
isbn = {9781450370240},
publisher = {Association for Computing Machinery},
address = {New York, NY, USA},
url = {https://doi.org/10.1145/3366424.3383555},
doi = {10.1145/3366424.3383555},
booktitle = {Companion Proceedings of the Web Conference 2020},
pages = {714–722},
numpages = {9},
location = {Taipei, Taiwan},
series = {WWW '20}
}

@inproceedings{G23.rahmattalabi2021fair,
  title={Fair influence maximization: A welfare optimization approach},
  author={Rahmattalabi, Aida and Jabbari, Shahin and Lakkaraju, Himabindu and Vayanos, Phebe and Izenberg, Max and Brown, Ryan and Rice, Eric and Tambe, Milind},
  booktitle={Proceedings of the AAAI Conference on Artificial Intelligence},
  volume={35},
  number={13},
  pages={11630--11638},
  year={2021}
}

@inproceedings{G24.10.1145/3580305.3599847,
author = {Feng, Yuting and Patel, Ankitkumar and Cautis, Bogdan and Vahabi, Hossein},
title = {Influence Maximization with Fairness at Scale},
year = {2023},
isbn = {9798400701030},
publisher = {Association for Computing Machinery},
address = {New York, NY, USA},
url = {https://doi.org/10.1145/3580305.3599847},
doi = {10.1145/3580305.3599847},
booktitle = {Proceedings of the 29th ACM SIGKDD Conference on Knowledge Discovery and Data Mining},
pages = {4046–4055},
numpages = {10},
location = {Long Beach, CA, USA},
series = {KDD '23}
}

@ARTICLE{G25.10161600,
  author={Zhang, Congcong and Zhou, Jingya and Wang, Jin and Fan, Jianxi and Shi, Yingdan},
  journal={IEEE Transactions on Computational Social Systems}, 
  title={Fairness-Aware Competitive Bidding Influence Maximization in Social Networks}, 
  year={2024},
  volume={11},
  number={2},
  pages={2147-2159},
  doi={10.1109/TCSS.2023.3285605}}

@article{G26.zhang2025fairness,
  title={Fairness-constrained multigroup influence maximization},
  author={Zhang, Zizhen and Li, Deying and Wang, Yongcai and Chen, Wenping and Zhu, Yuqing},
  journal={Knowledge and Information Systems},
  pages={1--25},
  year={2025},
  publisher={Springer}
}

@article{G27.10.1145/3701737,
author = {Wang, Zhixiao and Zhao, Jiayu and Sun, Chengcheng and Rui, Xiaobin and Yu, Philip S.},
title = {A General Concave Fairness Framework for Influence Maximization Based on Poverty Reward},
year = {2024},
issue_date = {January 2025},
publisher = {Association for Computing Machinery},
address = {New York, NY, USA},
volume = {19},
number = {1},
issn = {1556-4681},
url = {https://doi.org/10.1145/3701737},
doi = {10.1145/3701737},
journal = {ACM Trans. Knowl. Discov. Data},
month = dec,
articleno = {14},
numpages = {23}
}

@article{G28.xiao2025fairness,
  title={Fairness-Aware Influence Blocking Maximization in Social Networks},
  author={Xiao, Yu and Liu, Ding and Yang, Lan and Li, Zhiwu},
  journal={IEEE Transactions on Computational Social Systems},
  year={2025},
  publisher={IEEE}
}

@article{FAIRSNA.10.1145/3653711,
author = {Saxena, Akrati and Fletcher, George and Pechenizkiy, Mykola},
title = {FairSNA: Algorithmic Fairness in Social Network Analysis},
year = {2024},
issue_date = {August 2024},
publisher = {Association for Computing Machinery},
address = {New York, NY, USA},
volume = {56},
number = {8},
issn = {0360-0300},
url = {https://doi.org/10.1145/3653711},
doi = {10.1145/3653711},
journal = {ACM Comput. Surv.},
month = apr,
articleno = {213},
numpages = {45}
}

@article{S1.behaviour-aware,
  title={Influence maximization in social networks: a survey of behaviour-aware methods},
  author={Zareie, Ahmad and Sakellariou, Rizos},
  journal={Social Network Analysis and Mining},
  volume={13},
  number={1},
  pages={78},
  year={2023},
  publisher={Springer}
}

@article{S2.singh2021influence,
  title={Influence maximization on social networks: a study},
  author={Singh, Shashank S and Singh, Kuldeep and Kumar, Ajay and Biswas, Bhaskar},
  journal={Recent Advances in Computer Science and Communications (Formerly: Recent Patents on Computer Science)},
  volume={14},
  number={1},
  pages={13--29},
  year={2021},
  publisher={Bentham Science Publishers direct}
}

@article{S3.ye2022influence,
  title={Influence maximization in social networks: Theories, methods and challenges},
  author={Ye, Yuxin and Chen, Yunliang and Han, Wei},
  journal={Array},
  volume={16},
  pages={100264},
  year={2022},
  publisher={Elsevier}
}

@article{S4.jaouadi2024survey,
  title={A survey on influence maximization models},
  author={Jaouadi, Myriam and Romdhane, Lotfi Ben},
  journal={Expert Systems with Applications},
  volume={248},
  pages={123429},
  year={2024},
  publisher={Elsevier}
}

@article{S7.singh2022influence,
  title={Influence maximization frameworks, performance, challenges and directions on social network: A theoretical study},
  author={Singh, Shashank Sheshar and Srivastva, Divya and Verma, Madhushi and Singh, Jagendra},
  journal={Journal of King Saud University-Computer and Information Sciences},
  volume={34},
  number={9},
  pages={7570--7603},
  year={2022},
  publisher={Elsevier}
}

@article{S8.peng2016social,
  title={Social influence analysis in social networking big data: Opportunities and challenges},
  author={Peng, Sancheng and Wang, Guojun and Xie, Dongqing},
  journal={IEEE network},
  volume={31},
  number={1},
  pages={11--17},
  year={2016},
  publisher={IEEE}
}

@article{SNA1.singh2024social,
  title={Social network analysis: A survey on process, tools, and application},
  author={Singh, Shashank Sheshar and Muhuri, Samya and Mishra, Shivansh and Srivastava, Divya and Shakya, Harish Kumar and Kumar, Neeraj},
  journal={ACM Computing Surveys},
  volume={56},
  number={8},
  pages={1--39},
  year={2024},
  publisher={ACM New York, NY}
}

@article{SNA2.singh2025big,
  title={Big Data meets Social Networks: A Survey of Analytical Strategies and Research Challenges},
  author={Singh, Shashank Sheshar and Singh, Shashank and Singh, Kuldeep and Srivastava, Vishal and Shakya, Harish Kumar},
  journal={IEEE Access},
  year={2025},
  publisher={IEEE}
}

@article{SNA3.singh2024quantum,
  title={Quantum social network analysis: Methodology, implementation, challenges, and future directions},
  author={Singh, Shashank Sheshar and Kumar, Sumit and Meena, Sunil Kumar and Singh, Kuldeep and Mishra, Shivansh and Zomaya, Albert Y},
  journal={Information Fusion},
  pages={102808},
  year={2024},
  publisher={Elsevier}
}

@article{SNA4.10.1145/3539732,
author = {Singh, Shashank Sheshar and Srivastava, Vishal and Kumar, Ajay and Tiwari, Shailendra and Singh, Dilbag and Lee, Heung-No},
title = {Social Network Analysis: A Survey on Measure, Structure, Language Information Analysis, Privacy, and Applications},
year = {2023},
issue_date = {May 2023},
publisher = {Association for Computing Machinery},
address = {New York, NY, USA},
volume = {22},
number = {5},
issn = {2375-4699},
doi = {10.1145/3539732},
journal = {ACM Trans. Asian Low-Resour. Lang. Inf. Process.},
month = may,
articleno = {137},
numpages = {47}
}

@ARTICLE{11037728,
  author={Meena, Sunil Kumar and Singh, Shashank Sheshar and Singh, Kuldeep},
  journal={IEEE Transactions on Computational Social Systems}, 
  title={Exploring Influence Maximization in Criminal Social Networks to Identify Key Individuals and Patterns}, 
  year={2025},
  volume={},
  number={},
  pages={1-10},
  doi={10.1109/TCSS.2025.3571963}}

@article{10.1145/3707463,
author = {Singh, Shashank Sheshar and Muhuri, Samya and Kumar, Sumit and Barua, Jayendra},
title = {From Nodes to Knowledge: Exploring Social Network Analysis in Education},
year = {2025},
issue_date = {February 2025},
publisher = {Association for Computing Machinery},
address = {New York, NY, USA},
volume = {19},
number = {1},
issn = {1559-1131},
url = {https://doi.org/10.1145/3707463},
doi = {10.1145/3707463},
journal = {ACM Trans. Web},
month = jan,
articleno = {7},
numpages = {36}
}

@article{del2020social,
  title={Social networks as tools for the prevention and promotion of health among youth},
  author={Del Castillo, Jos{\'e} Antonio Garc{\'\i}a and del Castillo-L{\'o}pez, {\'A}lvaro Garc{\'\i}a and Dias, Paulo C{\'e}sar and Garc{\'\i}a-Castillo, Fernando},
  journal={Psicologia: Reflex{\~a}o e Cr{\'\i}tica},
  volume={33},
  pages={13},
  year={2020},
  publisher={SciELO Brasil}
}

@incollection{brass2012power,
  title={Power, politics, and social networks in organizations},
  author={Brass, Daniel J and Krackhardt, David M},
  booktitle={Politics in organizations},
  pages={355--375},
  year={2012},
  publisher={Routledge}
}

@article{tang2017profit,
  title={Profit maximization for viral marketing in online social networks: Algorithms and analysis},
  author={Tang, Jing and Tang, Xueyan and Yuan, Junsong},
  journal={IEEE Transactions on Knowledge and Data Engineering},
  volume={30},
  number={6},
  pages={1095--1108},
  year={2017},
  publisher={IEEE}
}

@article{hosain2020recruitment,
  title={E-recruitment: A social media perspective},
  author={Hosain, Sajjad and Manzurul Arefin, Abu Hena Mohammad and Hossin, Md Altab},
  journal={Asian Journal of Economics, Business and Accounting},
  volume={16},
  number={4},
  pages={51--62},
  year={2020}
}

@article{jaeger2007community,
  title={Community response grids: E-government, social networks, and effective emergency management},
  author={Jaeger, Paul T and Shneiderman, Ben and Fleischmann, Kenneth R and Preece, Jennifer and Qu, Yan and Wu, Philip Fei},
  journal={Telecommunications Policy},
  volume={31},
  number={10-11},
  pages={592--604},
  year={2007},
  publisher={Elsevier}
}

@article{patacchini2017social,
  title={Social ties and the demand for financial services},
  author={Patacchini, Eleonora and Rainone, Edoardo},
  journal={Journal of Financial Services Research},
  volume={52},
  pages={35--88},
  year={2017},
  publisher={Springer}
}

@inproceedings{budak2011limiting,
  title={Limiting the spread of misinformation in social networks},
  author={Budak, Ceren and Agrawal, Divyakant and El Abbadi, Amr},
  booktitle={Proceedings of the 20th international conference on World wide web},
  pages={665--674},
  year={2011}
}

@article{202.SEIERSTAD201144,
title = {For the few not the many? The effects of affirmative action on presence, prominence, and social capital of women directors in Norway},
journal = {Scandinavian Journal of Management},
volume = {27},
number = {1},
pages = {44-54},
year = {2011},
note = {Part Special Topic Forum on "Architectural Competitions"},
issn = {0956-5221},
doi = {https://doi.org/10.1016/j.scaman.2010.10.002},
url = {https://www.sciencedirect.com/science/article/pii/S0956522110001272},
author = {Cathrine Seierstad and Tore Opsahl}
}

@inproceedings{203.Shetty2004TheEE,
  title={The Enron Email Dataset Database Schema and Brief Statistical Report},
  author={Jitesh Shetty and Jafar Adibi},
  year={2004},
  url={https://api.semanticscholar.org/CorpusID:59919272}
}

@article{205.SNIJDERS201044,
title = {Introduction to stochastic actor-based models for network dynamics},
journal = {Social Networks},
volume = {32},
number = {1},
pages = {44-60},
year = {2010},
note = {Dynamics of Social Networks},
issn = {0378-8733},
doi = {https://doi.org/10.1016/j.socnet.2009.02.004},
author = {Tom A.B. Snijders and Gerhard G. {van de Bunt} and Christian E.G. Steglich}
}

@article{182.10.1371/journal.pcbi.1001109,
    doi = {10.1371/journal.pcbi.1001109},
    author = {Rocha, Luis E. C. AND Liljeros, Fredrik AND Holme, Petter},
    journal = {PLOS Computational Biology},
    publisher = {Public Library of Science},
    title = {Simulated Epidemics in an Empirical Spatiotemporal Network of 50,185 Sexual Contacts},
    year = {2011},
    month = {03},
    volume = {7},
    url = {https://doi.org/10.1371/journal.pcbi.1001109},
    pages = {1-9},
    number = {3},
}

@inproceedings{148.2999134.2999195,
author = {McAuley, Julian and Leskovec, Jure},
title = {Learning to discover social circles in ego networks},
year = {2012},
publisher = {Curran Associates Inc.},
address = {Red Hook, NY, USA},
booktitle = {Proceedings of the 26th International Conference on Neural Information Processing Systems - Volume 1},
pages = {539–547},
numpages = {9},
location = {Lake Tahoe, Nevada},
series = {NIPS'12}
}

@article{104.HOLME2004155,
title = {Structure and time evolution of an Internet dating community},
journal = {Social Networks},
volume = {26},
number = {2},
pages = {155-174},
year = {2004},
issn = {0378-8733},
doi = {https://doi.org/10.1016/j.socnet.2004.01.007},
url = {https://www.sciencedirect.com/science/article/pii/S0378873304000085},
author = {Petter Holme and Christofer R. Edling and Fredrik Liljeros}
}

@inproceedings{116.10.1145/5,
author = {Karimi, Fariba and Wagner, Claudia and Lemmerich, Florian and Jadidi, Mohsen and Strohmaier, Markus},
title = {Inferring Gender from Names on the Web: A Comparative Evaluation of Gender Detection Methods},
year = {2016},
isbn = {9781450341448},
publisher = {International World Wide Web Conferences Steering Committee},
address = {Republic and Canton of Geneva, CHE},
url = {https://doi.org/10.1145/2872518.2889385},
doi = {10.1145/2872518.2889385},
booktitle = {Proceedings of the 25th International Conference Companion on World Wide Web},
pages = {53–54},
numpages = {2},
location = {Montr\'{e}al, Qu\'{e}bec, Canada},
series = {WWW '16 Companion}
}

@misc{232.wang2021infoetwork,
      title={Information Access Equality on Network Generative Models}, 
      author={Xindi Wang and Onur Varol and Tina Eliassi-Rad},
      year={2022},
      journal={Applied Network Science}
}

@inproceedings{211.10.1145/3178876.3186140,
author = {Stoica, Ana-Andreea and Riederer, Christopher and Chaintreau, Augustin},
title = {Algorithmic Glass Ceiling in Social Networks: The effects of social recommendations on network diversity},
year = {2018},
isbn = {9781450356398},
publisher = {International World Wide Web Conferences Steering Committee},
address = {Republic and Canton of Geneva, CHE},
url = {https://doi.org/10.1145/3178876.3186140},
doi = {10.1145/3178876.3186140},
booktitle = {Proceedings of the 2018 World Wide Web Conference},
pages = {923–932},
numpages = {10},
location = {Lyon, France},
series = {WWW '18}
}

@article{193.0489-023-04928-3,
author = {Saxena, Akrati and Guti\'{e}rrez Bierbooms, Cristina and Pechenizkiy, Mykola},
title = {Fairness-aware fake news mitigation using counter information propagation},
year = {2023},
issue_date = {Nov 2023},
publisher = {Kluwer Academic Publishers},
address = {USA},
volume = {53},
number = {22},
issn = {0924-669X},
url = {https://doi.org/10.1007/s10489-023-04928-3},
doi = {10.1007/s10489-023-04928-3},
journal = {Applied Intelligence},
month = sep,
pages = {27483–27504},
numpages = {22}
}

@misc{snapnets,
                    author       = {Jure Leskovec and Andrej Krevl},
                    title        = {{SNAP Datasets}: {Stanford} Large Network Dataset Collection},
                    howpublished = {\url{http://snap.stanford.edu/data}},
                    month        = jun,
                    year         = 2014
}

@inproceedings{10.1145/956750.956769,
author = {Kempe, David and Kleinberg, Jon and Tardos, \'{E}va},
title = {Maximizing the spread of influence through a social network},
year = {2003},
isbn = {1581137370},
publisher = {Association for Computing Machinery},
address = {New York, NY, USA},
url = {https://doi.org/10.1145/956750.956769},
doi = {10.1145/956750.956769},
booktitle = {Proceedings of the Ninth ACM SIGKDD International Conference on Knowledge Discovery and Data Mining},
pages = {137–146},
numpages = {10},
location = {Washington, D.C.},
series = {KDD '03}
}

@article{2022_LMP_Bouyer_IM_Bridge,
title = {Influence maximization problem by leveraging the local traveling and node labeling method for discovering most influential nodes in social networks},
journal = {Physica A: Statistical Mechanics and its Applications},
volume = {592},
pages = {126841},
year = {2022},
issn = {0378-4371},
doi = {https://doi.org/10.1016/j.physa.2021.126841},
url = {https://www.sciencedirect.com/science/article/pii/S0378437121009973},
author = {Asgarali Bouyer and Hamid Ahmadi Beni}
}

@article{2020_Xareie_IM_NatureInsp_Wolf,
title = {Identification of influential users in social network using gray wolf optimization algorithm},
journal = {Expert Systems with Applications},
volume = {142},
pages = {112971},
year = {2020},
issn = {0957-4174},
doi = {https://doi.org/10.1016/j.eswa.2019.112971},
url = {https://www.sciencedirect.com/science/article/pii/S095741741930689X},
author = {Ahmad Zareie and Amir Sheikhahmadi and Mahdi Jalili}
}

@inproceedings{CELF_Leskovec_2007,
author = {Leskovec, Jure and Krause, Andreas and Guestrin, Carlos and Faloutsos, Christos and VanBriesen, Jeanne and Glance, Natalie},
title = {Cost-Effective Outbreak Detection in Networks},
year = {2007},
isbn = {9781595936097},
publisher = {Association for Computing Machinery},
address = {New York, NY, USA},
url = {https://doi.org/10.1145/1281192.1281239},
doi = {10.1145/1281192.1281239},
booktitle = {Proceedings of the 13th ACM SIGKDD International Conference on Knowledge Discovery and Data Mining},
pages = {420–429},
numpages = {10},
location = {San Jose, California, USA},
series = {KDD '07}
}

@INPROCEEDINGS{SIMPATH_2011_Goyal,
  author={Goyal, Amit and Lu, Wei and Lakshmanan, Laks V.S.},
  booktitle={2011 IEEE 11th International Conference on Data Mining}, 
  title={SIMPATH: An Efficient Algorithm for Influence Maximization under the Linear Threshold Model}, 
  year={2011},
  volume={},
  number={},
  pages={211-220},
  doi={10.1109/ICDM.2011.132}
}

@inproceedings{CELF_PP_2011_Goyal,
author = {Goyal, Amit and Lu, Wei and Lakshmanan, Laks V.S.},
title = {CELF++: Optimizing the Greedy Algorithm for Influence Maximization in Social Networks},
year = {2011},
isbn = {9781450306379},
publisher = {Association for Computing Machinery},
address = {New York, NY, USA},
url = {https://doi.org/10.1145/1963192.1963217},
doi = {10.1145/1963192.1963217},
booktitle = {Proceedings of the 20th International Conference Companion on World Wide Web},
pages = {47–48},
numpages = {2},
location = {Hyderabad, India},
series = {WWW '11}
}

@techreport{PAGERANK_1999,
number = {1999-66},
month = {November},
author = {Lawrence Page and Sergey Brin and Rajeev Motwani and Terry Winograd},
note = {Previous number = SIDL-WP-1999-0120},
title = {The PageRank Citation Ranking: Bringing Order to the Web.},
type = {Technical Report},
publisher = {Stanford InfoLab},
year = {1999},
institution = {Stanford InfoLab},
 url = {http://ilpubs.stanford.edu:8090/422/}
}

@inproceedings{DEGREE_DISCOUNT_Chen_2009,
author = {Chen, Wei and Wang, Yajun and Yang, Siyu},
title = {Efficient Influence Maximization in Social Networks},
year = {2009},
isbn = {9781605584959},
publisher = {Association for Computing Machinery},
address = {New York, NY, USA},
url = {https://doi.org/10.1145/1557019.1557047},
doi = {10.1145/1557019.1557047},
booktitle = {Proceedings of the 15th ACM SIGKDD International Conference on Knowledge Discovery and Data Mining},
pages = {199–208},
numpages = {10},
location = {Paris, France},
series = {KDD '09}
}

@article{MIA_PMIA_Wang_2012_Scalable,
	author = {Wang, Chi and Chen, Wei and Wang, Yajun},
	journal = {Data Mining and Knowledge Discovery},
	number = {3},
	year = {2012},
	month = {apr 1},
	pages = {545--576},
	publisher = {{Springer Science and Business Media LLC}},
	title = {Scalable influence maximization for independent cascade model in large-scale social networks},
	volume = {25},
}

@inproceedings{RIS2014Borgs,
author = {Christian Borgs and Michael Brautbar and Jennifer Chayes and Brendan Lucier},
title = {Maximizing Social Influence in Nearly Optimal Time},
booktitle = {Proceedings of the 2014  Annual ACM-SIAM Symposium on Discrete Algorithms (SODA)},
chapter = {},
pages = {946-957},
  year={2014},
doi = {10.1137/1.9781611973402.70}
}

@article{DYNAG_dynamic_top_k,
  title={Dynamic top-k influence maximization in social networks},
  author={Zhang, Binbin and Wang, Hao and U, Leong Hou},
  journal={GeoInformatica},
  pages={1--24},
  year={2020},
  publisher={Springer},
volume = {26},
issue={2},
issn  = {1573-7624},
url={https://doi.org/10.1007/s10707-020-00419-6},
doi={10.1007/s10707-020-00419-6}
}

@InProceedings{SP1M_SPM_2006,
author="Kimura, Masahiro
and Saito, Kazumi",
editor="F{\"u}rnkranz, Johannes
and Scheffer, Tobias
and Spiliopoulou, Myra",
title="Tractable Models for Information Diffusion in Social Networks",
booktitle="Knowledge Discovery in Databases: PKDD 2006",
year="2006",
publisher="Springer Berlin Heidelberg",
address="Berlin, Heidelberg",
pages="259--271",
isbn="978-3-540-46048-0"
}

@article{2018_TI_SC_CommunityBased_Based,
title = {TI-SC: top-k influential nodes selection based on community detection and scoring criteria in social networks},
journal = {Journal of Ambient Intelligence and Humanized Computing},
volume = {11},
issue={11},
pages = {119-130},
year = {2018},
issn = {1868-5145},
doi = {10.1007/s12652-020-01760-2},
url = {https://doi.org/10.1007/s12652-020-01760-2},
author = {Hamid Ahmadi Beni and Asgarali Bouyer}
}

@article{2019_C2IM_Shashank,
title = {C2IM: Community-based context-aware influence maximization in social networks},
journal = {Physica A: Statistical Mechanics and its Applications},
volume = {514},
pages = {796-818},
year = {2019},
issn = {0378-4371},
doi = {https://doi.org/10.1016/j.physa.2018.09.142},
url = {https://www.sciencedirect.com/science/article/pii/S0378437118312822},
author = {Shashank Sheshar Singh and Ajay Kumar and Kuldeep Singh and Bhaskar Biswas}
}

@article{2019_MIM2_IM_Shashank,
title = {MIM2: Multiple influence maximization across multiple social networks},
journal = {Physica A: Statistical Mechanics and its Applications},
volume = {526},
pages = {120902},
year = {2019},
issn = {0378-4371},
doi = {https://doi.org/10.1016/j.physa.2019.04.138},
url = {https://www.sciencedirect.com/science/article/pii/S037843711930500X},
author = {Shashank Sheshar Singh and Kuldeep Singh and Ajay Kumar and Bhaskar Biswas}
}

@article{2020_IM_NatureInsp_AntColony2,
  title={ACO-IM: maximizing influence in social networks using ant colony optimization},
  author={Singh, Shashank Sheshar and Singh, Kuldeep and Kumar, Ajay and Biswas, Bhaskar},
  journal={Soft Computing},
  volume={24},
  number={13},
  pages={10181--10203},
  year={2020},
  publisher={Springer},
  doi={https://doi.org/10.1007/s00500-019-04533-y}
}

@article{2020_Singh_IM_NatureInsp_IMSSO_Spider,
author = {Singh, Shashank Sheshar and Kumar, Ajay and Singh, Kuldeep and Biswas, Bhaskar},
title = {IM-SSO: Maximizing influence in social networks using social spider optimization},
journal = {Concurrency and Computation: Practice and Experience},
volume = {32},
number = {2},
pages = {e5421},
doi = {https://doi.org/10.1002/cpe.5421},
url = {https://onlinelibrary.wiley.com/doi/abs/10.1002/cpe.5421},
eprint = {https://onlinelibrary.wiley.com/doi/pdf/10.1002/cpe.5421},
note = {e5421 cpe.5421},
year = {2020}
}

@article{2019_Singh_IM_NatureInsp_LAPSO,
title = {LAPSO-IM: A learning-based influence maximization approach for social networks},
journal = {Applied Soft Computing},
volume = {82},
pages = {105554},
year = {2019},
issn = {1568-4946},
doi = {https://doi.org/10.1016/j.asoc.2019.105554},
url = {https://www.sciencedirect.com/science/article/pii/S1568494619303345},
author = {Shashank Sheshar Singh and Ajay Kumar and Kuldeep Singh and Bhaskar Biswas}
}

@article{2020_Weijia_PIMSN_SignedIM_FirstSignedUnderIC_ProbBasedOnPath_Based,
title = {A new algorithm for positive influence maximization in signed networks},
journal = {Information Sciences},
volume = {512},
pages = {1571-1591},
year = {2020},
issn = {0020-0255},
doi = {https://doi.org/10.1016/j.ins.2019.10.061},
url = {https://www.sciencedirect.com/science/article/pii/S0020025519310163},
author = {Weijia Ju and Ling Chen and Bin Li and Wei Liu and Jun Sheng and Yuwei Wang}
}

@article{2021_Guo_A_IMMA_SignedIM_UnwillingofInfNodes,
author = {Guo, Jianxiong and Wu, Weili},
title = {Adaptive Influence Maximization: If Influential Node Unwilling to Be the Seed},
year = {2021},
issue_date = {October 2021},
publisher = {Association for Computing Machinery},
address = {New York, NY, USA},
volume = {15},
number = {5},
issn = {1556-4681},
url = {https://doi.org/10.1145/3447396},
doi = {10.1145/3447396},
journal = {ACM Trans. Knowl. Discov. Data},
month = {may},
articleno = {84},
numpages = {23}
}

@ARTICLE{2013_FirstBudgeted_BIM_Nguyen_SOTA_BIM_WeightedDegree,
  author={Nguyen, Huy and Zheng, Rong},
  journal={IEEE Journal on Selected Areas in Communications}, 
  title={On Budgeted Influence Maximization in Social Networks}, 
  year={2013},
  volume={31},
  number={6},
  pages={1084-1094},
  doi={10.1109/JSAC.2013.130610}
}

@ARTICLE{2017_Nguyen_IM_BudgetedBIM_BCT_Cost_aware_Targeted_Viral_Marketing_CTVM_GeneralIMBIM_CostAndBudgetBIM,
  author={Nguyen, Hung T. and Thai, My T. and Dinh, Thang N.},
  journal={IEEE/ACM Transactions on Networking}, 
  title={A Billion-Scale Approximation Algorithm for Maximizing Benefit in Viral Marketing}, 
  year={2017},
  volume={25},
  number={4},
  pages={2419-2429},
  doi={10.1109/TNET.2017.2691544}
}

@book{2021_SocialNetworkAnalysis_Book_Tanmoy,
  author = {Chakraborty T},
  title = {Social Network Analysis},
  publisher = {Wiley India Pvt. Ltd.},
  year = {2021},
  isbn = {978-81-265-2007-7}
}

@article{2022_DynamicDGA_GeneticAlgorithmForDynamicNetwork,
title = {An improved influence maximization method for social networks based on genetic algorithm},
journal = {Physica A: Statistical Mechanics and its Applications},
volume = {586},
pages = {126480},
year = {2022},
issn = {0378-4371},
doi = {https://doi.org/10.1016/j.physa.2021.126480},
url = {https://www.sciencedirect.com/science/article/pii/S0378437121007536},
author = {Jalil {Jabari Lotf} and Mohammad {Abdollahi Azgomi} and Mohammad Reza {Ebrahimi Dishabi}}
}

@article{centrality_Freeman_1978,
title = {Centrality in social networks conceptual clarification},
journal = {Social Networks},
volume = {1},
number = {3},
pages = {215-239},
year = {1978},
issn = {0378-8733},
doi = {https://doi.org/10.1016/0378-8733(78)90021-7},
url = {https://www.sciencedirect.com/science/article/pii/0378873378900217},
author = {Linton C. Freeman}
}

@article{2020_Song_BudgetedIM_IMAGE_FrameworkForBiggernetwork,
author = {Bian, Song and Guo, Qintian and Wang, Sibo and Yu, Jeffrey Xu},
title = {Efficient algorithms for budgeted influence maximization on massive social networks},
year = {2020},
issue_date = {May 2020},
publisher = {VLDB Endowment},
volume = {13},
number = {9},
issn = {2150-8097},
url = {https://doi.org/10.14778/3397230.3397244},
doi = {10.14778/3397230.3397244},
journal = {Proc. VLDB Endow.},
month = {may},
pages = {1498–1510},
numpages = {13}
}

@article{2024_MineCSODynamicNetwork,
author = {Kumar Meena, Sunil and Sheshar Singh, Shashank and Singh, Kuldeep},
title = {Cuckoo Search Optimization-Based Influence Maximization in Dynamic Social Networks},
year = {2024},
issue_date = {November 2024},
publisher = {Association for Computing Machinery},
address = {New York, NY, USA},
volume = {18},
number = {4},
issn = {1559-1131},
url = {https://doi.org/10.1145/3690644},
doi = {10.1145/3690644},
journal = {ACM Trans. Web},
month = oct,
articleno = {49},
numpages = {25}
}

@ARTICLE{2020_Fuzzy1_CNL_UsingStructuralPropoerty_Logistic_Sparsification_Disadv_AI_Regularisation,
  author={Shen, Xiao and Mao, Sitong and Chung, Fu-lai},
  journal={IEEE Transactions on Fuzzy Systems}, 
  title={Cross-Network Learning With Fuzzy Labels for Seed Selection and Graph Sparsification in Influence Maximization}, 
  year={2020},
  volume={28},
  number={9},
  pages={2195-2208},
  doi={10.1109/TFUZZ.2019.2931272}
}

@article{2020_RNVGA_CompetitiveIM_Follower_based,
title = {Recurrent neural variational model for follower-based influence maximization},
journal = {Information Sciences},
volume = {528},
pages = {280-293},
year = {2020},
issn = {0020-0255},
doi = {https://doi.org/10.1016/j.ins.2020.04.023},
url = {https://www.sciencedirect.com/science/article/pii/S0020025520303297},
author = {Huimin Huang and Zaiqiao Meng and Shangsong Liang}
}

@article{DynamicIM_TTG_TTH_TopicBased_Min_2020,
  title={Topic based time-sensitive influence maximization in online social networks},
  author={Min, Huiyu and Cao, Jiuxin and Yuan, Tangfei and Liu, Bo},
  journal={World Wide Web},
  volume={23},
  number={3},
  pages={1831--1859},
  year={2020},
  publisher={Springer},
  doi={https://doi.org/10.1007/s11280-020-00792-0}
}

@inproceedings{2016_Sun_MIM_Greedy,
author = {Sun, Honglin and Gao, Xiaofeng and Chen, Guihai and Gu, Jian and Wang, Yongjian},
title = {Multiple Influence Maximization in Social Networks},
year = {2016},
isbn = {9781450341424},
publisher = {Association for Computing Machinery},
address = {New York, NY, USA},
url = {https://doi.org/10.1145/2857546.2857591},
doi = {10.1145/2857546.2857591},
booktitle = {Proceedings of the 10th International Conference on Ubiquitous Information Management and Communication},
articleno = {44},
numpages = {8},
location = {Danang, Viet Nam},
series = {IMCOM '16}
}

@article{2021_CompetitiveIM_GCW_GameTheoryBased_AwarenessModel,
author = {Tsaras, Dimitris and Trimponias, George and Ntaflos, Lefteris and Papadias, Dimitris},
title = {Collective influence maximization for multiple competing products with an awareness-to-influence model},
year = {2021},
issue_date = {March 2021},
publisher = {VLDB Endowment},
volume = {14},
number = {7},
issn = {2150-8097},
url = {https://doi.org/10.14778/3450980.3450981},
doi = {10.14778/3450980.3450981},
journal = {Proc. VLDB Endow.},
month = {mar},
pages = {1124–1136},
numpages = {13}
}

@article{2021_CompetitiveIM_DRHGA_DeepLearingCompetitiveProduct,
title = {Competitive and complementary influence maximization in social network: A follower’s perspective},
journal = {Knowledge-Based Systems},
volume = {213},
pages = {106600},
year = {2021},
issn = {0950-7051},
doi = {https://doi.org/10.1016/j.knosys.2020.106600},
url = {https://www.sciencedirect.com/science/article/pii/S0950705120307292},
author = {Huimin Huang and Zaiqiao Meng and Hong Shen}
}

@ARTICLE{2018_LocationIM_LIBM_HC_BlockingInf,
  author={Zhu, Wenlong and Yang, Wu and Xuan, Shichang and Man, Dapeng and Wang, Wei and Du, Xiaojiang},
  journal={IEEE Access}, 
  title={Location-Aware Influence Blocking Maximization in Social Networks}, 
  year={2018},
  volume={6},
  number={},
  pages={61462-61477},
  doi={10.1109/ACCESS.2018.2876141}
}

@inproceedings{2020_LocationIM_LazySower_IntroGeodemographic,
author = {Zhang, Kaichen and Zhou, Jingbo and Tao, Donglai and Karras, Panagiotis and Li, Qing and Xiong, Hui},
title = {Geodemographic Influence Maximization},
year = {2020},
isbn = {9781450379984},
publisher = {Association for Computing Machinery},
address = {New York, NY, USA},
url = {https://doi.org/10.1145/3394486.3403327},
doi = {10.1145/3394486.3403327},
booktitle = {Proceedings of the 26th ACM SIGKDD International Conference on Knowledge Discovery \& Data Mining},
pages = {2764–2774},
numpages = {11},
location = {Virtual Event, CA, USA},
series = {KDD '20}
}

@InProceedings{2019_LocationIM_DSSA_TheoreticalGurantee,
author="Li, Shasha
and Han, Kai
and Zhang, Jiahao",
editor="Gao, Honghao
and Wang, Xinheng
and Yin, Yuyu
and Iqbal, Muddesar",
title="Exploring Influence Maximization in Location-Based Social Networks",
booktitle="Collaborative Computing: Networking, Applications and Worksharing",
year="2019",
publisher="Springer International Publishing",
address="Cham",
pages="92--111",
isbn="978-3-030-12981-1"
}

@ARTICLE{2020_TopicIM_NEW_UserIntrestAndCharacterstics,
  author={Chen, Shiyu and Yin, Xiaochun and Cao, Qi and Li, Qianmu and Long, Huaqiu},
  journal={IEEE Access}, 
  title={Targeted Influence Maximization Based on Cloud Computing Over Big Data in Social Networks}, 
  year={2020},
  volume={8},
  number={},
  pages={45512-45522},
  doi={10.1109/ACCESS.2020.2978010}}

@ARTICLE{DynamicTopic_TCIC_CFDI_FeatureBased_2022_Qin,
  author={Qin, Xi and Zhong, Cheng and Yang, Qingshan},
  journal={IEEE Transactions on Network Science and Engineering}, 
  title={An Influence Maximization Algorithm Based on Community-Topic Features for Dynamic Social Networks}, 
  year={2022},
  volume={9},
  number={2},
  pages={608-621},
  doi={10.1109/TNSE.2021.3127921}
}

@article{2024_MultiplexIM_MPRPF_UsingMetaPaths,
title = {Heterogeneous network influence maximization algorithm based on multi-scale propagation strength and repulsive force of propagation field},
journal = {Knowledge-Based Systems},
volume = {291},
pages = {111580},
year = {2024},
issn = {0950-7051},
doi = {https://doi.org/10.1016/j.knosys.2024.111580},
url = {https://www.sciencedirect.com/science/article/pii/S0950705124002156},
author = {Chang Guo and Weimin Li and Jingchao Wang and Xiao Yu and Xiao Liu and Alex Munyole Luvembe and Can Wang and Qun Jin}
}

@ARTICLE{DynamicAndMultiNWIM_TONE_DMNIMP,
  author={Meng, Yanhong and Chen, Na and Yi, Yunhui and Wang, Shuanghong and Pei, Changxing},
  journal={IEEE Access}, 
  title={Research on the Dynamic Multisocial Networks Influence Maximization Problem Based on Common Users}, 
  year={2021},
  volume={9},
  number={},
  pages={127407-127419},
  doi={10.1109/ACCESS.2021.3112344}
}

@article{2020_MultiplexIM_DeepIM_PathBased,
title = {Influence maximization across heterogeneous interconnected networks based on deep learning},
journal = {Expert Systems with Applications},
volume = {140},
pages = {112905},
year = {2020},
issn = {0957-4174},
doi = {https://doi.org/10.1016/j.eswa.2019.112905},
url = {https://www.sciencedirect.com/science/article/pii/S0957417419306232},
author = {Mohammad Mehdi Keikha and Maseud Rahgozar and Masoud Asadpour and Mohammad Faghih Abdollahi}
}

@article{10.1109/TNET.2016.2563397,
author = {Tong, Guangmo and Wu, Weili and Tang, Shaojie and Du, Ding-Zhu},
title = {Adaptive Influence Maximization in Dynamic Social Networks},
year = {2017},
issue_date = {February 2017},
publisher = {IEEE Press},
volume = {25},
number = {1},
issn = {1063-6692},
url = {https://doi.org/10.1109/TNET.2016.2563397},
doi = {10.1109/TNET.2016.2563397},
journal = {IEEE/ACM Trans. Netw.},
month = feb,
pages = {112–125},
numpages = {14}
}

@article{10.1145/3772061,
author = {Singh, Shashank Sheshar and Srivastava, Vishal and Kishor, Avadh and Barua, Jayendra and Kumar, Neeraj},
title = {Influence Maximization on Signed Social Networks: A Survey},
year = {2025},
issue_date = {March 2026},
publisher = {Association for Computing Machinery},
address = {New York, NY, USA},
volume = {58},
number = {4},
issn = {0360-0300},
url = {https://doi.org/10.1145/3772061},
doi = {10.1145/3772061},
journal = {ACM Comput. Surv.},
month = nov,
articleno = {106},
numpages = {32}
}

@article{XIAO2026131192,
title = {Fairness-Aware Influence Blocking Maximization: A Centrality-Enhanced Adversarial Graph Embedding Framework},
journal = {Expert Systems with Applications},
pages = {131192},
year = {2026},
issn = {0957-4174},
doi = {https://doi.org/10.1016/j.eswa.2026.131192},
url = {https://www.sciencedirect.com/science/article/pii/S0957417426001065},
author = {Yu Xiao and Lan Yang and Ding Liu and Zhiwu Li}
}

@article{Meena.10.1145/3779058,
author = {Meena, Sunil Kumar and Singh, Shashank Sheshar and Singh, Kuldeep},
title = {Exploring Influence Maximization: State-of-the-Art Methods, Taxonomies, and Trends},
year = {2026},
issue_date = {February 2026},
publisher = {Association for Computing Machinery},
address = {New York, NY, USA},
volume = {20},
number = {2},
issn = {1556-4681},
url = {https://doi.org/10.1145/3779058},
doi = {10.1145/3779058},
journal = {ACM Trans. Knowl. Discov. Data},
month = jan,
articleno = {23},
numpages = {50}
}

@Article{StructuralCasualModelsSCMsForCasualFairness,
AUTHOR = {Lin, Han and Liu, Wenhe},
TITLE = {Symmetry-Aware Causal-Inference-Driven Web Performance Modeling: A Structure-Aware Framework for Predictive Analysis and Actionable Optimization},
JOURNAL = {Symmetry},
VOLUME = {17},
YEAR = {2025},
NUMBER = {12},
ARTICLE-NUMBER = {2058},
URL = {https://www.mdpi.com/2073-8994/17/12/2058},
ISSN = {2073-8994},
DOI = {10.3390/sym17122058}
}

@INPROCEEDINGS{2021_MultiObjectiveSubmodularOptimization_IM,
  author={Fu, Xiaoyun and Bhatt, Rishabh Rajendra and Basu, Samik and Pavan, A.},
  booktitle={2021 IEEE International Conference on Big Data (Big Data)}, 
  title={Multi-Objective Submodular Optimization with Approximate Oracles and Influence Maximization}, 
  year={2021},
  volume={},
  number={},
  pages={328-334},
  doi={10.1109/BigData52589.2021.9671756}}

@inproceedings{SensitiveAttributeOrthogonal,
author = {Sarhan, Mhd Hasan and Navab, Nassir and Eslami, Abouzar and Albarqouni, Shadi},
title = {Fairness by Learning Orthogonal Disentangled Representations},
year = {2020},
isbn = {978-3-030-58525-9},
publisher = {Springer-Verlag},
address = {Berlin, Heidelberg},
url = {https://doi.org/10.1007/978-3-030-58526-6_44},
doi = {10.1007/978-3-030-58526-6_44},
booktitle = {Computer Vision – ECCV 2020: 16th European Conference, Glasgow, UK, August 23–28, 2020, Proceedings, Part XXIX},
pages = {746–761},
numpages = {16},
location = {Glasgow, United Kingdom}
}

@ARTICLE{FairGraphMining_ASurvey,
  author={Dong, Yushun and Ma, Jing and Wang, Song and Chen, Chen and Li, Jundong},
  journal={IEEE Transactions on Knowledge and Data Engineering}, 
  title={Fairness in Graph Mining: A Survey}, 
  year={2023},
  volume={35},
  number={10},
  pages={10583-10602},
  doi={10.1109/TKDE.2023.3265598}}

@article{ML_SurveyOnBiasAndFairness,
author = {Mehrabi, Ninareh and Morstatter, Fred and Saxena, Nripsuta and Lerman, Kristina and Galstyan, Aram},
title = {A Survey on Bias and Fairness in Machine Learning},
year = {2021},
issue_date = {July 2022},
publisher = {Association for Computing Machinery},
address = {New York, NY, USA},
volume = {54},
number = {6},
issn = {0360-0300},
url = {https://doi.org/10.1145/3457607},
doi = {10.1145/3457607},
journal = {ACM Comput. Surv.},
month = jul,
articleno = {115},
numpages = {35}
}

@inproceedings{IndvidualFairness,
author = {Dwork, Cynthia and Hardt, Moritz and Pitassi, Toniann and Reingold, Omer and Zemel, Richard},
title = {Fairness through awareness},
year = {2012},
isbn = {9781450311151},
publisher = {Association for Computing Machinery},
address = {New York, NY, USA},
url = {https://doi.org/10.1145/2090236.2090255},
doi = {10.1145/2090236.2090255},
booktitle = {Proceedings of the 3rd Innovations in Theoretical Computer Science Conference},
pages = {214–226},
numpages = {13},
location = {Cambridge, Massachusetts},
series = {ITCS '12}
}

@inproceedings{MeasureIndvidualFairness,
author = {Zemel, Richard and Wu, Yu and Swersky, Kevin and Pitassi, Toniann and Dwork, Cynthia},
title = {Learning fair representations},
year = {2013},
publisher = {JMLR.org},
booktitle = {Proceedings of the 30th International Conference on International Conference on Machine Learning - Volume 28},
pages = {III–325–III–333},
location = {Atlanta, GA, USA},
series = {ICML'13}
}

@inproceedings{DistributionalFairness,
author = {Dong, Yushun and Liu, Ninghao and Jalaian, Brian and Li, Jundong},
title = {EDITS: Modeling and Mitigating Data Bias for Graph Neural Networks},
year = {2022},
isbn = {9781450390965},
publisher = {Association for Computing Machinery},
address = {New York, NY, USA},
url = {https://doi.org/10.1145/3485447.3512173},
doi = {10.1145/3485447.3512173},
booktitle = {Proceedings of the ACM Web Conference 2022},
pages = {1259–1269},
numpages = {11},
location = {Virtual Event, Lyon, France},
series = {WWW '22}
}

@article{MultipleFairness_IndvidualAndGroup,
author = {Lahoti, Preethi and Gummadi, Krishna P. and Weikum, Gerhard},
title = {Operationalizing individual fairness with pairwise fair representations},
year = {2019},
issue_date = {December 2019},
publisher = {VLDB Endowment},
volume = {13},
number = {4},
issn = {2150-8097},
url = {https://doi.org/10.14778/3372716.3372723},
doi = {10.14778/3372716.3372723},
journal = {Proc. VLDB Endow.},
month = dec,
pages = {506–518},
numpages = {13}
}

@inproceedings{MultipleFairness_TradeOff1,
author = {Burkholder, Kai and Kwock, Kenny and Xu, Yuesheng and Liu, Jiaxin and Chen, Chao and Xie, Sihong},
title = {Certification and Trade-off of Multiple Fairness Criteria in Graph-based Spam Detection},
year = {2021},
isbn = {9781450384469},
publisher = {Association for Computing Machinery},
address = {New York, NY, USA},
url = {https://doi.org/10.1145/3459637.3482325},
doi = {10.1145/3459637.3482325},
booktitle = {Proceedings of the 30th ACM International Conference on Information \& Knowledge Management},
pages = {130–139},
numpages = {10},
location = {Virtual Event, Queensland, Australia},
series = {CIKM '21}
}

@article{FairnessWithRegularisation, title={Fair Representation Learning for Heterogeneous Information Networks}, volume={15}, url={https://ojs.aaai.org/index.php/ICWSM/article/view/18111}, DOI={10.1609/icwsm.v15i1.18111}, abstractNote={Recently, much attention has been paid to the societal impact of AI, especially concerns regarding its fairness. A growing body of research has identified unfair AI systems and proposed methods to debias them, yet many challenges remain. Representation learning methods for Heterogeneous Information Networks (HINs), fundamental building blocks used in complex network mining, have socially consequential applications such as automated career counseling, but there have been few attempts to ensure that it will not encode or amplify harmful biases, e.g. sexism in the job market. To address this gap, we propose a comprehensive set of de-biasing methods for fair HINs representation learning, including sampling-based, projection-based, and graph neural networks (GNNs)-based techniques. We systematically study the behavior of these algorithms, especially their capability in balancing the trade-off between fairness and prediction accuracy. We evaluate the performance of the proposed methods in an automated career counseling application where we mitigate gender bias in career recommendation. Based on the evaluation results on two datasets, we identify the most effective fair HINs representation learning techniques under different conditions.}, number={1}, journal={Proceedings of the International AAAI Conference on Web and Social Media}, author={Zeng, Ziqian and Islam, Rashidul and Keya, Kamrun Naher and Foulds, James and Song, Yangqiu and Pan, Shimei}, year={2021}, month={May}, pages={877-887} }

@inproceedings{saxena2015understanding,
  title={Understanding spreading patterns on social networks based on network topology},
  author={Saxena, Akrati and Iyengar, SRS and Gupta, Yayati},
  booktitle={Proceedings of the 2015 IEEE/ACM International Conference on Advances in Social Networks Analysis and Mining 2015},
  pages={1616--1617},
  year={2015}
}

@book{anderson1992infectious,
  title={Infectious diseases of humans: dynamics and control},
  author={Anderson, Roy M and May, Robert M},
  year={1992},
  publisher={Oxford university press}
}

@article{allen1994some,
  title={Some discrete-time SI, SIR, and SIS epidemic models},
  author={Allen, Linda JS},
  journal={Mathematical biosciences},
  volume={124},
  number={1},
  pages={83--105},
  year={1994},
  publisher={Elsevier}
}

@article{hethcote1995sis,
  title={An SIS epidemic model with variable population size and a delay},
  author={Hethcote, Herbert W and Van den Driessche, P},
  journal={Journal of mathematical biology},
  volume={34},
  number={2},
  pages={177--194},
  year={1995},
  publisher={Springer}
}

@article{evans2018opinion,
  title={Opinion formation on dynamic networks: identifying conditions for the emergence of partisan echo chambers},
  author={Evans, Tucker and Fu, Feng},
  journal={Royal Society open science},
  volume={5},
  number={10},
  pages={181122},
  year={2018},
  publisher={The Royal Society}
}

@article{he2022reinforcement,
  title={Reinforcement Learning-Based Rumor Blocking Approach in Directed Social Networks},
  author={He, Qiang and Lv, Yingjie and Wang, Xingwei and Huang, Min and Cai, Yuliang},
  journal={IEEE Systems Journal},
  year={2022},
  publisher={IEEE}
}

@article{li2022piano,
  title={PIANO: Influence Maximization Meets Deep Reinforcement Learning},
  author={Li, Hui and Xu, Mengting and Bhowmick, Sourav S and Rayhan, Joty Shafiq and Sun, Changsheng and Cui, Jiangtao},
  journal={IEEE Transactions on Computational Social Systems},
  year={2022},
  publisher={IEEE}
}

@article{saxena2025dq4fairim,
  title={DQ4FairIM: Fairness-aware Influence Maximization using Deep Reinforcement Learning},
  author={Saxena, Akrati and Yadav, Harshith Kumar and Rutten, Bart and Jha, Shashi Shekhar},
  journal={arXiv preprint arXiv:2512.00545},
  year={2025}
}

@article{saxena2025homophily,
  title={Homophily in complex networks: Measures, models, and applications},
  author={Saxena, Akrati and Kumar, Gaurav and Meena, Chandrakala},
  journal={arXiv preprint arXiv:2509.18289},
  year={2025}
}

@inproceedings{saxena2020mitigating,
  title={Mitigating Misinformation in Online Social Network with Top-k Debunkers and Evolving User Opinions},
  author={Saxena, Akrati and Hsu, Wynne and Lee, Mong Li and Leong Chieu, Hai and Ng, Lynette and Teow, Loo Nin},
  booktitle={Companion Proceedings of the Web Conference 2020},
  pages={363--370},
  year={2020}
}

@article{10.1145/3652854,
author = {Yang, Hao and Wu, Xian and Qiu, Zhaopeng and Zheng, Yefeng and Chen, Xu},
title = {Distributional Fairness-aware Recommendation},
year = {2024},
issue_date = {September 2024},
publisher = {Association for Computing Machinery},
address = {New York, NY, USA},
volume = {42},
number = {5},
issn = {1046-8188},
url = {https://doi.org/10.1145/3652854},
doi = {10.1145/3652854},
journal = {ACM Trans. Inf. Syst.},
month = apr,
articleno = {131},
numpages = {28}
}

@inproceedings{kumar2018community,
  title={Community interaction and conflict on the web},
  author={Kumar, Srijan and Hamilton, William L and Leskovec, Jure and Jurafsky, Dan},
  booktitle={Proceedings of the 2018 World Wide Web Conference on World Wide Web},
  pages={933--943},
  year={2018},
  organization={International World Wide Web Conferences Steering Committee}
}

@inproceedings{kumar2016edge,
  title={Edge weight prediction in weighted signed networks},
  author={Kumar, Srijan and Spezzano, Francesca and Subrahmanian, VS and Faloutsos, Christos},
  booktitle={Data Mining (ICDM), 2016 IEEE 16th International Conference on},
  pages={221--230},
  year={2016},
  organization={IEEE}
}

@article{LIU2024101466,
title = {Large-scale evolutionary optimization: A review and comparative study},
journal = {Swarm and Evolutionary Computation},
volume = {85},
pages = {101466},
year = {2024},
issn = {2210-6502},
doi = {https://doi.org/10.1016/j.swevo.2023.101466},
author = {Jing Liu and Ruhul Sarker and Saber Elsayed and Daryl Essam and Nurhadi Siswanto}
}

@article{10.1007/s00224-021-10066-5,
author = {Bazgan, Cristina and Ruzika, Stefan and Thielen, Clemens and Vanderpooten, Daniel},
title = {The Power of the Weighted Sum Scalarization for Approximating Multiobjective Optimization Problems},
year = {2022},
issue_date = {Feb 2022},
publisher = {Springer-Verlag},
address = {Berlin, Heidelberg},
volume = {66},
number = {1},
issn = {1432-4350},
url = {https://doi.org/10.1007/s00224-021-10066-5},
doi = {10.1007/s00224-021-10066-5},
journal = {Theor. Comp. Sys.},
month = feb,
pages = {395–415},
numpages = {21}
}

@article{2008_CommunityDetection_FirstBaseLouvain,
doi = {10.1088/1742-5468/2008/10/P10008},
url = {https://dx.doi.org/10.1088/1742-5468/2008/10/P10008},
year = {2008},
month = {oct},
publisher = {},
volume = {2008},
number = {10},
pages = {P10008},
author = {Vincent D Blondel and Jean-Loup Guillaume and Renaud Lambiotte and Etienne Lefebvre},
title = {Fast unfolding of communities in large networks},
journal = {Journal of Statistical Mechanics: Theory and Experiment}
}

@article{DiversifiedEffiecient,
title = {Efficient diversified influence maximization with adaptive policies},
journal = {Knowledge-Based Systems},
volume = {213},
pages = {106692},
year = {2021},
issn = {0950-7051},
doi = {https://doi.org/10.1016/j.knosys.2020.106692},
url = {https://www.sciencedirect.com/science/article/pii/S0950705120308212},
author = {Can Wang and Qihao Shi and Weizhao Xian and Yan Feng and Chun Chen}
}

@article{khalil2017learning,
  title={Learning combinatorial optimization algorithms over graphs},
  author={Khalil, Elias and Dai, Hanjun and Zhang, Yuyu and Dilkina, Bistra and Song, Le},
  journal={Advances in neural information processing systems},
  volume={30},
  year={2017}
}

@inproceedings{moens2022reproducibility,
  title={Reproducibility study of the Fairness-enhanced Node Representation Learning},
  author={Moens, Gijs Joppe and de Witte, Job and G{\"o}bel, Tobias Pieter and van den Oever, Meggie Anna Antwanette},
  booktitle={ML Reproducibility Challenge 2022}
}

@inproceedings{khajehnejad2022crosswalk,
  title={Crosswalk: Fairness-enhanced node representation learning},
  author={Khajehnejad, Ahmad and Khajehnejad, Moein and Babaei, Mahmoudreza and Gummadi, Krishna P and Weller, Adrian and Mirzasoleiman, Baharan},
  booktitle={Proceedings of the AAAI Conference on Artificial Intelligence},
  volume={36},
  number={11},
  pages={11963--11970},
  year={2022}
}

@article{cui2018survey,
  title={A survey on network embedding},
  author={Cui, Peng and Wang, Xiao and Pei, Jian and Zhu, Wenwu},
  journal={IEEE transactions on knowledge and data engineering},
  volume={31},
  number={5},
  pages={833--852},
  year={2018},
  publisher={IEEE}
}

@article{hu2024imne,
  title={IMNE: Maximizing influence through deep learning-based node embedding in social network},
  author={Hu, Qian and Jiang, Jiatao and Xu, Hongfeng and Kassim, Murizah},
  journal={Swarm and Evolutionary Computation},
  volume={88},
  pages={101609},
  year={2024},
  publisher={Elsevier}
}

@article{kumar2022influence,
  title={Influence maximization in social networks using graph embedding and graph neural network},
  author={Kumar, Sanjay and Mallik, Abhishek and Khetarpal, Anavi and Panda, Bhawani Sankar},
  journal={Information Sciences},
  volume={607},
  pages={1617--1636},
  year={2022},
  publisher={Elsevier}
}

@inproceedings{EfficientFairAlgo1,
 author = {Rui, Xiaobin and Wang, Zhixiao and Zhao, Jiayu and Sun, Lichao and Chen, Wei},
 booktitle = {Advances in Neural Information Processing Systems},
 editor = {A. Oh and T. Naumann and A. Globerson and K. Saenko and M. Hardt and S. Levine},
 pages = {66675--66691},
 publisher = {Curran Associates, Inc.},
 title = {Scalable Fair Influence Maximization},
 url = {https://proceedings.neurips.cc/paper_files/paper/2023/file/d242dafdb2c5407ae420bc54c9325fdf-Paper-Conference.pdf},
 volume = {36},
 year = {2023}
}

@misc{EfficientFairAlgo2,
      title={Efficient Approximation Algorithms for Fair Influence Maximization under Maximin Constraint}, 
      author={Xiaobin Rui and Qiangpeng Fang and Chen Peng and Jilong Shi and Zhixiao Wang and Wei Chen},
      year={2026},
      eprint={2509.26579},
      archivePrefix={arXiv},
      primaryClass={cs.DS},
      url={https://arxiv.org/abs/2509.26579}, 
}
